\documentclass[paper,traditabstract]{aa} 
\usepackage{graphicx}
\usepackage{txfonts}
\usepackage{natbib}
\def\revised{}
\def\rev{}

\usepackage{longtable}
\usepackage{rotating}
\usepackage{pdflscape}
\usepackage{multirow}

\begin{document}
    \title{Formation and evolution of   interstellar filaments
    }
   \subtitle{Hints from  velocity dispersion measurements\thanks{Based on observations carried out with the IRAM 30m telescope. IRAM is supported by INSU/CNRS (France), MPG (Germany), and IGN (Spain).} 
   } 

   \author{
     D. Arzoumanian\inst{1,2}
      \and
            Ph. Andr\'e\inst{1}
                 \and
            N. Peretto\inst{1}
         \and
         V. K\"onyves\inst{1,2}         
            }

   \institute{Laboratoire AIM, CEA/DSM--CNRS--Universit\'e Paris Diderot, IRFU/Service d'Astrophysique, C.E.A. Saclay,
              Orme des Merisiers, 91191 Gif-sur-Yvette, France
    \and
    IAS, CNRS (UMR 8617), Universit\'e Paris-Sud, B\^atiment 121, 91400 Orsay, France\\
              \email{doris.arzoumanian@ias.u-psud.fr, pandre@cea.fr}\\ \\
              Received -- / Accepted --
                  }
   \date{}

   \abstract{
{We investigate the gas velocity dispersions of a sample of filaments recently detected as part of  the $Herschel$ Gould Belt Survey in the IC5146, Aquila, and Polaris interstellar  clouds. 
To measure these velocity dispersions, we use $^{13}$CO, C$^{18}$O,  and N$_2$H$^+$ line observations obtained with the 
IRAM 30m telescope. Correlating our velocity dispersion measurements with the filament column densities derived from $Herschel$ data, 
we show that interstellar filaments can be divided into two regimes: thermally subcritical filaments, which have transonic 
velocity dispersions ($c_s \la \sigma_{tot} < 2\, c_s $) 
independent of column density, and are gravitationally unbound; 
and thermally supercritical filaments, which have higher velocity dispersions scaling roughly 
as the square root of column density ($ \sigma_{\rm tot} \propto {\Sigma_{0}}^{0.5}$), and are self-gravitating. 
The higher velocity dispersions of supercritical filaments may not directly arise from supersonic interstellar turbulence but may be driven by gravitational contraction/accretion. 
Based on our observational results, 
we propose an evolutionary scenario whereby supercritical filaments undergo gravitational contraction and 
increase in mass per unit length through accretion of  background material while remaining in rough virial balance. 
We further suggest that this accretion process allows supercritical filaments to keep {\revised their approximately constant inner widths ($\sim 0.1$~pc) 
 while contracting}.
 }

             }

   \maketitle
%

\section{Introduction}

\nocite{Palmeirim2013}

Interstellar filaments have recently received special attention, thanks to the high quality and dynamic range of  $Herschel$\footnote{$Herschel$ is an ESA space observatory 
with science instruments provided by European-led Principal Investigator consortia and with important participation from NASA \citep{Pilbratt2010}.}  observations \citep{Andre2010,Men'shchikov2010,molinari2010,Henning2010,hill2011}. 
The submillimeter dust continuum images of nearby molecular clouds (MCs) taken with the SPIRE \citep{Griffin2010} and PACS \citep{Poglitsch2010} cameras on board
$Herschel$ provide key information on both dense cores on small scales ($<$~0.1~pc) and the filamentary structure of the parent clouds on large 
scales ($>$~1~pc),  making it possible to investigate  the physical connection between these two components of the cold interstellar medium (ISM) and to draw 
a more complete picture of star formation.

Already before $Herschel$, filamentary structures were  known to be present in MCs  {\revised \citep[e.g.][]{Schneider1979,Myers2009}} and  
appeared to be  easily produced by any numerical simulation of MC evolution that  includes hydrodynamic (HD) or 
magneto-hydrodynamic (MHD)  turbulence \citep[e.g.,][]{Federrath2010,Hennebelle2008,MacLow2004}. 
{\revised Individual interstellar  filaments have also been recently   studied with molecular line observations \citep[e.g.,][]{Pineda2011,Hacar2011,Miettinen2012,
Bourke2012, Li2012}.} 
$Herschel$  observations 
now demonstrate that these filaments are truly ubiquitous in MCs  
and have a direct link with the  formation process of prestellar cores  \citep{Andre2010,Palmeirim2013}.

Filaments are detected with $Herschel$ in both non-star-forming regions, such as the Polaris translucent cloud \citep{Men'shchikov2010, Miville2010, Ward-Thompson2010}, 
and in star-forming regions \citep{Konyves2010,Bontemps2010}, 
where they are associated with the presence of prestellar cores and protostars  \citep{Andre2010}.  
These findings support the view that filament formation precedes any star forming activity in MCs. 

{\revised In a previous work \citep{Arzoumanian2011}, we characterized the physical properties of the  filaments detected in three regions observed as 
part of the $Herschel$ Gould Belt survey   \citep[HGBS --][]{Andre2010}  
and found  that all  filaments share  a roughly constant  inner width of $\sim$0.1~pc   
regardless of their central column density and environment. This result has been  
confirmed in other fields of  the HGBS \citep{Peretto2012,Palmeirim2013}. }
Observationally, only  thermally supercritical filaments for which  the mass per unit length exceeds  the  critical value ($M_{\rm line} > M_{\rm line,crit} $)show evidence of prestellar cores and the presence of star formation activity, 
whereas thermally subcritical filaments ($M_{\rm line} < M_{\rm line,crit} $) are generally devoid of  $Herschel$ 
prestellar cores and protostars  \citep{Andre2010}. The critical mass per unit length, $M_{\rm line,crit}~=~2 c^{2}_{\rm s}/G$ 
(where $c_{\rm s}$ is the istothermal sound speed and $G$ is the gravitational constant),  
is the critical value required for a filament to be 
gravitationally unstable to radial contraction and fragmentation along its length  \citep{Inutsuka1997}. 
Remarkably, this critical line mass $M_{\rm line,crit} $ only depends on gas temperature \citep{Ostriker1964}. 
It is approximately equal to $\sim 16~M_{\odot}$/pc for gas filaments at T= 10~K, corresponding to 
$c_{\rm s} \sim 0.2$~km/s.
{\revised Assuming that interstellar filaments have Gaussian radial column density profiles, an estimate of the mass per unit length is given by 
$M_{\rm line}~\approx~\Sigma_{0} \times W_{\rm fil}$ where $W_{\rm fil}$ is  the typical filament width  (see Appendix~A of Andr\'e et al. 2010 
and Arzoumanian et al. 2011)  and $\Sigma_{0}= \mu{\rm m}_{\rm H}N^{0}_{{\rm H}_{2}}$ is the central gas surface density of the filament. %
For a typical filament width of $\sim$ 0.1~pc, the theoretical value of $M_{\rm line,crit} \sim$ 16~M$_\odot$/pc (for $T \approx 10$~K)  
corresponds %
to a central column density $ N^{0}_{{\rm H}_{2}}  \sim 8\times10^{21}$~cm$^{-2}$ (or visual extinction $A_V^0 \sim 8$).  
To first order, therefore, interstellar filaments may be divided into thermally supercritical and thermally subcritical filaments depending 
on whether their central column density $ N^{0}_{{\rm H}_{2}} $ 
is higher or lower than $\sim 8\times10^{21}$~cm$^{-2}$, respectively.
Thermally supercritical filaments, with average column densities $\gtrsim 8\times10^{21}$~cm$^{-2}$, are expected to be {\it globally} unstable 
to radial gravitational collapse and fragmentation into prestellar cores along their lengths. The prestellar cores formed in this way are themselves 
expected to collapse {\it locally} into protostars. 
Interestingly, the critical column density $ \sim 8\times10^{21}$~cm$^{-2}$ of interstellar filaments is very close 
to the star formation threshold at a background extinction $A_V \sim \,$6--9 above which the bulk of prestellar cores and young stellar objects 
are observed in nearby clouds  \citep[e.g.,][]{Johnstone2004,Goldsmith2008,Heiderman2010,Lada2010,Andre2010,Andre2011}.  }

The  
{\revised results summarized above on the properties of  interstellar filaments } were made possible thanks to the sensitivity and resolution of  the HGBS 
observations, which allowed us to  detect  structures down to A$_{\rm V}~\sim$~0.1 
and to resolve individual filaments in nearby regions (d$<$0.5~kpc), down to a typical Jeans length of $\sim$ 0.02~pc. 
While the $Herschel$ images already are  a powerful  tool to assess  the importance of  different physical processes involved in the formation and evolution of  interstellar filaments,
they do not provide any information on the underlying kinematics of the gas.

In the present  work, we complement our first results obtained from $Herschel$ imaging  data with molecular line observations to gain insight into  
the velocity dispersion of the interstellar filaments and reinforce our understanding of filament formation and evolution.
 We present  $^{13}$CO, C$^{18}$O, 
and N$_2$H$^+$  line observations, taken with the IRAM 30m-telescope, toward a sample of   filaments   detected with $Herschel$ in  the IC5146,   Aquila Rift, and Polaris flare regions. %
In Sect.  2 we present the observations and the data reduction. Sect.  3 and 4 summarize the data analysis, Sect. 5  presents the results and  Sect. 6 discusses their implications in the proposed  scenario for interstellar filament formation and evolution.

\begin{figure}
    \begin{minipage}{1\linewidth}
       \resizebox{\hsize}{!}{\includegraphics[angle=0]{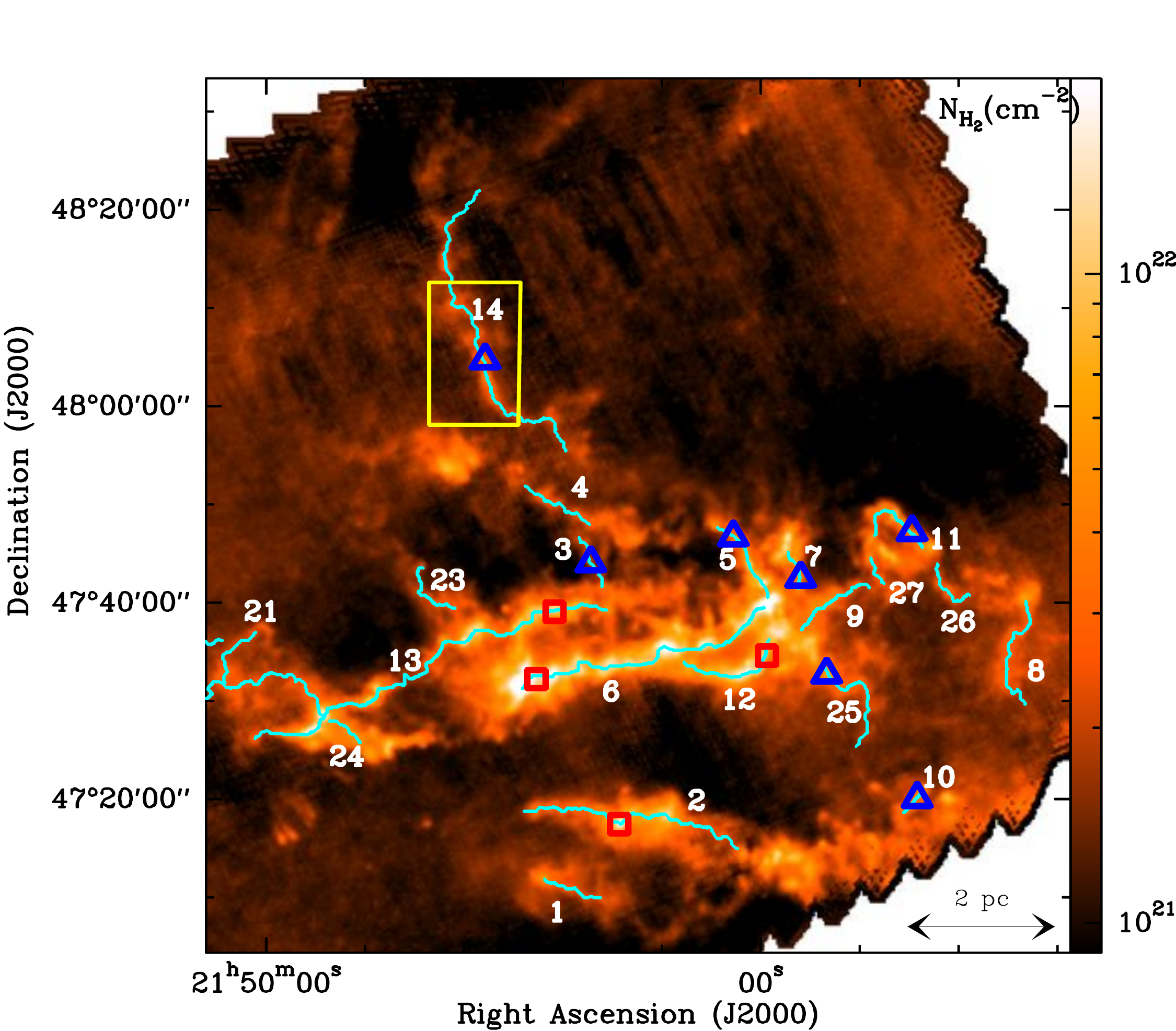}}
       \end{minipage}   
          \caption{$Herschel$ column density map  of the northern and southern streamers of IC5146 with the network of filaments highlighted in blue \citep[adapted from Fig.~3b of ][] {Arzoumanian2011}. The positions of the observed spectra are plotted as red squares and  blue triangles for N$_{2}$H$^{+}$ and C$^{18}$O, respectively (the corresponding spectra of the positions observed for filament 2 and filament 14 are shown in Fig.~\ref{IC5146_spectra}). The yellow rectangle shows the subcritical filament 14 which was mapped in   C$^{18}$O(2-1) and $^{13}$CO(2-1); cf. Fig.~\ref{SubColdensCO}.%
                    } 
  \label{IC5146SpectraPos}
   \end{figure}


\section{IRAM observations and data reduction}

Molecular line observations of  a sample of filaments in IC5146, Aquila, and Polaris  were carried on at the IRAM-30m telescope at Pico Veleta,  Spain,  during two observing runs in March  and August 2011.  

In March 2011, we observed  the   C$^{18}$O(2-1) and $^{13}$CO(2-1)  spectral lines with HERA   \citep[the  HEterodyne Receiver Array, ][]{Schuster2004}  and  the N$_{2}$H$^{+}$(1-0) line  with EMIR \citep[the {\it Eight MIxer Receiver}, a multi-band mm-wave receiver,][]{Carter2012} toward  
 the   filaments of IC5146. In addition to single-position spectra taken in the on-off  observing mode, we mapped  filament 14  in IC5146 (yellow rectangle in Fig.~\ref{IC5146SpectraPos}) in the C$^{18}$O(2-1) and $^{13}$CO(2-1)  transitions using the on-the-fly mapping mode with HERA. %
The  Aquila rift  and the Polaris clouds were observed in August 2011 in the  
C$^{18}$O(1-0) and N$_{2}$H$^{+}$(1-0)  transitions with EMIR. %

The spatial resolutions of the observations correspond to the 12$\arcsec$, 23$\arcsec$ and 28$\arcsec$ FWHM beamwidths of the 30~m antenna at 219.6~GHz [C$^{18}$O(2-1)], 109.8~GHz [C$^{18}$O(1-0)], and 93.2~GHz [N$_{2}\rm H^{+}$(1-0)], respectively. 
 We used the VESPA autocorrelator  as backend with a spectral resolution of 20~KHz. This spectral resolution translates to velocity resolutions of 
0.027, 0.055, and 0.064~km/s at the frequencies of the C$^{18}$O(2-1), C$^{18}$O(1-0), and N$_{2}\rm H^{+}$(1-0) transitions, respectively. 
 
We used the  frequency-switching mode for the  CO observations and the %
position-switching mode for the  N$_{2}$H$^{+}$ observations,  using a reference position offseted  by $\sim10\arcmin$--$15\arcmin$ from each target   position. The off position was  selected from the $Herschel$ images   and the lack of  N$_{2}$H$^{+}$  emission was checked  using observations in the  frequency-switching mode.  
During the observations, calibration was achieved by measuring the emission from the sky, an ambient load and a cold load  every $\sim 15$~minutes. The telescope pointing was checked and adjusted every  $\sim 2$ hours and  the pointing accuracy  was found to be better than 3$\arcsec$. The typical rms of each single position (on-off mode) spectrum is $\sim$0.05~K (in T$_{\rm A}^{*}$ scale). All of the data were reduced with the GILDAS/CLASS  software package (the Grenoble Image and Line Data Analysis System, a software provided and actively developed by IRAM -- http://www.iram.fr/IRAMFR/ GILDAS).

\subsection{Filament sample}
A total number  of 100 positions toward  {\revised 70 filaments  or segments of filaments\footnote{Due to the presence of apparent intersections of  filamentary structures  in the plane of the sky, there is sometimes some ambiguity  in the definition of an ``individual filament".}} were observed in  the three regions (IC5146, Aquila, Polaris). 
Most of the  densest filaments,  {\revised with $N_{\rm H_{2}} \gtrsim 1 \times 10^{22}$~cm$^{-2}$}, were  detected in N$_{2}$H$^{+}$ (1-0), which is a low optical depth  tracer of dense gas,  as well as in C$^{18}$O and $^{13}$CO. The  lower column density  filaments  were  only  detected in  C$^{18}$O and $^{13}$CO, {\revised while the faintest filaments were not detected}\footnote{ {\revised A few filaments observed in IC5146 (5 filaments) and Polaris (1 filament) were only detected in $^{13}$CO. The velocity dispersions ($\sim$0.3~km/s on average) of these filaments were not included in the filament sample due to larger uncertainties in the optical depth of this transition.  }}. 
   
 \begin{figure}
   \centering
     \resizebox{9.cm}{!}{
     \includegraphics[angle=0]{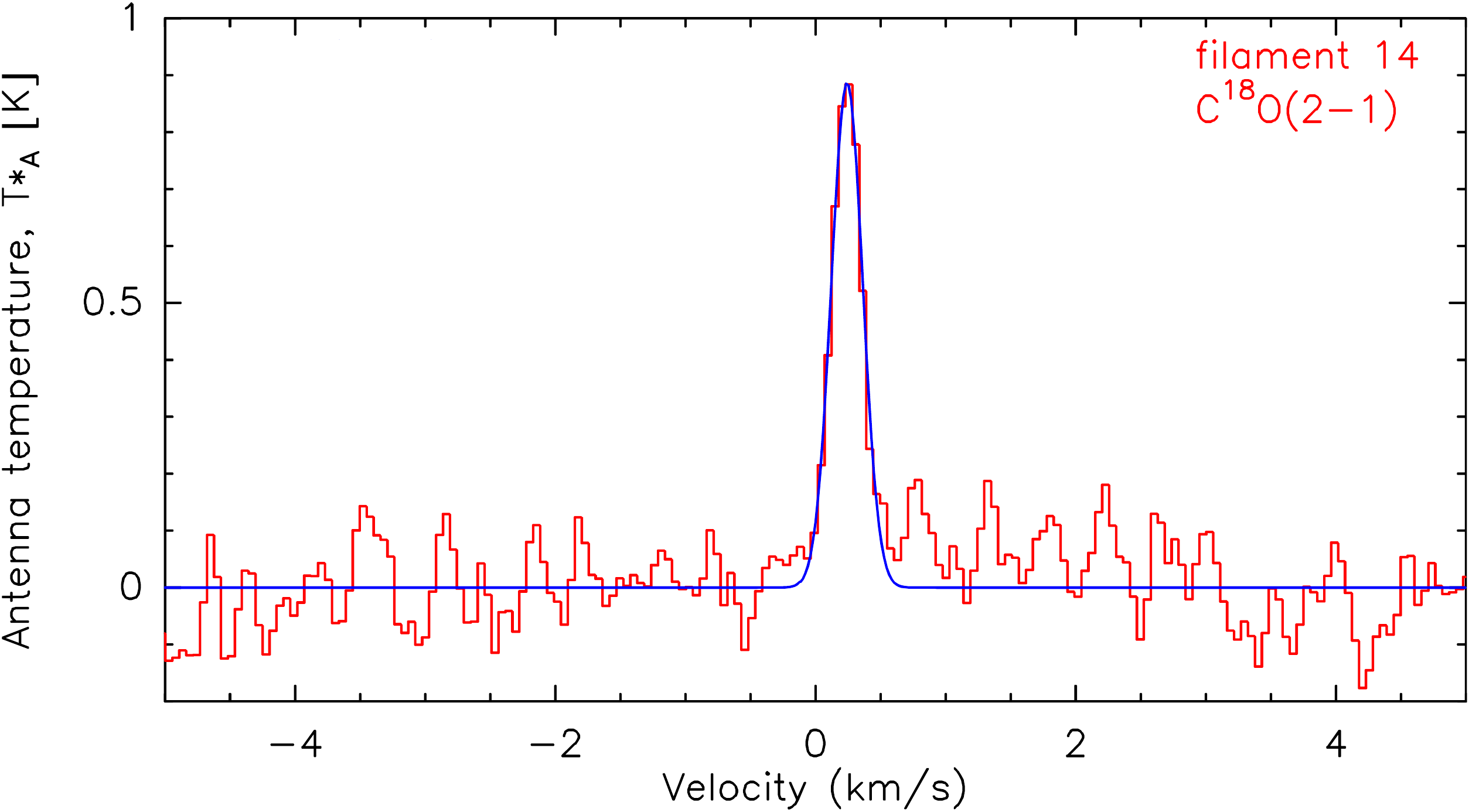}} 
      \hspace{0.5mm}
  \resizebox{9.cm}{!}{
   \includegraphics[angle=0]{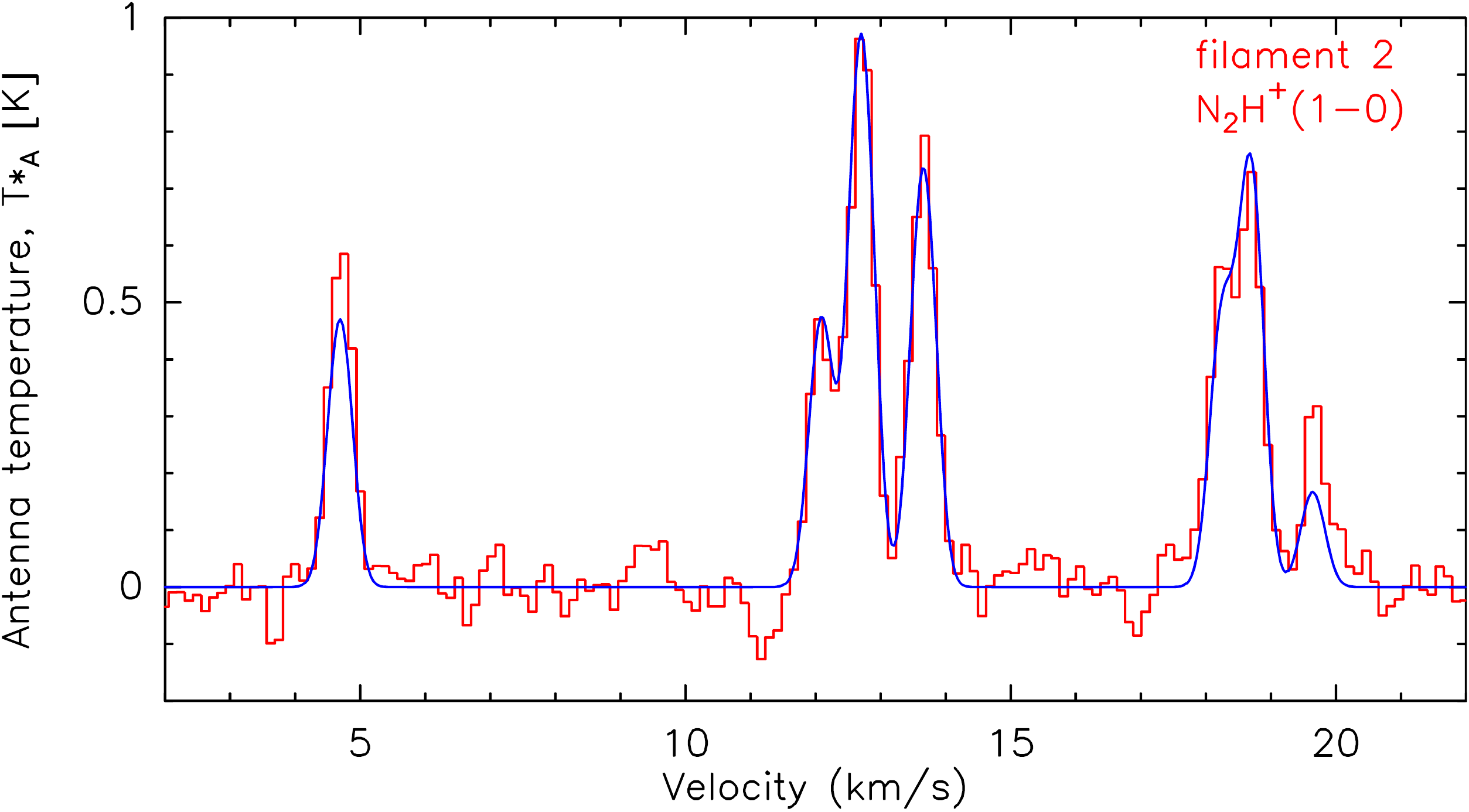}}
  \caption{ Spectra  observed     in the C$^{18}$O(2-1) and the N$_{2}$H$^{+}$(1-0)  transitions, toward filament 14 (top) and 2 (bottom), respectively.   The best  Gaussian fits to the spectra are overlaid in blue. The fitted parameters are given in Table~1.  }
              \label{IC5146_spectra}
    \end{figure}

The final sample discussed in the remainder  of this paper corresponds {\revised to the 46  filaments detected in either  N$_{2}$H$^{+}$ or C$^{18}$O.} 
This sample consists of   
44 {\revised individual filaments  detected} with the IRAM 30m telescope in {\revised IC5146 (11 filaments), Aquila rift (26 filaments), and Polaris flare (7 filaments)}, plus  2 filaments corresponding to the NGC2264C  elongated clump   \citep{Peretto2006} and the DR21 ridge \citep{Schneider2010,Hennemann2012}. 

The  positions of the filaments observed in IC5146 are shown in  Fig.~\ref{IC5146SpectraPos}, overlaid on the $Herschel$ column density map. The filaments observed  in Aquila and Polaris are shown in online Fig.~\ref{Aqu}/\ref{Aqu_SW} and Fig.~\ref{Pol}, respectively. A N$_{2}$H$^{+}$ (1--0) spectrum and  a C$^{18}$O(2--1) spectrum toward two filaments 
observed in IC5146   are shown in Fig.~\ref{IC5146_spectra}. The positions and derived  parameters of  all the  filaments  used for the analysis presented in this paper  are summarized in Table~1 %
and the corresponding  spectra  are shown in online Appendix \ref{VelComp}. %

\section{Line-of-sight velocity dispersions}\label{VelDisp}

In order to combine  the velocity information obtained with various line tracers,   we estimated the total velocity dispersion of the mean free particle   in molecular clouds. 

We estimated the non--thermal velocity dispersion of the gas by subtracting the thermal velocity dispersion  from the  linewidth measured for each species,   assuming that the two contributions are independent of each other and could be added in quadrature \citep{Myers1983}. 
The thermal velocity dispersion of each species  is given by
\begin{equation}
\sigma_{\rm T}(\mu_{\rm obs}) = \sqrt{ \frac{k_{\rm B} T}{\mu_{\rm obs}m_{\rm H}}},
\end{equation}
where $k_{\rm B}$ is the Boltzmann constant,  $\mu_{\rm obs}$ is the atomic weight  of the observed molecule, i.e.  $\mu_{\rm obs} = 29$ for N$_{2}$H$^{+}$ and $\mu_{\rm obs} = 30$ for C$^{18}$O. We adopted a gas  temperature  of $10$~K which is the typical temperature of starless molecular clouds and is close to the  dust temperature measured with $Herschel$ for the present sample of  filaments.  The non--thermal velocity dispersion is then equal to 
\begin{equation}
\sigma_{\rm NT} = \sqrt{\sigma_{\rm obs}^{2}-\sigma^{2}_{ \rm T}(\mu_{\rm obs})}\, ,
\end{equation}
where  $\sigma_{\rm obs}=\Delta V / \sqrt{8\ln 2} $ and $\Delta V$ is the measured FWHM linewidth of the observed spectra. The linewidths were  estimated by single Gaussian fits to the C$^{18}$O spectra (cf. Appendix~\ref{VelComp} for more details) and by multiple Gaussian fits to the seven components of the hyperfine multiplet  of the  N$_{2}$H$^{+}$ transition, using the Gaussian  HyperFine Structure (HFS)  fitting routine of the CLASS software package. This routine derives the line optical depth by assuming the same excitation temperature for all hyperfine components, and therefore yields an estimate of the intrinsic linewidth.

The total velocity dispersion of the mean free particle   of molecular weight $\mu$=2.33 is finally   given by:
\begin{equation}
\sigma_{\rm tot}(\mu) = \sqrt{\sigma_{\rm NT}^{2}+\sigma^{2}_{\rm T}(\mu)}\, ,
\end{equation}\label{veltot}
where $\sigma_{\rm T}(\mu) = \sqrt{ \frac{k_{\rm B} T}{\mu \rm m_{\rm H}}} \sim 0.2$~km/s  for $T~=~10$~K.
{\revised The typical gas temperature in the center of the dense filaments of our sample is approximately $T_{\rm gas} \sim 10$~K 
since the dust temperature derived from $Herschel$ data is $T_{\rm dust}\sim10$~K and  $T_{\rm gas}$ 
is expected to be well coupled to $T_{\rm dust}$ at densities $\gtrsim 3\times10^{4}$~cm$^{-3}$ \citep[see][]{Galli2002}.
As to the lower-density filaments in our sample, CO observations directly 
show that their gas temperature is also close to 10~K \citep[e.g.,][]{Goldsmith2008,Heyer2009}. 
}


  \begin{figure*}
   \centering
     \resizebox{17.cm}{!}{
   \includegraphics[angle=0]{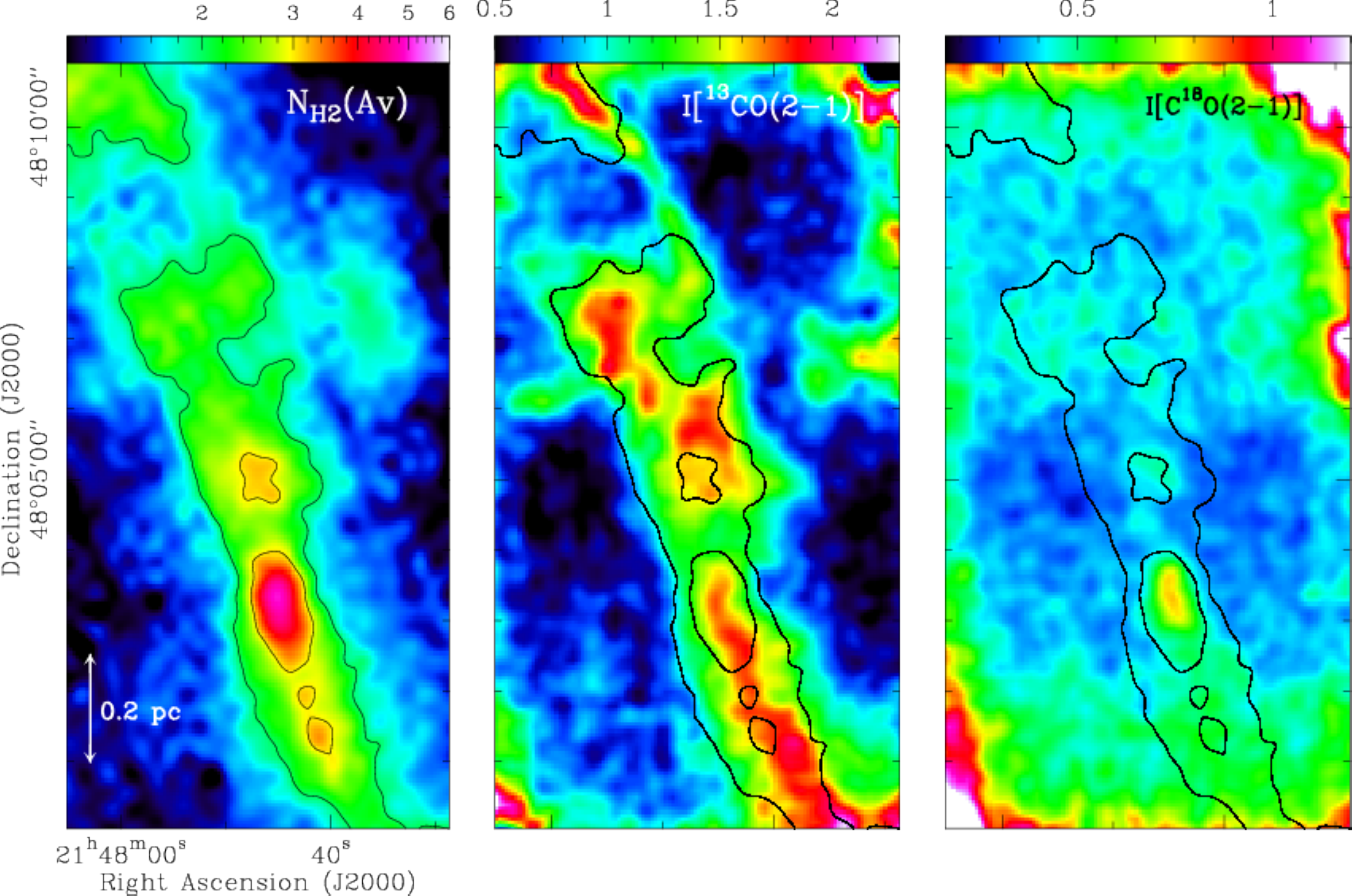}}  
  \caption{{\bf Left:} $Herschel$ column density map of filament 14 in IC5146 %
  corresponding to the yellow rectangle in Fig.~\ref{IC5146SpectraPos}, in units of visual extinction (A$_{\rm V}$) where $N_{\rm H_{2}}~\sim~\rm A_{V}~\times~10^{21}$~cm$^{-2}$. The resolution of the map is 36.9$\arcsec$. The black contours correspond to  column densities of  1 and 2~$\times 10^{21}$~cm$^{-2}$. These contours are   reproduced in the middle and right panels.  {\bf Middle:} $^{13}$CO(2-1) integrated intensity map (in units of {\revised K km/s})  over the  LSR  velocity range  $-$2.5 to 1~km/s. %
  The map has been  smoothed to a resolution of 18.5$\arcsec$, which corresponds to half  the resolution of the  $Herschel$ column density map. The  $^{13}$CO(2-1) integrated  emission of the filament  is concentrated   within the $10^{21}$~cm$^{-2}$  column density contour, which traces the elongated   structure of the filament. %
  {\bf Right:} Same as the middle panel for the C$^{18}$O(2-1) transition.}
              \label{SubColdensCO}
    \end{figure*}

\section{Velocity dispersion along  filament 14 in IC5146 \label{velSubSec}}

The velocity dispersions of the filaments in our sample  have been estimated from single-position observations toward each filament.
To assess  the reliability of inferring the  velocity  dispersion  of a filament from a single position spectrum,  
we  investigated the variations  of the velocity dispersion  along the crest of  filament 14 in IC5146 which we mapped in both  $^{13}$CO(2-1)  and C$^{18}$O(2-1) -- see Fig.~\ref{SubColdensCO}.

From the  $^{13}$CO(2-1)/C$^{18}$O(2-1) ratio map,  we first  estimated the  optical depth   to check whether or not the C$^{18}$O(2-1) emission is optically thin.

Assuming uniform excitation temperature along the line of sight, $^{13}$CO(2-1)  and C$^{18}$O(2-1) line emissions 
  can  be  expressed as a function of  optical depth as follows:
\begin{equation}
 T_{\rm A}^{*,i}(\upsilonup ) = T_{\rm ex}^{i} [1 - \exp (-\tau^{i}(\upsilonup ))],
\end{equation}
 where {\it i} corresponds to  either C$^{18}$O(2-1) or $^{13}$CO(2-1),  $T_{\rm A}^{*,i}(\upsilonup )$ is the detected antenna temperature for each velocity channel $\upsilonup $, %
 $T^{i}_{ex}$ is the excitation temperature which we assume to be the same for the  two isotopomers and  $\tau_{21}^{i}$ is the optical depth of the (2--1) transition of C$^{18}$O or   $^{13}$CO.
 

  \begin{figure*}[!h]
   \centering
     \resizebox{15.cm}{!}{
     \includegraphics[angle=0]{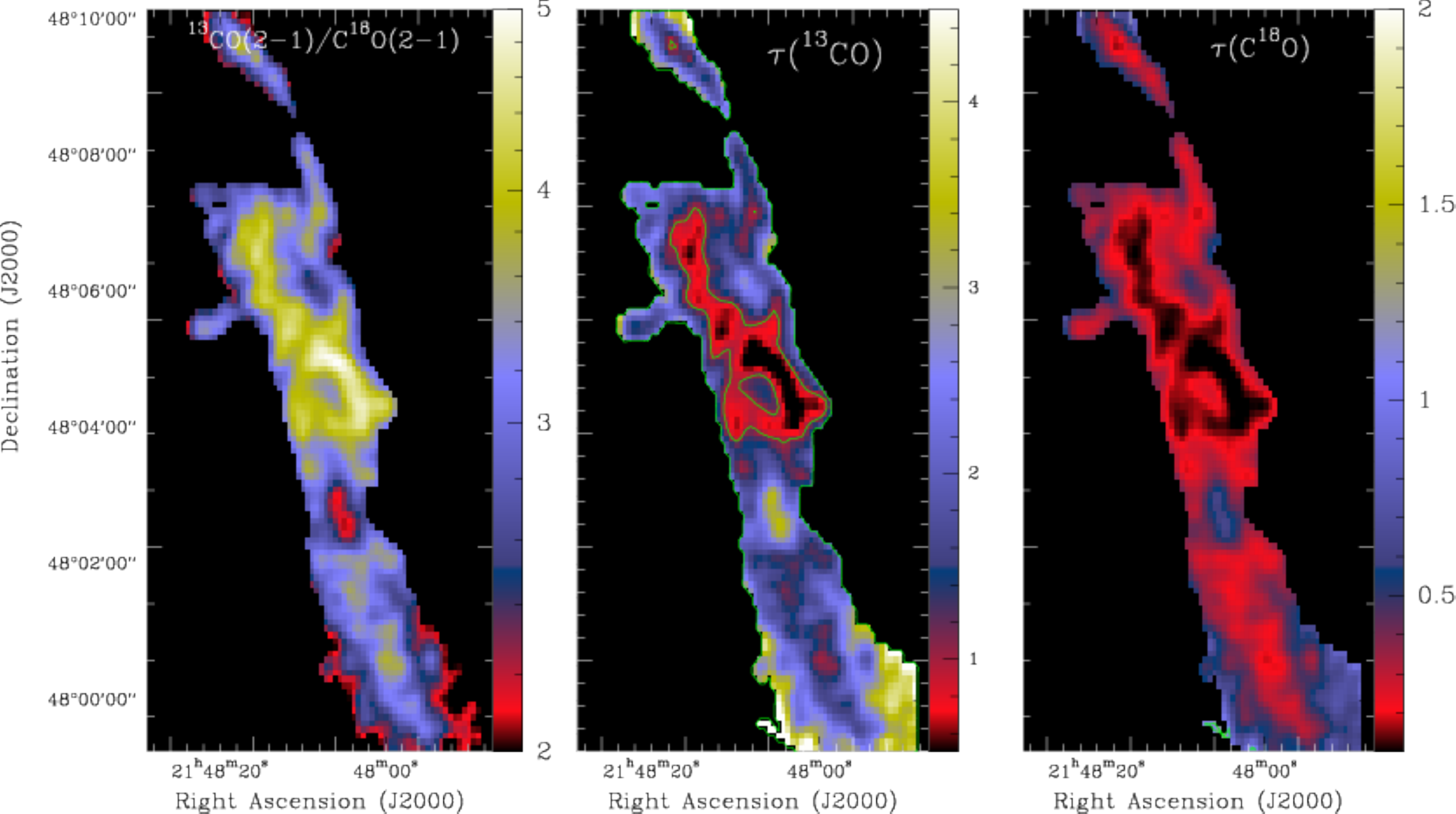}}  
  \caption{ {\bf Left:}  Integrated intensity ratio map  $\int_{-2.5}^{1}T_{\rm A}^{*,13}(\upsilonup )\rm d{\it \upsilonup }/ \int_{-2.5}^{1}\rm T_{\rm A}^{*,18}({\it \upsilonup })\rm d{\it \upsilonup }$ for  filament 14 in  IC5146. Only those pixels where the   column density derived from {\it Herschel} is   higher than$10^{21}$~H$_{2}$.cm$^{-2}$ have been considered (black contour in Fig.~\ref{SubColdensCO}). {\bf Middle:} Map of the mean optical depth of  the $^{13}$CO(2-1) transition. The contours in green correspond to a value of 1. {\bf Right:} Map of the mean optical depth of  the C$^{18}$O(2-1) transition. The mean optical depth of the C$^{18}$O transition is $\tau$(C$^{18}$O)~$<$~1 over  the entire filament.}
              \label{Fil14_tauMaps}
    \end{figure*}

  \begin{figure*}[!h]
   \centering
        \resizebox{8.7cm}{!}{
     \includegraphics[angle=0]{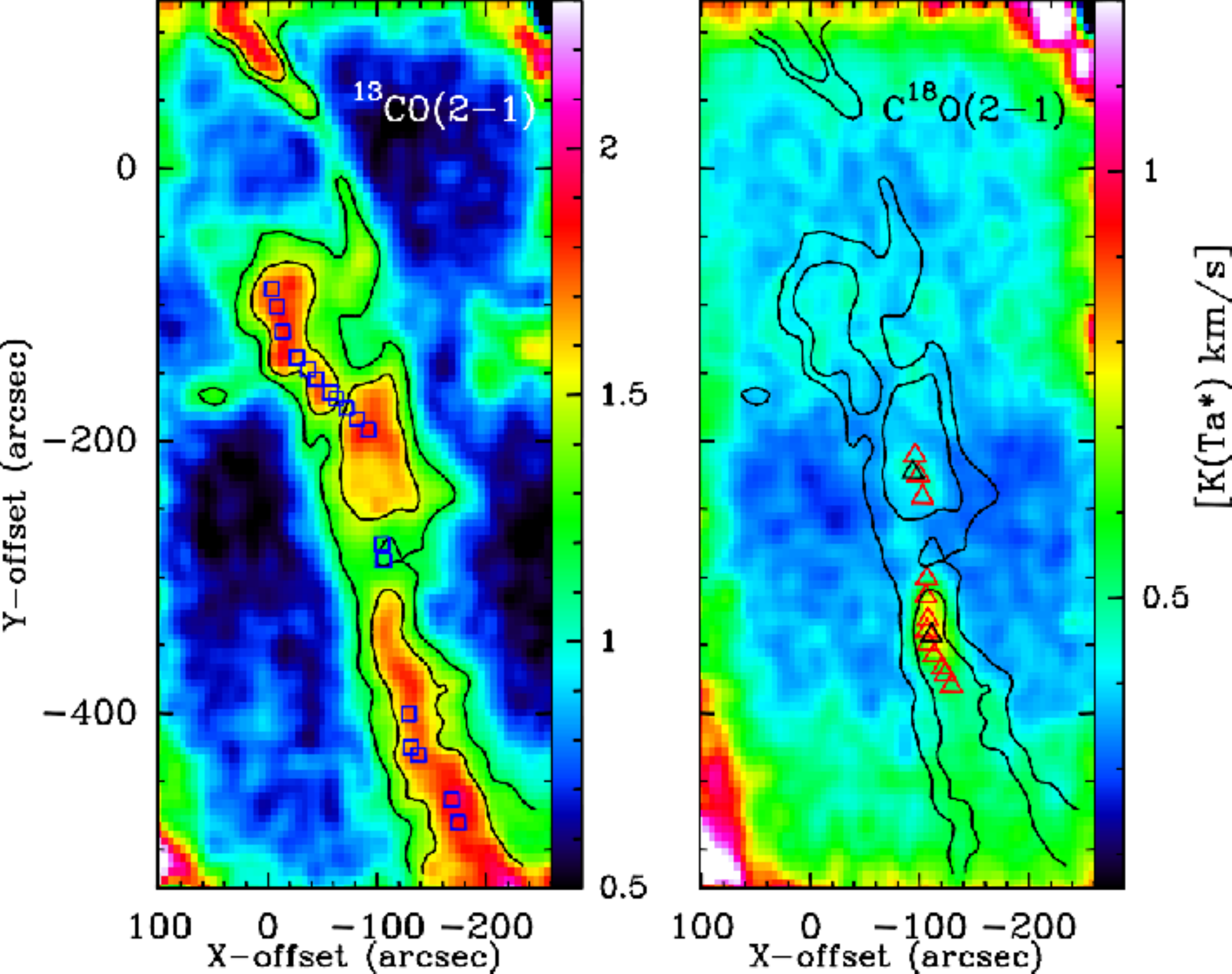}}
      \hspace{0.4mm}
     \resizebox{9.2cm}{!}{
     \includegraphics[angle=0]{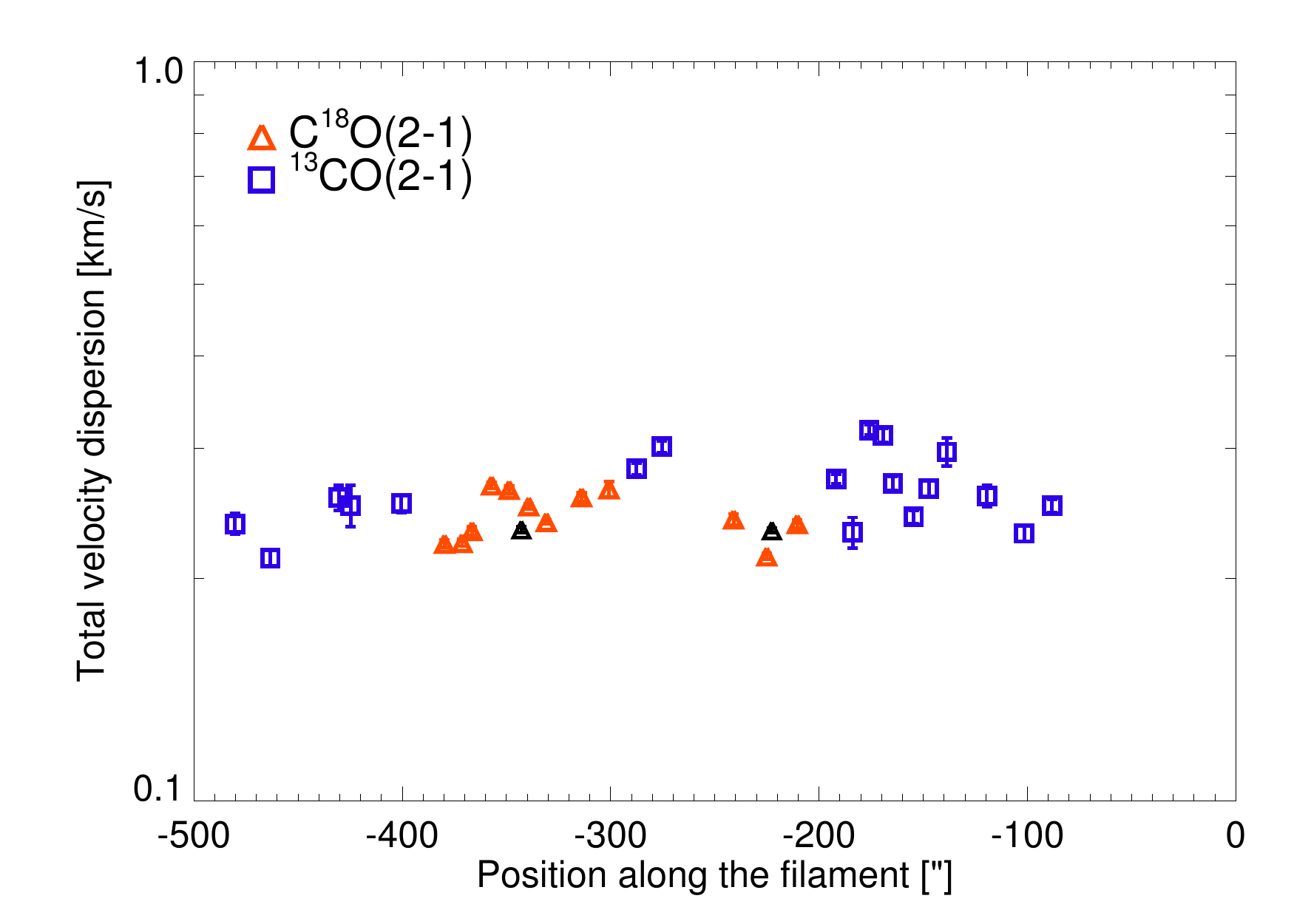}}
  \caption{ {\bf Left:}  Integrated  $^{13}$CO(2-1) and  C$^{18}$O(2-1) intensity maps of  filament 14 in IC5146 (cf. Fig. \ref{SubColdensCO}). 
  Two    $^{13}$CO(2-1) contours are overlaid: 1.2 and 1.5~K.km/s. The blue squares and  red triangles correspond to the positions along the filament where the line-of-sight velocity dispersion has been  measured from the $^{13}$CO(2-1) and  C$^{18}$O(2-1) maps respectively. The two black triangles on the C$^{18}$O(2-1) map correspond to the positions of the central pixel of the array receiver of HERA observed in a  single pointing. {\rev The labels  indicate the offsets in arcsec relative to the centre of the map at Ra(J2000) = 21:47:56 and Dec(J2000) = 48:08:55.} The black triangle  at  position  ($-94\arcsec~,~-223\arcsec$)  is the point which represents  filament 14 in the column density vs velocity dispersion diagram (black triangle in figure~\ref{VelDisp_coldens}).   {\bf Right:} Total velocity dispersions, $\sigma_{\rm tot}$ [see Eq.~(\ref{veltot})], measured along the same filament. The x-axis goes from the south  to the north of the filament.%
    The red triangle and  blue squares correspond to measurements made on the C$^{18}$O(2-1) and  $^{13}$CO(2-1) maps, respectively. Individual spectra were constructed by spatially averaging the signal located within 10$\arcsec$ of the central position to increase the signal to noise ratio of the data.  The linewidths of the $^{13}$CO(2-1) spectra were corrected for the broadening due to optical depth effects (see explanations in Sect.~\ref{velSubSec}).  The median value of $\sigma_{\rm tot}$ is 0.25~km/s with a narrow dispersion of 0.03~km/s.
  }
         \label{veloFil14}%
    \end{figure*}

\begin{table*}[!h]  
\label{tab:FilParamTable}    
\centering
 \caption{Summary of derived parameters of the observed spectra toward  the filaments of our sample.}
\begin{tabular}{|c|c|c|c||c|c||c|cc|c|cc| }   
\hline\hline   
Field / tracer&Filament& Ra & Dec&$N_{\rm H_{2}} $  &$M_{\rm line}$&  $V_{\rm LSR}$ & $T_{\rm peak}$ & rms&$\sigma_{\rm obs}$&$\sigma_{\rm tot}$&$\sigma_{\rm tot}^{err}$  \\
 && (J2000) & (J2000) &[$10^{21}$ c$\rm m^{-2}$]&M$_{\odot}$/pc &[km/s]&   \multicolumn{2}{|c|}{[K(T$_{\rm A}$*)]}&[km/s]& \multicolumn{2}{|c|}{[km/s]} \\
  (1)&(2)& (3) & (4) &(5) &(6) &(7) &  (8) &(9) &(10) & (11) &(12) \\
\hline  
        &    1 &18:29:39.5&-01:50:14 & 213.4&  252&7.70 & 1.43   & 0.67 &0.60 &0.64 &0.01 \\
       &    2 &18:32:38.5&-02:50:24 & 40.9&  176&7.85 & 0.67 & 0.22&0.30 &0.36 &0.01\\
       &    3 &18:33:15.2&-02:48:14 & 13.0&   35&6.42 &  0.65 &   0.13&0.11 &0.23 &0.01 \\
       &    4 &18:33:06.7&-02:47:50 & 11.6&   32&7.04 & 0.26 &  0.08&0.11 &0.23 &0.01 \\
       &    5 &18:33:22.5&-02:42:30 & 12.8&   20&6.84 & 0.16 &  0.05&0.15 &0.25 &0.01 \\
       &    6 &18:29:36.6 &-01:51:55 & 13.2&   39&7.63 &0.12 & 0.05 &0.11 &0.23 &0.01 \\
       &    7 &18:29:50.4 &-01:54:47 & 37.9&  123&7.67 & 1.18 &   0.25&0.18 &0.27 &0.01 \\
 Aquila      &    8 &18:30:27.6&-01:55:32 & 35.2&  139&7.42 &0.41 &  0.13 &0.20 &0.28 &0.01 \\
 N$_{2}$H$^{+}$(1--0)          &    9 &18:30:19.1 &-01:53:09 & 33.9&   90&7.72 &0.18 &     0.11 &0.41 &0.45 &0.01 \\
   &   10 &18:29:08.9 &-01:42:40 & 29.9&  121&7.05 &0.76 &     0.18&0.21 &0.29 &0.01 \\
       &   11 &18:29:19.9 &-01:52:25 & 25.1&   68&7.27 &0.32 &  0.09 &0.23 &0.31 &0.01 \\
       &   12 &18:30:03.9&-02:04:29 & 20.3&   73&7.12 &0.68&     0.19 &0.19 &0.27 &0.06 \\
       &   13 &18:29:55.8&-02:00:18 & 74.5&  182&7.09 & 1.39  &  0.60&0.42 &0.47 &0.01 \\
       &   14 &18:29:58.1&-01:57:38 & 55.8&  265&7.41 &0.45 &   0.21 &0.44 &0.48 &0.01 \\
       &   15 &18:28:50.1&-01:34:52 & 114.1&  288&7.90 &0.79 &  0.23 &0.36 &0.41 &0.01 \\
       &   16 &18:30:03.5 &-02:02:26 & 10.9&   35&7.36 & 0.37&   0.07 &0.10 &0.22 &0.01 \\
       &   17 &18:29:55.4&-01:37:02 & 12.0&   40&8.29 & 0.53 &    0.12&0.14 &0.24 &0.01 \\
       &   18 &18:29:34.6&-01:55:15 & 9.2&	 38&7.34 & 0.10 &  0.04 &0.15 &0.25 &0.01 \\
	  \hline
       &    2 &21:46:28.1&47:17:34& 25.3&   80&4.69 &0.61 & 0.17&0.19 &0.27 &0.01 \\
IC5146       &  6 &21:47:17.2&47:33:01& 17.0&  109&3.69 & 0.32 &  0.11&0.25 &0.32 &0.01 \\
   N$_{2}$H$^{+}$(1--0)    &    12 &21:44:58.4&47:34:15& 14.7&   41&4.11 & 0.40 &  0.13&0.19 &0.28 &0.01 \\
      &    13 &21:47:07.9&47:39:30& 17.1&   50&4.35 & 0.25 &  0.10&0.22 &0.30 &0.01 \\
         
	   \hline
Polaris   &    1 &02:00:29.0 &87:41:54& 8.0&   16&-4.62 &0.18 &0.01 &0.13 &0.24 &0.01 \\
N$_{2}$H$^{+}$(1--0)    &    &&& & &&&&&&\\
	   \hline\hline
      &     19 &18:29:53.5 &-03:43:39& 7.9&	10&4.88 &1.22 &0.06 &0.23 &0.31 &0.01 \\
      &     20 &18:28:58.2 &-01:22:51& 7.8&	28&7.17 &1.10 &0.04 &0.30 &0.36 &0.01 \\
      &     21 &18:31:06.4 &-01:18:58& 9.9&	20&7.77 &1.10 &0.05 &0.26 &0.33 &0.01 \\
 Aquila     &     22 &18:33:15.8 &-02:47:15& 7.2&	51&6.47 &1.26 &0.03 &0.13 &0.24 &0.01 \\
 C$^{18}$O(1--0)     &     23 &18:22:51.1 &-03:26:38& 6.3&	12&5.78 &0.36 &0.03 &0.28 &0.34 &0.01 \\
      &     24 &18:23:31.3 &-03:03:43& 7.2&	15&6.87 &0.89 &0.03 &0.27 &0.34 &0.01 \\
      &     25 &18:29:45.6 &-02:32:21& 7.3&	22&6.52 &0.94 &0.03 &0.15 &0.25 &0.01 \\
      &     26&18:32:53.1 &-01:29:42& 7.0&	21&7.28 &0.28 &0.03 &0.16 &0.26 &0.01 \\
	  \hline
	   &     3 &21:46:42.1&47:44:26& 2.8&    3&1.89 &0.80 &0.08 &0.17 &0.26 &0.01 \\
	    &     5 &21:45:15.3&47:46:52& 3.0&   11&1.42 &0.63 &0.09 &0.24 &0.31 &0.02 \\
IC5146 	    &     7 &21:44:37.1&47:42:39& 5.2&   11&3.71 &0.76 &0.09 &0.29 &0.35 &0.02 \\
	 C$^{18}$O(2--1)           &     10 &21:43:29.1&47:19:29& 4.3&   16&4.21 &0.79 &0.06 &0.13 &0.24 &0.01 \\
	        &    11 &21:43:30.9&47:47:12& 3.1&    9&3.29 &0.42 &0.08 &0.11 &0.23 &0.01 \\
       &     14 &21:47:48.1&48:04:53&  3.9&   15&0.24 &0.89 &0.07 &0.12 &0.23 &0.01 \\
        &     25 &21:44:18.8&47:32:25& 5.7&   17&2.32 &0.47 &0.09 &0.27 &0.34 &0.03 \\
	   \hline
       &   2 &01:58:59.7&87:39:33& 9.6&	16&-4.29 &2.65 &0.04 &0.22 &0.29 &0.01 \\
       &    3 &01:40:31.4&87:46:19& 2.4&	15&-4.14 &0.28 &0.04 &0.22 &0.30 &0.02 \\
 Polaris      &    4 &01:31:21.7&87:45:56& 2.7&	25&-4.48 &0.27 &0.03 &0.19 &0.27 &0.02 \\
 C$^{18}$O(1--0)   &    5 &02:00:49.1&87:35:18& 3.4&	15&-4.08 &0.62 &0.04 &0.17 &0.30 &0.01 \\
       &    6 &01:50:48.0&87:43:47& 1.5&	 3&-3.46 &0.19 &0.04 &0.19 &0.28 &0.02 \\
       &    7 &03:36:47.4&88:08:13& 2.3&	16&-3.39 &0.12 &0.05 &0.08 &0.26 &0.01 \\ 
         \hline         
         \hline  
                  \end{tabular}
\vspace*{-0.45ex}
\begin{list}{}{}
\item[]{ {\bf Notes:} Columns 3 and 4: Equatorial coordinates providing the positions of the observed spectra. Columns 5 and 6: Projected column density and mass per unit length for each filament measured from {\it Herschel} column density maps.  The listed values have not been corrected by the 60$\%$ systematic overestimation of true column densities and masses per unit length. All the filaments detected in N$_{2}$H$^{+}$ are supercritical while most of the filaments observed in C$^{18}$O(1--0) and (2--1) and not detected in N$_{2}$H$^{+}$ are subcritical.
  Columns 7, 8 and 10: Parameters of  single and multiple Gaussian fits to the C$^{18}$O and N$_{2}\rm H^{+}$ observed spectra, respectively. For the N$_{2}\rm H^{+}$ spectra $T_{\rm peak}$ (Col.~8) corresponds to the peak temperature of the isolated component of the N$_{2}\rm H^{+}$ multiplet. Column 9 gives  the corresponding rms. {\revised  The reference positions used for the sposition-switched  N$_{2}$H$^{+}$ observations   
were  selected from the $Herschel$ images and correspond to 21:47:22.6 47:45:59 for IC5146, 01:56:30.26 87:43:28 for Polaris, 18:28:23.9 -02:00:22 for Aquila, except for filaments 3, 4 and 5 for which a reference  position at 18:33:27.75 -02:51:22 was used. } Columns 10 and 11: $\sigma_{\rm obs}$ and $\sigma_{\rm tot}$ are derived as explained in Sect.~\ref{VelDisp}.   }
 \end{list}     
\end{table*}

The optical depth values at the systemic velocity ($\upsilonup _{\rm sys}$) of the cloud  for each transition were estimated from the ratio of the   $^{13}$CO(2-1)  and C$^{18}$O(2-1) line intensities:
\begin{equation}
\frac{T_{\rm A}^{*,13}(\upsilonup _{\rm sys})}{T_{\rm A}^{*,18}(\upsilonup _{\rm sys})} =   \frac{ T_{\rm ex}^{13}[1-\exp(-\tau_{21}^{13}(\upsilonup _{\rm sys}))]}{ T_{\rm ex}^{18}[1-\exp(-\tau_{21}^{18}(\upsilonup _{\rm sys}))]}\,\,,
\label{tau}
\end{equation}
 where $T_{\rm ex}^{13}=T_{\rm ex}^{18}$ and $\tau_{21}^{13} = X  \tau_{21}^{18}$ , with $X = [^{13}\rm CO]/[\rm C^{18}\rm O] = 5.5$  \citep[the assumed mean value of the abundance ratio  in the local ISM, e.g.][]{Wilson1994}. 
 
  Maps of the mean optical depth   of filament 14 were constructed from the observed integrated intensity ratio map of the  $^{13}$CO(2-1)  and C$^{18}$O(2-1) transitions,  over the velocity range \, $-$2.5 to  $+$1~km/s, 
$$R_{13/18}= \int_{-2.5}^{1} T_{\rm A}^{*,13}({\it \upsilonup })\rm d{\it \upsilonup }\,\,/\,\,\int_{-2.5}^{1} {\it T}_{\rm A}^{*,18}({\it \upsilonup })\rm d{\it \upsilonup }\, .$$
We estimated  the mean values of  $\tau_{21}^{18}$ and $\tau_{21}^{13}$ (corresponding to the optical depth values integrated over the velocity range $-$2.5 to $+$1~km/s)  %
from the $R_{13/18}$ ratio map by solving a velocity-integrated  form of Eq.~\ref{tau} on a pixel by pixel basis. Fig.~\ref{Fil14_tauMaps}  shows the  mean optical depth  maps of the  $^{13}$CO(2-1) and C$^{18}$O(2-1) transitions for filament 14 (middle and right panels, respectively) along with the $R_{13/18}$ ratio  map (left panel). 
While the bulk of the C$^{18}\rm O$ emission is optically thin ($\tau_{21}^{18}~<~0.6$) along the main axis of the filament,   the  $^{13}$CO emission  is significantly optically thick ($\tau_{21}^{13}~>~2$) in the southern half part of the filament and only marginally optically thin ($\tau_{21}^{13}\sim 0.5 - 2$) in the northern part of the filament.

Fig.~\ref{veloFil14} shows the positions along  filament 14 where spectra were extracted from the maps (left-hand side of the figure) and the velocity dispersions measured along the filament (right). The velocity dispersions derived from the $^{13}$CO(2-1)  spectra were corrected for the broadening effect due to the finite optical depth of the $^{13}$CO(2-1) transition. 

To do so, we measured  the variations of the  FWHM  linewidth, $\Delta V_{\rm obs}$,   of a simple Gaussian line  model,  $T_{\tau_{0}}^{\rm model}(\upsilonup )$,  
as a function of optical depth. 
 The model line spectrum had the form 
$T_{\tau_{0}}^{\rm model}(\upsilonup ) = T_{\rm ex} [1 - \exp^{-\tau(\upsilonup )}]$, where  $T_{\rm ex}$ is a  
 uniform  excitation temperature,  $\tau(\upsilonup )=\tau_{0}\,\exp\left[-\frac{4\ln2\,(\upsilonup -\upsilonup _{0})^{2}}{\Delta V_{\rm int}^{2}}\right]$ corresponds to an intrinsically  Gaussian distribution of  optical depths  with velocity, $\upsilonup $, 
 and  $\Delta V_{\rm int}$ is the intrinsic FWHM linewidth of the model. By fitting Gaussian profiles to synthetic spectra, we estimated the FWHM width 
$\Delta V_{\rm obs}$ of the model as a function of $\tau_{0}$, and derived correction factors $\Delta V_{\rm int}~/~\Delta V_{\rm obs}~\approx~$0.77 -- 0.92, 
for $^{13}$CO(2-1)  optical depths  ranging between 0.5 and 2.5. {\revised The most extreme  correction would be a factor of 0.74 for a $^{13}$CO(2-1) optical depth of 3,  
if the [$^{13}\rm CO]/[\rm C^{18}\rm O$]  abundance ratio has a value of 7 \citep[e.g.,][]{Wu2012}, i.e., higher than the standard value in the local ISM.}

Inspection of the linewdiths measured  
   along  filament 14 shows that the variations of the  internal  velocity dispersion are   small along the crest of the filament (cf.  right panel of 
   Fig.~\ref{veloFil14}), with an average velocity dispersion ($0.25~\pm~0.03$)~km/s. %
 This is consistent with  previous observations of a few  low-density filaments  showing that these structures have velocity dispersions close to the thermal velocity dispersion $\sim$0.2~km/s for T=10~K %
\citep[cf.][]{HilyBlant2004,HilyBlant2009}  and which do not  vary much  along their length \citep{Hacar2011,Pineda2011}. 
 Recently, \citet{Li2012} studied  the Taurus B213 filament and found that it is characterized by a coherent  velocity dispersion of about $\sim$0.3~km/s. 
 
 These results  suggest that  the  velocity dispersion  observed at  a single position toward a filament provides  a reasonably good estimate of the  velocity dispersion of the entire filament\footnote{{\revised An interstellar filament is an elongated structure characterized by small column density variations along its crest (less than a factor of 3; see 
 Fig.~\ref{SubColdensCO}-left). 
Therefore, the velocity dispersion variations induced by the trend $ \sigma_{\rm tot} \propto {\Sigma_{0}}^{0.5}$ found in Sect.~5 below for supercritical 
filaments remain small ($<< 2$) along a given filament.}}.
Nevertheless  mapping  observations of a broader sample of filaments in several regions would be valuable to confirm the robustness of this assumption.

\section{Two regimes for interstellar  filaments  }

In this section, we investigate the variation of  internal velocity dispersion with  column density for  our  sample of filaments. 

\begin{figure}
    \centering
     \resizebox{9.5cm}{!}{
     \includegraphics[angle=0]{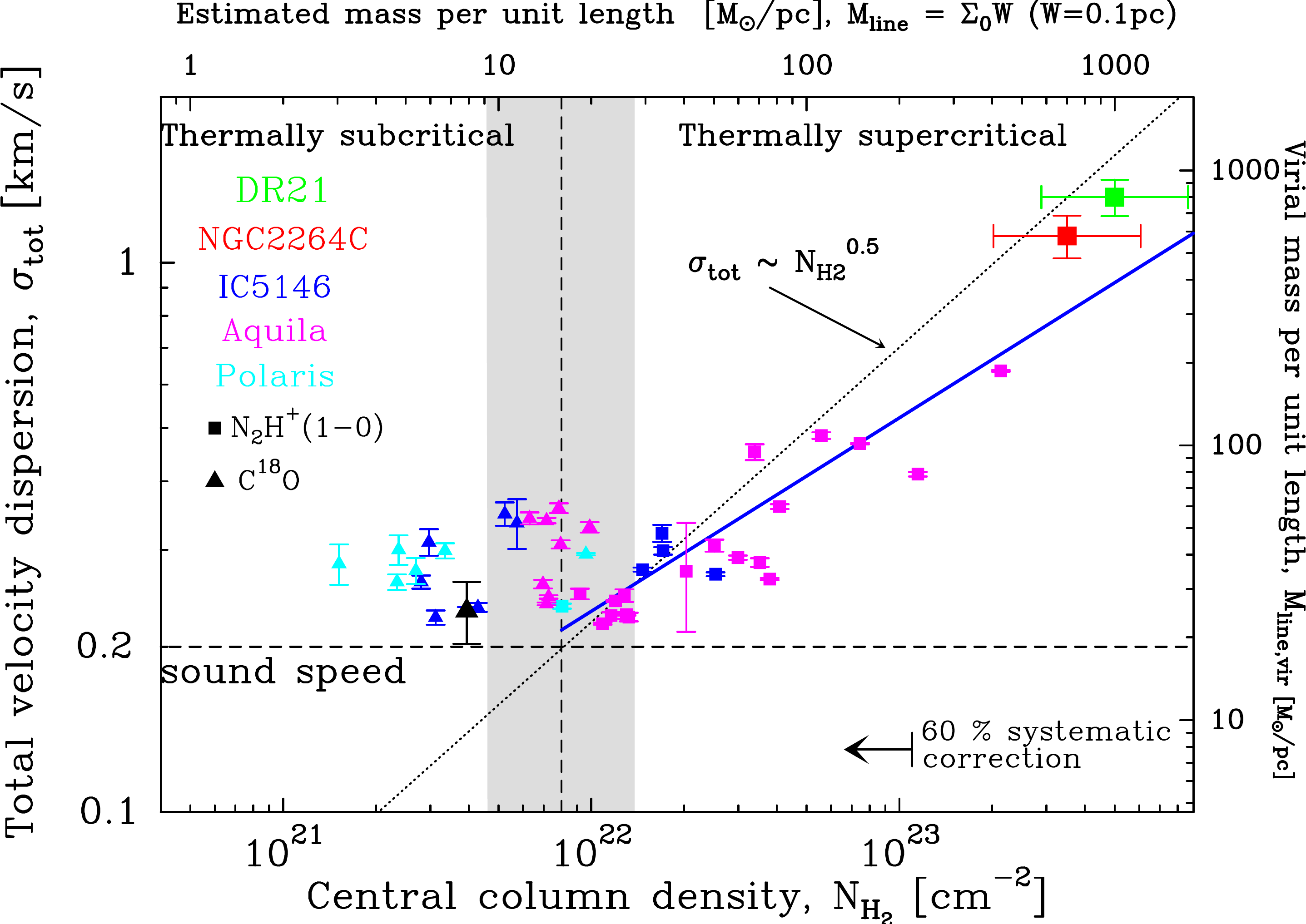}}
          \caption{Total velocity dispersion versus  observed central  column density (uncorrected for inclination effects):  
                    blue for IC5146, magenta for Aquila, cyan for Polaris, red for NGC2264C, and green for DR21 filaments.
                    The squares and triangles  correspond to  velocity dispersions measured from  $\rm N_{2}\rm H^{+}$ spectra and  C$^{18}$O spectra, respectively. The vertical error bars come from the uncertainties in the linewidths derived from Gaussian fits to the observed spectra. 
                    See text for details and uncertainties in the column densities. 
 The horizontal {\revised dashed} line shows the value of the sound speed $\sim 0.2$ km/s for  $T$~=~10~K. The vertical dashed line marks the  boundary
 between thermally subcritical and thermally supercritical filaments where the estimated mass per unit length $ M_{\rm line} $ is approximately 
 equal to the critical value $ M_{\rm line,crit} \sim$ 16~M$_{\odot}$/pc for $T$~=~10~K, {\revised equivalent to a column density of $8\times10^{21}$~cm$^{-2}$.  The grey band shows a dispersion of a factor of 3 around this nominal value.}
The {\revised dotted} line running from the bottom left to the top right  corresponds to $ \sigma_{\rm tot} \propto { N_{\rm H_2}}^{0.5} $   
normalized to 0.2~km/s at the subcritical/supercritical boundary. %
{\revised The blue  solid  line shows the best power-law fit $ \sigma_{\rm tot} \propto {N_{\rm H_2}}^{0.35~\pm~0.14}$
to the 
supercritical filaments. } %
} 
   \label{VelDisp_coldens}
    \end{figure}

Figure~\ref{VelDisp_coldens} shows the total velocity dispersion as a function of  central column density for the 46 filaments. Values for the column density and velocity dispersion of the NGC2264C elongated clump (red square in the following plots)  are taken from  \citet{Peretto2006}. 
{\revised Peretto et al. estimated the column density from 1.2~mm dust continuum mapping observations 
and the velocity dispersion of NGC2264C 
from N$_{2}$H$^{+}$(1--0) observations, both obtained with the IRAM 30m telescope. The velocity dispersion was averaged over the entire NCG2264C clump and reflects the velocity dispersion of the filamentary clump as opposed to the embedded protostellar cores.} The velocity dispersion of  the DR21 filament (green square in  the following plots) is taken from  \citet{Schneider2010} {\revised who derived it from N$_{2}$H$^{+}$(1--0) observations}, while the column density of DR21 comes  from  $Herschel$ observations 
\citep{Hennemann2012} obtained as part of the HOBYS key project  \citep{Motte2010}. The black triangle corresponds to filament 14 in IC5146 and the associated error bar comes from the dispersion of the total velocity dispersion along this filament as extracted from the   map shown in   Fig.~\ref{veloFil14}. The vertical error bars of the other points correspond to  the errors   in estimating the linewidth from   Gaussian fits to  the observed spectra, which are on the order of $\sim$0.01~km/s. 
The error on the measured velocity dispersion is $\sim$ 0.03~km/s  for  the filaments observed in Aquila and IC5146 and 
and 0.1~km/s for DR21 and NGC2264C (representing the variation of $\sigma_{\rm tot}$  along the filament crests). 
The errors  on the filament central column densities  %
come  from the uncertainties  in deriving column densities\footnote{\revised {The column densities were derived consistently for all filaments observed  with  $Herschel$ in Polaris, IC5146, Aquila, and DR21. The same dust opacity law was used to derive the column density of NGC2264C from MAMBO 1.2 mm continuum observations. The column density  estimates  for NGC2264C and DR21 are more  uncertain  due to the presence of prominent    cores  along the filament crests \citep[cf.][]{Peretto2006,Motte2007}.}} from {\it Herschel} dust continuum maps  and are typically a factor of $\sim$2 mainly due to uncertain assumptions concerning the 
adopted dust opacity law \citep[see][]{Konyves2010, Arzoumanian2011}. %
{\revised A recent study  comparing $Herschel$ and near-IR extinction data in Orion A reported a weak trend between dust opacity and column density, at least in the 
regime $1\  \la A_V  \la 10$, which has been interpreted as evidence of dust grain evolution with environment \citep{Roy2013}. But the uncertainty induced by 
this effect remains small  for most of the filaments in our sample.  
}

An additional systematic effect on the column density %
comes  from  the random  inclination angles of the filaments  to the plane of the sky,  which implies an overestimation   of the intrinsic column densities    (and masses per unit length) by $\sim$60~$\%$ on average for  a large sample  of filaments \citep[for more details see Appendix A of][]{Arzoumanian2011}. {\revised This systematic correction factor is shown as a horizontal arrow pointing to the left in Fig.~\ref{VelDisp_coldens}}.

{\revised The results of our velocity dispersion measurements confirm that the filaments of our sample can be divided into two physically distinct groups based on their central column density.} 
Filaments with column densities  $\leq 8\times10^{21}$~cm$^{-2}$  seem to have roughly constant velocity dispersions close to or slightly larger than  the sound  speed  ($c_s \la \sigma_{tot} < 2\, c_s $), while denser filaments with column densities $\geq  8\times10^{21}$~cm$^{-2}$ have velocity dispersions which  increase as a function of column density  (cf. Fig.~\ref{VelDisp_coldens}). 
The upper x-axis of Fig.~\ref{VelDisp_coldens} shows an approximate mass-per-unit-length scale derived from the bottom x-axis scale 
by multiplying the  central column density of each filament by a characteristic filament width $W_{\rm fil} \sim$ 0.1~pc 
(i.e., $M_{\rm line}~=~\Sigma_{0} \times W_{\rm fil}$ -- see Sect.~1). 

It can be seen from Fig.~\ref{VelDisp_coldens} that the critical mass per unit length $M_{\rm line,crit} \sim$ 16~M$_\odot$/pc for $T \approx 10$~K 
introduced in Sect.~1 corresponds to a column density boundary which divides the filaments into two regimes where 
thermally subcritical filaments ($M_{\rm line}<M_{\rm line, crit}$) have roughly constant velocity dispersions with a mean value of  $0.26 \pm 0.05$~km/s, while thermally supercritical filaments ($M_{\rm line}>M_{\rm line, crit}$) have velocity dispersions which increase as a function of  projected column density as a {\revised power law    
$ \sigma_{\rm tot} \propto {N_{\rm H_2, obs}}^{0.35~\pm~0.14}$.  
The latter relation becomes  $\sigma_{\rm tot} \propto { N_{\rm H_2, corr}}^{0.41~\pm~0.15}$ if ``intrinsic' column density values are used} 
(after correcting the observed values for a  $\sim 60\%$ overestimation on average due to inclination effects). 
{\revised The division of the filament sample into two regimes does not correspond to a sharp boundary but to a narrow border zone
represented by the grey band in Fig.~\ref{VelDisp_coldens} %
which results from 1) a spread of a factor of $\la 2$ in the effective width of the filaments around the nominal value of 0.1~pc 
used to convert the critical mass per unit length to a critical column density, 
2) uncertainties in the intrinsic column densities of the filaments 
due to a random distribution of viewing angles, %
and 3) a small dispersion in the effective gas temperature of the filaments, implying a small spread in the theoretical value of the 
effective critical mass per unit length (see below).}

To derive more  accurate filament masses  per unit length, we constructed  radial column density profiles  from the $Herschel$ column density maps,  
taking  a perpendicular cut  to each  filament at the  position     observed   with the 30m telescope. %
 The observed mass per unit length, $ M_{\rm line}^{\rm obs}$, of each filament was then derived by  integrating the measured 
 column density profile over radius. {\revised The mass per unit length of DR21 corresponds to the average value along the crest of the filament 
 derived by \citet{Hennemann2012}, while the value for NGC2264C was obtained from its estimated mass (1650~M$_\odot$) and length (0.8~pc).} 
These $ M_{\rm line}^{\rm obs}$ estimates of the mass per unit length are used in the following discussion instead of the simpler estimates shown 
at the top of Figure~\ref{VelDisp_coldens}. 
 
\begin{figure}[ht!]
   \centering
   \begin{minipage}{1\linewidth}
     \resizebox{8.3cm}{!}{
    \includegraphics[angle=0]{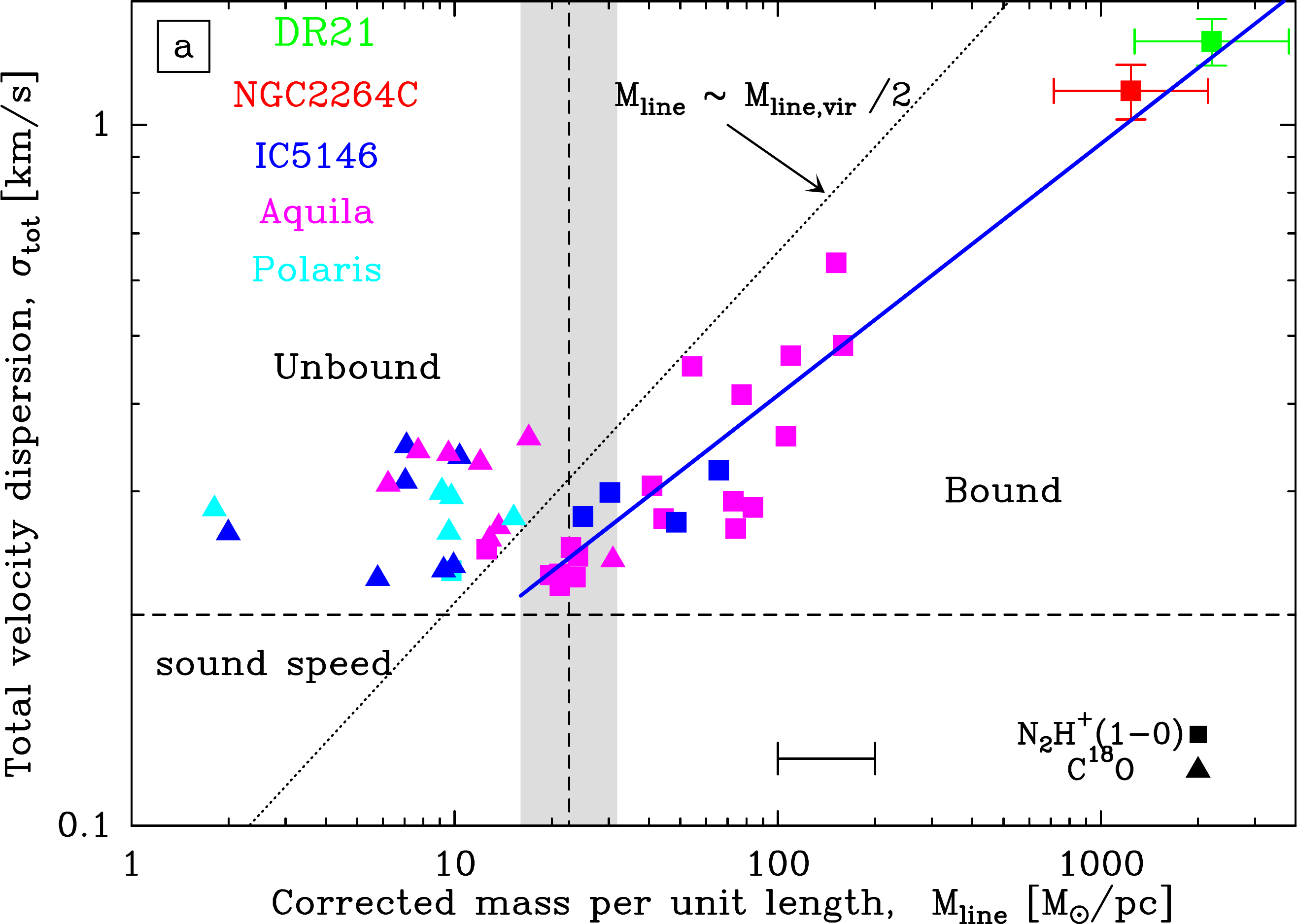}}
      \hspace{0.3cm}
  \resizebox{9.cm}{!}{
   \includegraphics[angle=0]{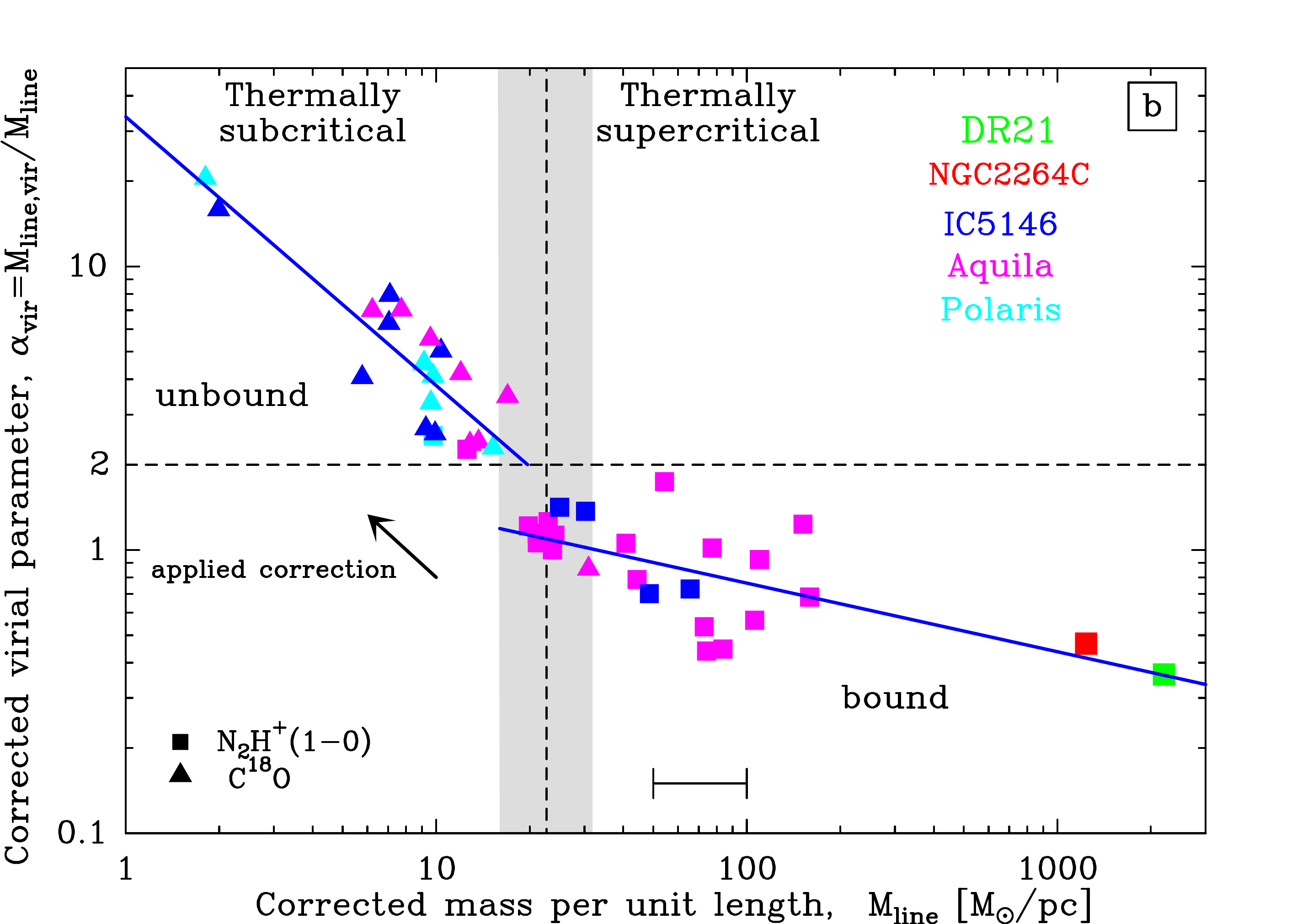}} 
   \end{minipage}
  \caption{ {\bf (a)} Total velocity dispersion   
 versus 
intrinsic mass per unit length (corrected for inclination effects) 
for the sample of 46  filaments. The symbols and  colors, as well as the y-axis of the plot, are the same as in Fig.~\ref{VelDisp_coldens}. 
{\revised The horizontal segment at the bottom of the plot shows the typical error bar on the estimated masses per unit length.} 
Specific  error bars are shown  for NGC2264C and DR21 since they are larger$^{5}$.
The horizontal {\revised dashed  line shows the value of the thermal sound speed $\sim 0.2$ km/s for $T_{\rm gas}$ = 10~K. 
The vertical grey band marks the theoretical position of the effective critical mass per unit length $M_{\rm line,crit}^{\rm eff} $ 
for effective temperatures in the range $10 \lesssim T_{\rm eff} \lesssim 20$~K (see text).}
The {\revised dotted} line running from  bottom left to top right corresponds to  $M_{\rm line}~=~M_{\rm line,vir}/2$, which marks the boundary between unbound 
and bound filaments located on the left and right  hand side of the border line, respectively. {\revised  The blue solid line,  $ \sigma_{\rm tot} \propto M_{\rm line}^{0.36~\pm~0.07}$, shows the best power-law fit for the bound filaments. The best-fit  exponent remains almost unchanged   $(0.31~\pm~0.08)$ when the  fitting is performed after removing  DR21 and NGC2264C from the sample.}  
{\bf (b)} Corrected virial parameter ($\alpha_{\rm line,vir}$) versus 
intrinsic mass per unit length for the same  filaments (x-axis is the same as in panel {\bf a}). %
{\revised The vertical grey band is the same as in  panel {\bf a} and }
separates 
subcritical filaments ($M^{\rm obs}_{\rm line}~<$~{\revised $M_{\rm line,crit}^{\rm eff}$}) on the left 
from   
 supercritical filaments ($M^{\rm obs}_{\rm line}~>$~{\revised $M_{\rm line,crit}^{\rm eff}$}) on the right. 
 The horizontal dashed line ($\alpha_{\rm line,vir} = 2$)  shows the boundary between gravitationally  unbound  ($\alpha_{\rm line,vir} > 2$) 
 and bound ($\alpha_{\rm line,vir} < 2$)  filaments, where  $\alpha_{\rm line,vir} =  M_{\rm line,vir} /  M_{\rm line}^{\rm obs} $. {\revised 
 The two blue solid lines show the best fits for the subcritical and supercritical filaments, respectively (see text for details).}  
 }
              \label{VelIntMass_Virparam}
    \end{figure}

Fig.~\ref{VelIntMass_Virparam}a shows the total velocity dispersion of each filament (same y-axis as in  Fig.~\ref{VelDisp_coldens}) as a function 
of the {\revised estimated intrinsic mass per unit length obtained from the observed value after applying a 60\% correction factor }
to account for the average overestimation of the intrinsic column density 
due to projection effects. 
This panel can be divided into two parts by comparing the {\revised intrinsic} mass per unit length to the virial mass per unit length   $M_{\rm line,vir} = 2 \sigma_{\rm tot}^{2}/G$  \citep{Fiege2000}, where $\sigma_{tot}$ is the observed total velocity dispersion (instead of the thermal sound speed used in the expression of  $M_{\rm line,crit}$). 
The dividing line is defined by   $M_{\rm line,vir}=2\,M_{\rm line}^{\rm obs}$ equivalent to $ \alpha_{\rm line,vir}~=~2$, where
\begin{equation}
 \alpha_{\rm line,vir} =M_{\rm line,vir}/M_{\rm line}^{\rm obs}
\end{equation}
is the  {\it virial parameter}  of a filament\footnote{For a sphere of uniform density  $\alpha_{\rm vir}=M_{\rm vir}/ M $ where $ M_{\rm vir} \sim  5R \sigma^{2} / G$ and $M$ is the mass of the object \citep[e.g.,][]{Bertoldi1992}.}.

Fig.~\ref{VelIntMass_Virparam}b shows how the  virial parameter depends on the {\revised intrinsic} mass per unit length for our sample of  filaments. This plot (Fig.~\ref{VelIntMass_Virparam}b) can be divided into four quadrants. 
The {\revised effective  critical mass per unit length $M_{\rm line,crit}^{\rm eff} = 2 {\sigma_{\rm tot}^{\rm eff}}^2/G $} divides the sample  
into 
subcritical filaments on the left-hand side and 
supercritical filaments on the right-hand side\footnote{{\rev In this paper, we use the terms `supercritical' and `subcritical' to characterize 
the mass per unit length of a filament, independently of the magnetic field strength  (but see Fiege \& Pudritz 2000
for how $M_{\rm line,crit}$ is modified in the presence of magnetic fields).
This should not be confused with the terms `(magnetically) supercritical and subcritical' 
used to characterize clouds with supercritical and subcritical mass-to-flux ratios, respectively.}}.
{\revised Conceptually, we define $\sigma_{\rm tot}^{\rm eff}$ as the total velocity dispersion of a filament on the verge of global radial collapse.
Observationally, the range of internal velocity dispersions $\sigma_{\rm tot} $ measured here for thermally subcritical and nearly critical filaments 
suggests that $\sigma_{\rm tot}^{\rm eff}$  corresponds to effective temperatures 
$T_{\rm eff} \equiv \mu \, m_{\rm H}\, {\sigma_{\rm tot}^{\rm eff}}^2/k_{\rm B} $ in the range $10 \la T_{\rm eff} \la 20$~K. 
This implies an {\it effective} critical mass per unit 
length $M_{\rm line,crit}^{\rm eff} $ in the range between $\sim$ 16 and 32~M$_{\odot}$/pc, which 
is shown as a grey band in the plots of Fig.~\ref{VelIntMass_Virparam}a and Fig.~\ref{VelIntMass_Virparam}b.
}
The virial parameter, $\alpha_{\rm line,vir}$, %
divides the sample of filaments into   unbound ($\alpha_{\rm line,vir}~>~2$) structures in the upper part of  the diagram and gravitationally bound structures ($\alpha_{\rm line,vir}~<~2$) in the lower part of the diagram.  {\revised The virial parameter  decreases as $\alpha_{\rm line,vir} \propto M_{\rm line}^{-0.95\pm 0.12}$ for the subcritical filaments,  while this dependence is much shallower,  $\alpha_{\rm line,vir} \propto M_{\rm line}^{-0.24 \pm 0.29}$,  for the supercritical filaments of our sample. 
The best-fit  power law exponent for the supercritical filaments remains almost unchanged   $(-0.28 \pm 0.30)$ when the linear fitting is performed 
after removing  DR21 and NGC2264C from the sample.} 
Figure~\ref{VelIntMass_Virparam}b indicates  that  thermally supercritical filaments tend to be  gravitationally bound with {\revised $ \alpha_{\rm line,vir} = 1.0~\pm~0.5$ on average}, and that, conversely, all  thermally subcritical filaments are unbound with $ \alpha_{\rm line,vir}~>~2$. %

  \section{Discussion and conclusions}

The results of our 
$^{13}$CO, C$^{18}$O, and N$_{2}$H$^{+}$ line observations toward a sample of  filaments previously detected with $Herschel$ 
show that, to an excellent approximation, interstellar filaments may be divided into bound and unbound structures depending on 
whether their mass per unit length $M_{\rm line} $ is larger or smaller than the {\it thermal} value of the critical mass per unit length 
$M_{\rm line,crit}~=~2c^{2}_{\rm s}/G$. 
Indeed, thermally supercritical filaments are found to be self-gravitating  
structures in rough virial balance with $ M_{\rm line}^{\rm obs} \sim M_{\rm line,vir}$, 
while thermally subcritical filaments appear to be unbound structures with $ M_{\rm line}^{\rm obs} \leq M_{\rm line,vir}/2$
and transonic line-of-sight velocity dispersions ($c_s \la \sigma_{tot} < 2\, c_s $).
This confirms the usefulness of the simple gravitational instability criterion based on $M_{\rm line,crit} $
which was adopted by \citet{Andre2010} in their analysis of the first results from the HGBS.

The detection, in the same cloud (e.g. Aquila), of both subcritical and supercritical filaments, with transonic and supersonic velocity dispersions respectively, 
suggests that the internal velocity dispersion within a filament is not directly related to the level of large-scale turbulence in the parent cloud. 
Instead, we propose that the internal velocity dispersion partly reflects the evolutionary state of a filament (see below). 
 
   \subsection{An evolutionary scenario for interstellar filaments?}

In the light of the results presented in this paper, we may revisit the issue of the formation and evolution of interstellar filaments 
and improve the  
scenario proposed in our earlier work based on HGBS observations \citep[e.g.][]{Arzoumanian2011}.

 The observed omnipresence of filamentary structures in gravitationally unbound complexes 
such as the Polaris translucent cloud \citep{Men'shchikov2010, Miville2010, Ward-Thompson2010} suggests
that large-scale turbulence rather than large-scale gravity play the dominant role in {\it forming} interstellar filaments. 
The finding of a characteristic  filament width $\sim 0.1$~pc \citep{Arzoumanian2011}, corresponding to better than a factor of $\sim 2$ 
to the sonic scale below which interstellar turbulence becomes subsonic in  diffuse molecular gas  \citep[cf. ][]{Goodman1998,Falgarone2009,Federrath2010}, 
also supports the view that filaments may form by turbulent compression of interstellar gas \citep{Padoan2001}.
In such a picture, filaments coincide with stagnation gas associated with regions of locally converging turbulent motions, where compression is at a maximum and relative velocity differences 
at a minimum, and are thus expected to have relatively low (transonic) internal velocity dispersions \citep[cf.][]{Klessen2005}. 
The line-of-sight velocity dispersions measured here for subcritical filaments, which all are in the range $c_s \la \sigma_{tot} \la 2\, c_s $, are consistent with this picture.
Moreover, our $^{13}$CO(2--1) and  C$^{18}$O(2--1) mapping results for filament 14 in IC5146 
(cf. Fig.~\ref{SubColdensCO} and Fig.~\ref{veloFil14} -- see also Hacar $\&$ Tafalla 2011 for other filaments in Taurus)
suggest that subcritical filaments are velocity-coherent structures in MCs, with small and roughly constant velocity dispersion along their length
\citep[cf.][for the definition of the concept of coherence in the context of dense cores]{Goodman1998,Heyer2009}. 
 
Since subcritical filaments are unbound, they may be expected to widen and disperse on a turbulent crossing time.  Given the mean velocity dispersion $\overline{\sigma}_{\rm tot}  \sim 0.3$~km/s measured here for subcritical filaments, the typical crossing time of a $\sim 0.1$-pc-wide filament is $\sim$~0.3~Myr. 
Subcritical filaments may be expected to disperse on this timescale unless they are confined by some external pressure as suggested by \citet{Fischera2012}. 
These authors modeled interstellar filaments as quasi-equilibrium isothermal structures in pressure balance 
with a typical ambient ISM pressure $P_{\rm ext} {\sim} 2$$-$5$\times$$10^4 \, \rm{K\, cm}^{-3} $ \citep[][]{Fischera2012}. 
The predicted widths of their model filaments are in rough agreement with the observed filament width of $\sim$0.1~pc. 
Moreover, the agreement between the expected and the observed widths improves when polytropic models of pressure-confined filaments,  
obeying a non-isothermal equation of state $P \propto \rho^\gamma $ with $\gamma < 1$   are considered ({\rev S. Inutsuka, private communication}). 

The observed increase in velocity dispersion with increasing central column density for thermally supercritical filaments,  
roughly consistent with the scaling $ \sigma_{\rm tot} \propto {N_{\rm H_2}}^{0.5}$  (cf. Fig.~\ref{VelDisp_coldens}), 
provides interesting clues to the evolution of dense, star-forming filaments. 
Since the filaments on the right-hand side of Fig.~\ref{VelDisp_coldens} have $\alpha_{\rm line,vir} \sim 1$ and are self-gravitating 
(see Sect.~5), they should be unstable to radial collapse and thus have a natural tendency to contract with time \citep[cf.][]{Inutsuka1992,Kawachi1998}. 
The shape of their radial column density profiles {\revised with a high degree of symmetry about the filament major axis 
and a well-defined power law regime at large radii, approximately consistent 
with a radial density structure $\rho \propto r^{-2}$ \citep[][]{Arzoumanian2011,Palmeirim2013,Hill2012Artemis}, 
is consistent with a theoretical model of a nearly isothermal collapsing cylinder \citep[][]{Kawachi1998,Palmeirim2013}}, and therefore 
also suggestive of gravitational contraction. 
If supercritical filaments are indeed contracting, then their central column density is expected to increase with time 
\citep[cf.][]{Kawachi1998}. 
It is thus tempting to suggest that the right-hand side of Fig.~\ref{VelDisp_coldens} may, at least partly, represent
an evolutionary sequence for supercritical filaments, in which central column density corresponds to  an evolutionary indicator, 
increasing with time. 
As the internal velocity dispersion of a filament provides a direct measure of its virial mass per unit length and supercritical filaments have 
{\revised $ M_{\rm line}  \ga M_{\rm line,vir}$}, the trend seen on the right-hand side of Fig.~\ref{VelDisp_coldens} suggests that 
the mass per unit length of supercritical filaments increases with time through accretion of background material. 
More direct evidence of this accretion process and growth in mass per unit length for supercritical filaments exists in 
several cases in the form of low-density striations or subfilaments observed perpendicular to the main filaments 
and apparently feeding them from the side. Examples include the B211/B213 filament in Taurus  %
{\revised where  the  typical velocities expected for the infalling material  are consistent with the existing kinematical constraints 
from  CO observations (Goldsmith et al. 2008; Palmeirim et al. 2013),}
the Musca filament (Cox et al. 2013, in preparation), %
and the DR21 ridge in Cygnus~X \citep[][]{Schneider2010}.

\begin{figure}
    \centering
     \resizebox{9.cm}{!}{
     \includegraphics[angle=0]{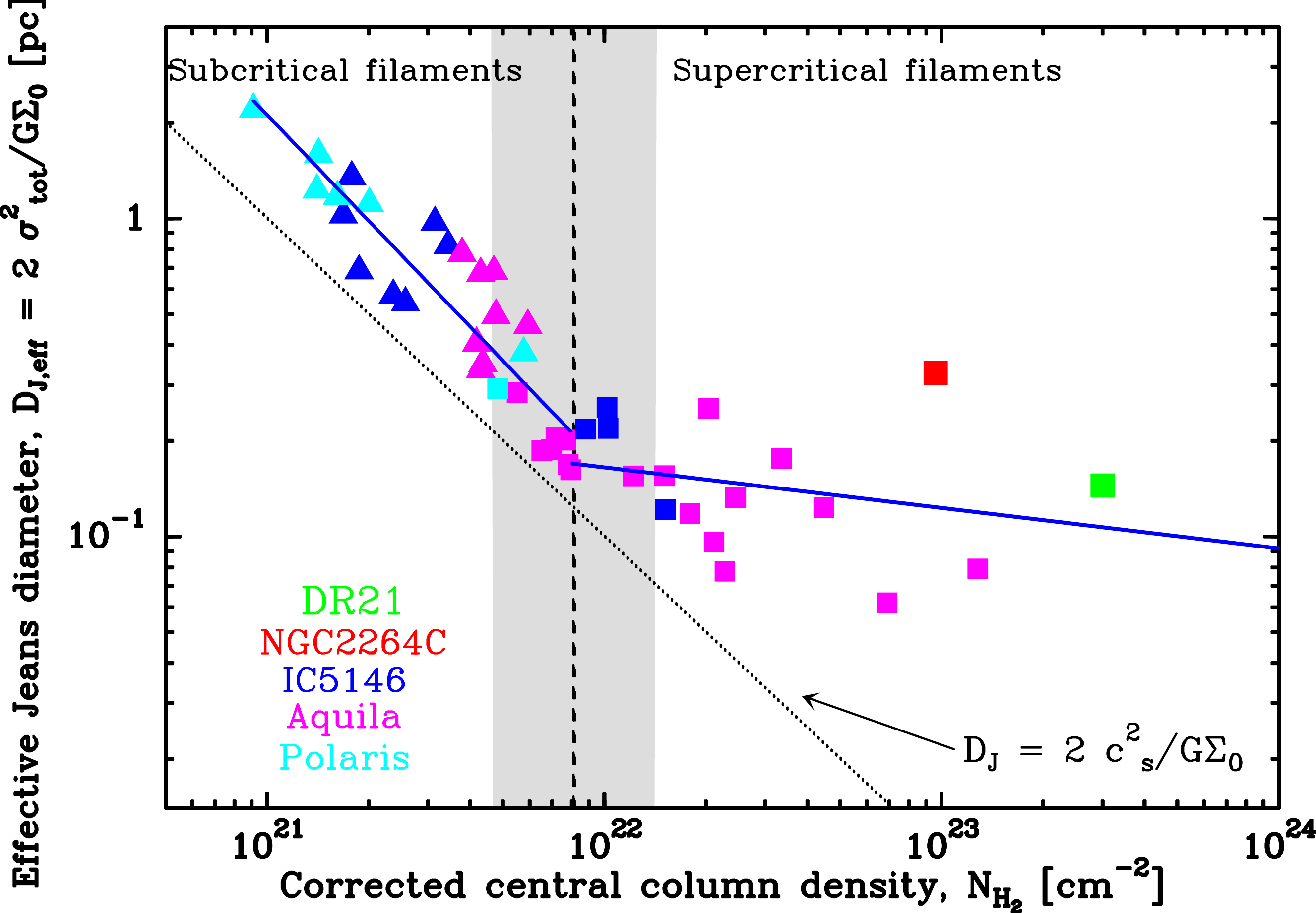}}
          \caption{Effective Jeans diameter ($D_{\rm{J,eff}} = 2\sigma^{2}_{\rm tot}/G\Sigma_{0}$)  versus   filament {\revised  intrinsic} central  column density. %
The vertical dashed line {\revised and the grey band mark} the {\revised border zone} between subcritical and  supercritical filaments (cf. Fig.~\ref{VelDisp_coldens}). 
The {\revised dotted} line running from the top left to the bottom right  corresponds to  the thermal Jeans diameter (2$R_{\rm{J}}$). {\revised The blue solid lines correspond to the best fits for the subcritical ($D_{\rm J,eff}^{\rm sub} \propto {N_{\rm H_2}}^{-1.1~\pm~0.1}$) and supercritical ($D_{\rm J,eff}^{\rm sup}  \propto {N_{\rm H_2}}^{-0.2~\pm~0.1}$) filament subsamples, respectively. The exponent of the power-law fit  for the supercritical filaments becomes $-0.4 \pm 0.1$ after excluding DR21 and NGC2264C from the sample.}  In contrast to the thermal Jeans length, the effective Jeans diameter of the supercritical filaments remains roughly constant with an average  value of 
 $0.14 \pm 0.07$~pc. %
} 
   \label{LJeff_coldens}%
    \end{figure}

If such an accretion process accompanied by an increase in internal velocity dispersion does indeed occur in supercritical filaments, it may explain how these filaments can
maintain a roughly constant inner width while contracting. In collapse models of nearly-isothermal cylindrical filaments %
\citep[cf.][]{Inutsuka1992,Kawachi1998}, 
the flat inner portion of the radial density profile has a radius corresponding to the instantaneous central Jeans length $R_{\rm J} \sim c_{\rm s}^2/G\Sigma_0 $, where $\Sigma_0 $ is again the central 
mass column density of the model filament. In the presence of a nonthermal component to the velocity dispersion, we may thus expect the central 
diameter of a contracting filament to be roughly given by twice the {\it effective} Jeans length, i.e.,  $2\, R_{\rm J,eff} \equiv D_{\rm J,eff} \sim 2\, \sigma_{\rm tot}^2/G\Sigma_0 $. 
This effective Jeans diameter is plotted in Fig.~\ref{LJeff_coldens} as a function of {\revised corrected} central column density for the sample of filaments studied in the present paper.
It can be seen in Fig.~\ref{LJeff_coldens} that $D_{\rm J,eff}$ is approximately constant for the thermally supercritical filaments in our  sample, 
with an average {\revised value of $0.14 \pm 0.07\,$pc}.

In this scenario, the enhanced velocity dispersions measured for thermally supercritical filaments 
would not directly arise from large-scale interstellar turbulence but from the accretion of ambient cloud gas and 
the gravitational amplification of initial velocity fluctuations through the conversion 
of gravitational energy into kinetic energy during the accretion and 
contraction process (see, e.g., the numerical simulations presented by Peretto et al. 2007 to explain the observed characteristics of NGC2264C).  
{\revised This would also be consistent with the concept of accretion-driven turbulence proposed by  \citet{Klessen2010} to explain the origin 
of turbulent motions in a wide range of astrophysical objects.} 
\nocite{Peretto2007}

At the same time as they contract and accrete material from the background cloud, supercritical filaments are expected to 
fragment into cores and form (proto)stars  \citep{Inutsuka1997,Pon2011}. 
Indeed, for filamentary clouds, fragmentation into cores and subsequent local core collapse occur faster than global cloud collapse \citep{Kawachi1998,Pon2011,Toala2012}.
As pointed out by \citet{Larson2005}, the filamentary geometry is thus especially favorable for fragmentation and core formation.
This is also supported by observations of supercritical filaments which are usually found to harbor 
several prestellar cores and Class 0/Class I protostars along their length \citep[cf.][]{Andre2010}.  
Core formation is  particularly well documented in the case of the B211/B213 filament in Taurus \citep{Schmalzl2010,Onishi2002} or  
the Serpens South filament in the Aquila Rift \citep{Gutermuth2008,Maury2011}.

To assess the reliability of this tentative scenario for filament formation and evolution,  
comparison with dedicated numerical simulations would be very valuable. More extensive  molecular line mapping of a larger sample of filaments are also very desirable,  
in order to set stronger observational constraints on the dynamics of these structures. 

\begin{acknowledgements}
DA wishes to thank the staff at the IRAM 30m telescope at Pico Veleta for the support  during the observations  and Q. Nguy$\tilde{\hat{\rm e}}$n Lu{\hskip-0.65mm\small'{}\hskip-0.5mm}o{\hskip-0.65mm\small'{}\hskip-0.5mm}ng for his help in reducing the IRAM 30m data using GILDAS/CLASS. 
This work has benefited from the support of the European Research Council under the European Union's Seventh 
Framework Programme (Grant Agreement no. 291294) and of the French National Research Agency (Grant no. ANR--11--BS56--0010).

\end{acknowledgements}

\bibliography{AA}

\Online

    \begin{appendix}
      
    \section{Velocity components of the observed filaments}\label{VelComp}

\begin{figure*}
   \centering
     \resizebox{6.cm}{!}{
     \includegraphics[angle=0]{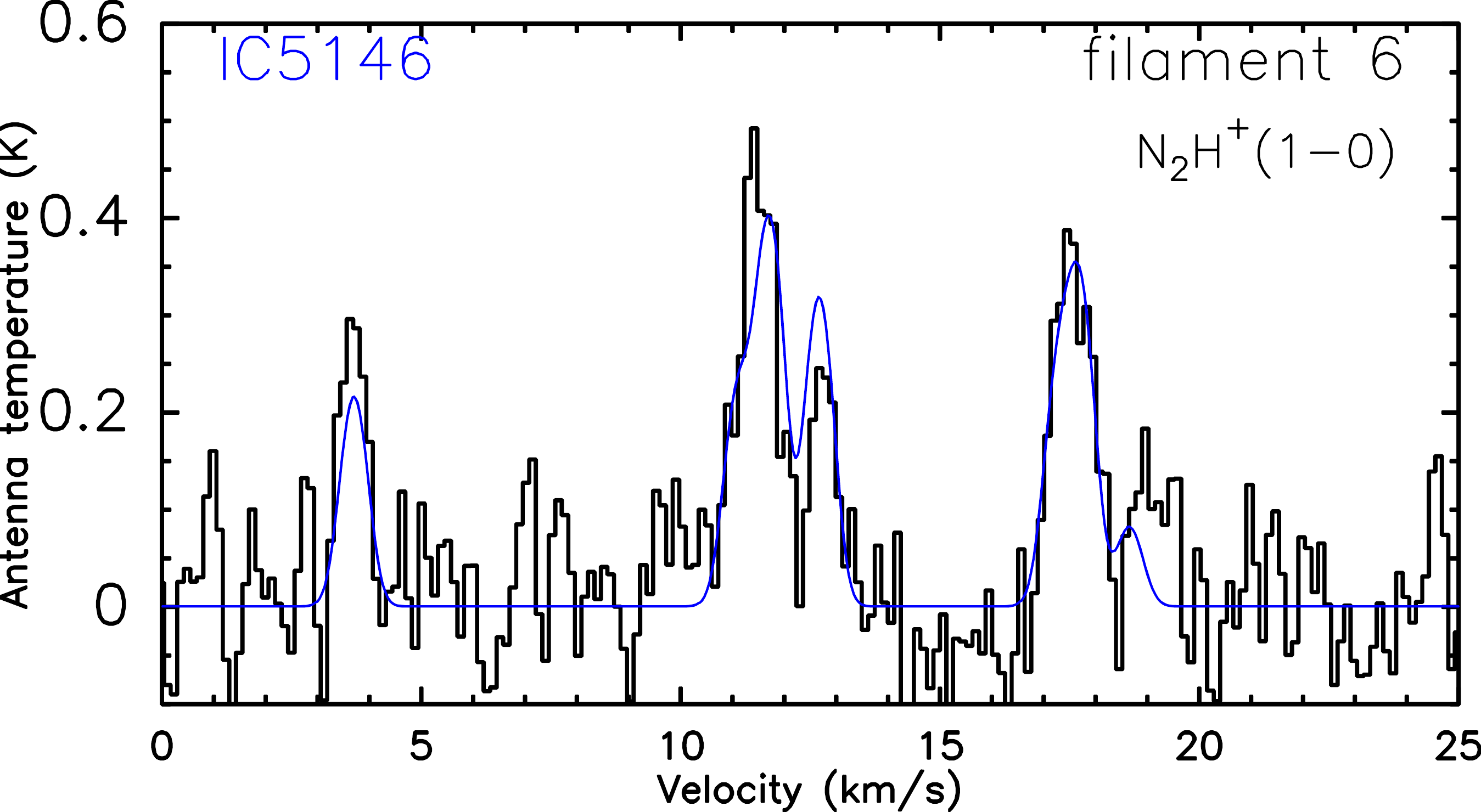}}  
         \resizebox{6.cm}{!}{
     \includegraphics[angle=0]{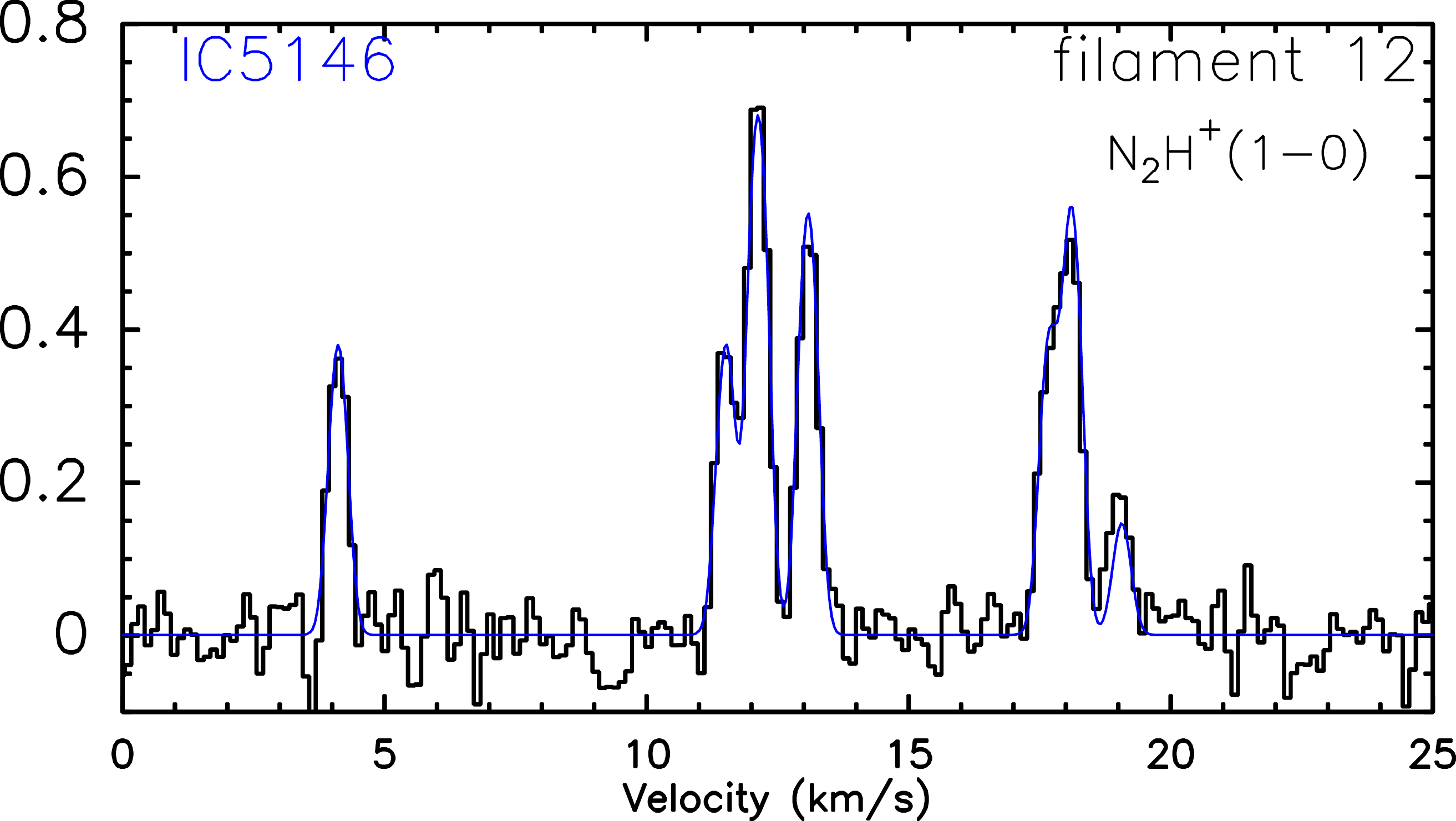}}   
              \resizebox{6.cm}{!}{
     \includegraphics[angle=0]{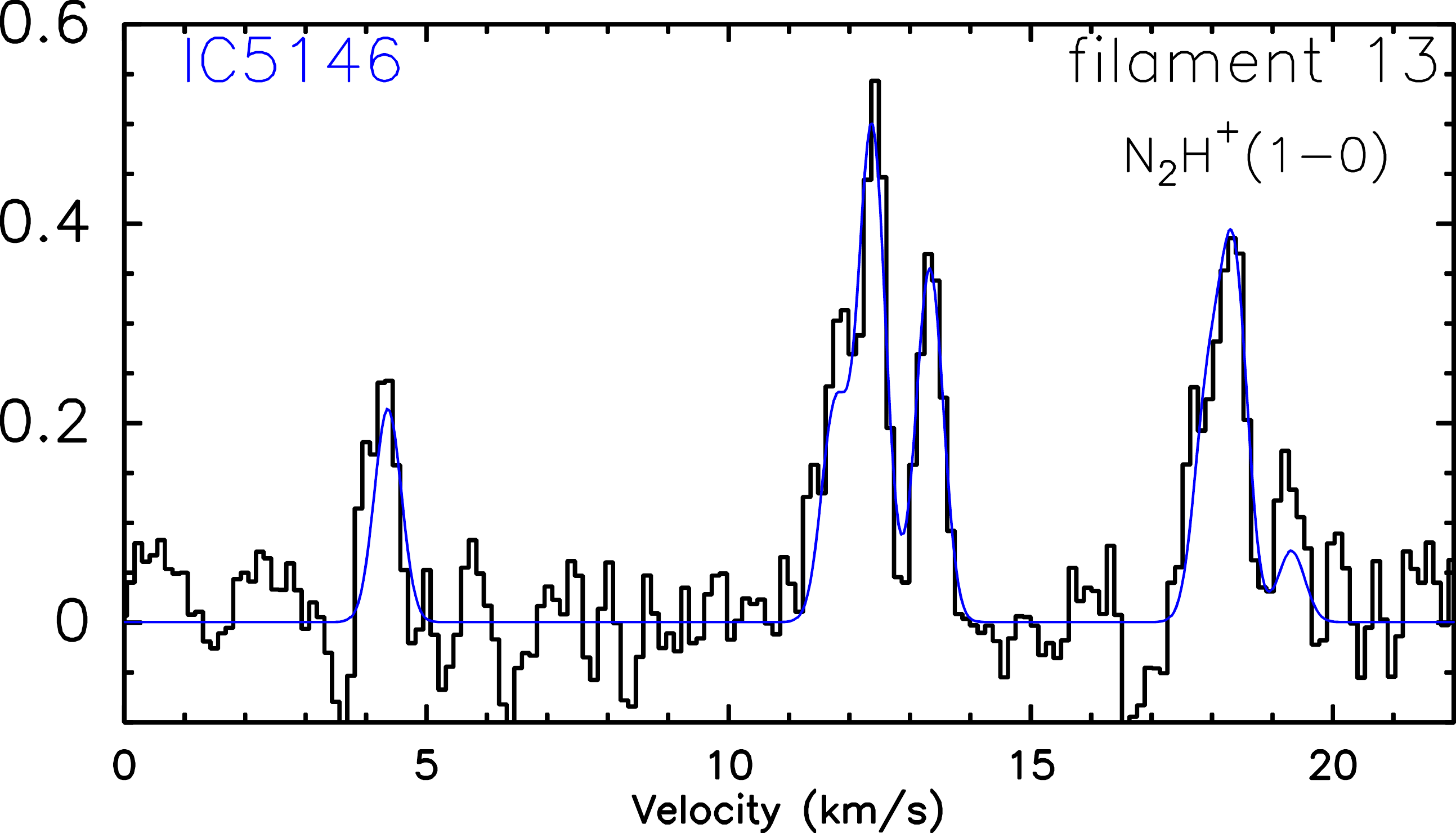}}  
  \caption{ N$_{2}$H$^{+}$(1--0) spectra observed toward  filaments 6, 12, 13 in IC5146. The filament numbers are indicated at the upper right of each panel.  The corresponding one velocity hyperfine structure Gaussian fits are highlighted in blue.}
              \label{ICN2H+spectra}
    \end{figure*}

\begin{figure*}
   \centering
         \resizebox{6.cm}{!}{
     \includegraphics[angle=0]{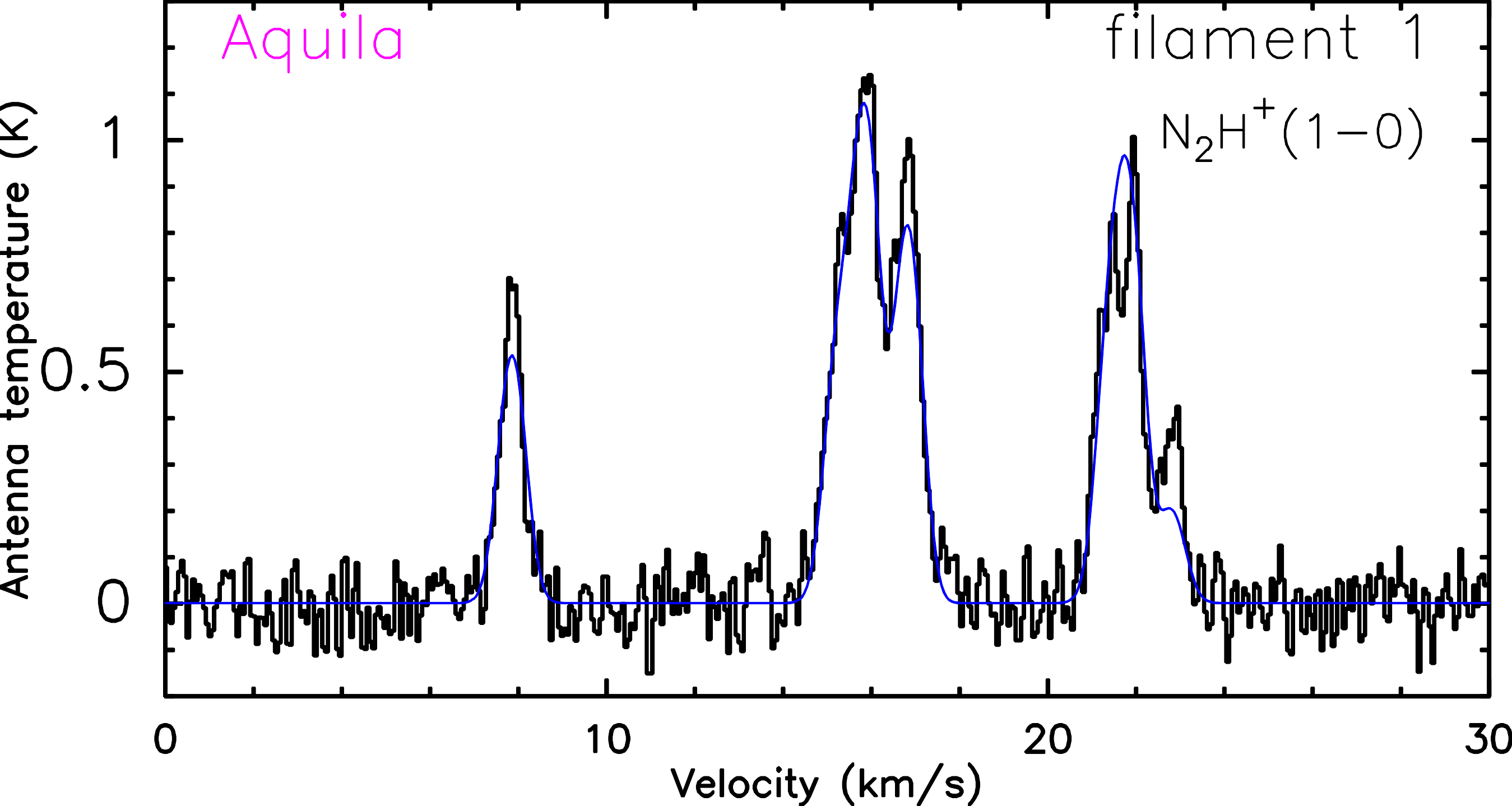}}   
              \resizebox{6.cm}{!}{
     \includegraphics[angle=0]{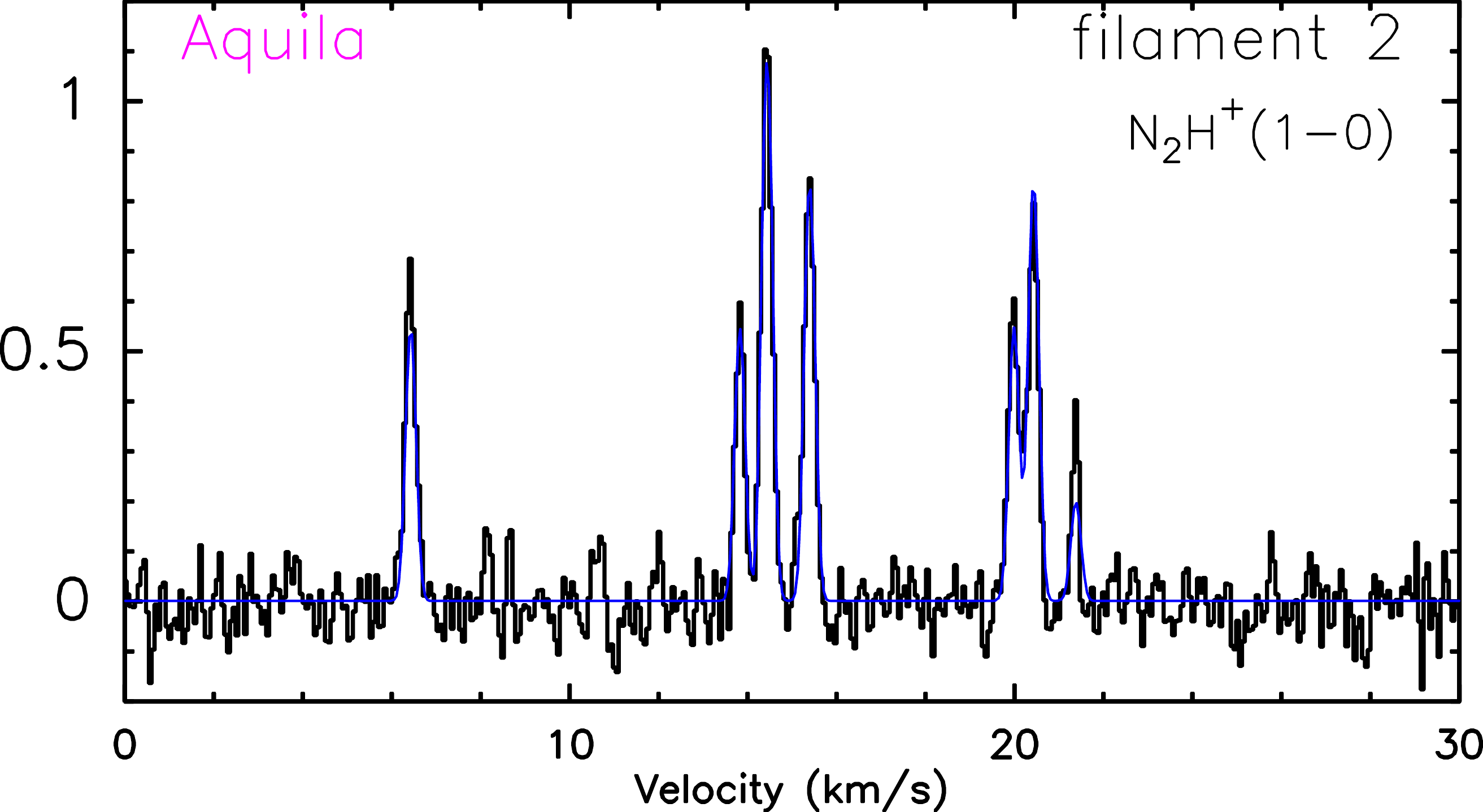}}  
                   \resizebox{6.cm}{!}{
     \includegraphics[angle=0]{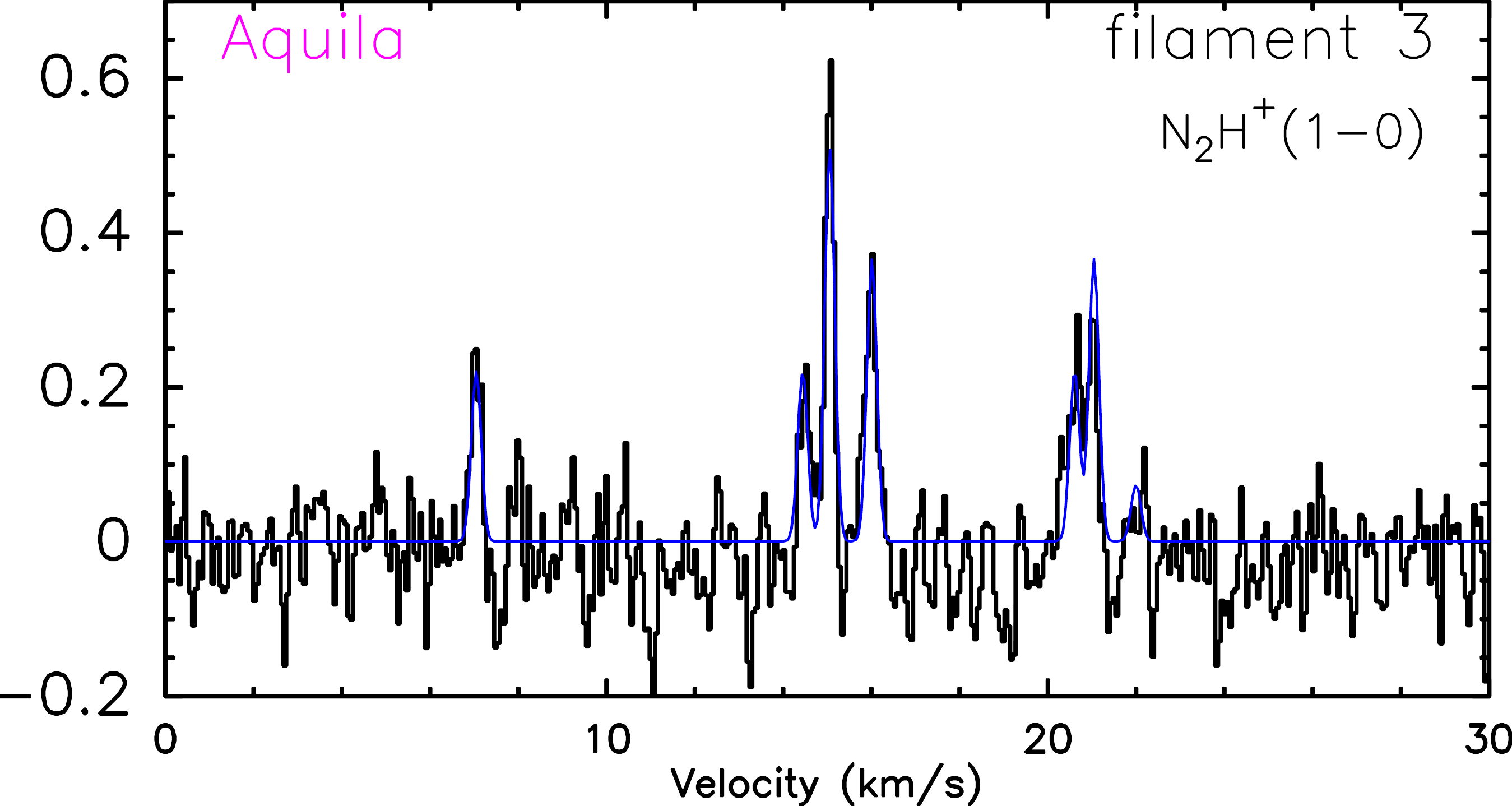}}  
                   \resizebox{6.cm}{!}{
     \includegraphics[angle=0]{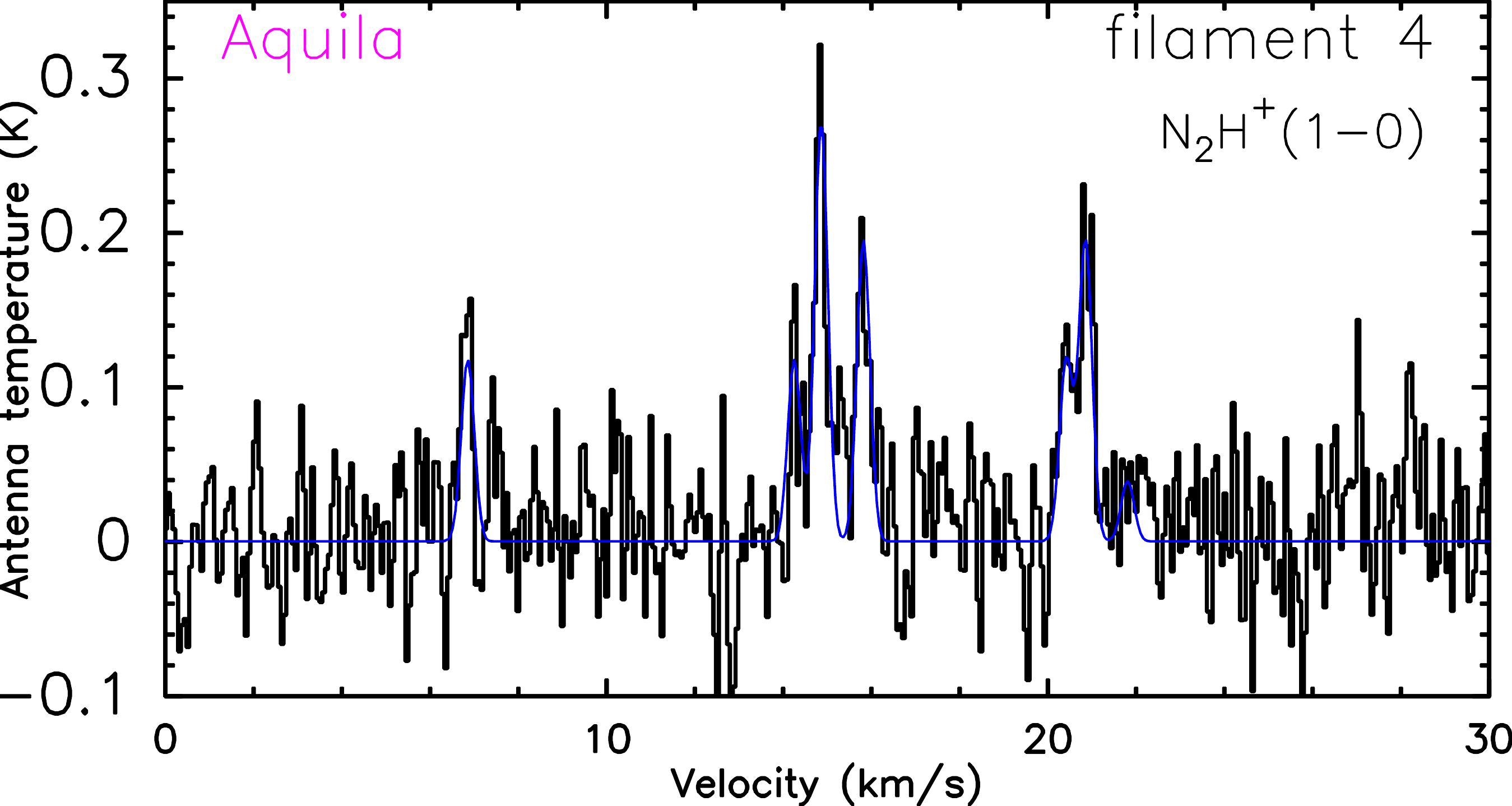}}  
                   \resizebox{6.cm}{!}{
     \includegraphics[angle=0]{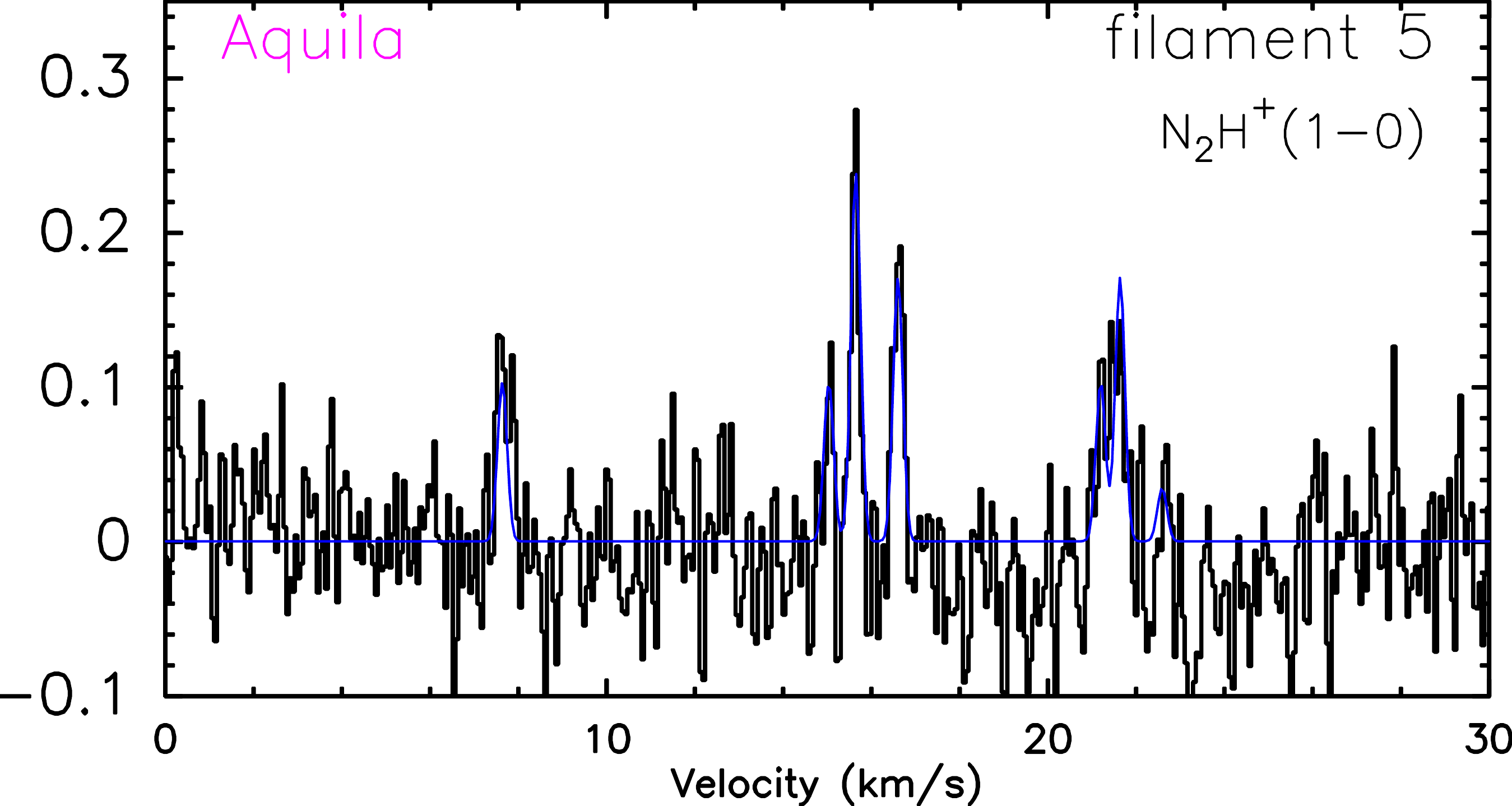}}  
                   \resizebox{6.cm}{!}{
     \includegraphics[angle=0]{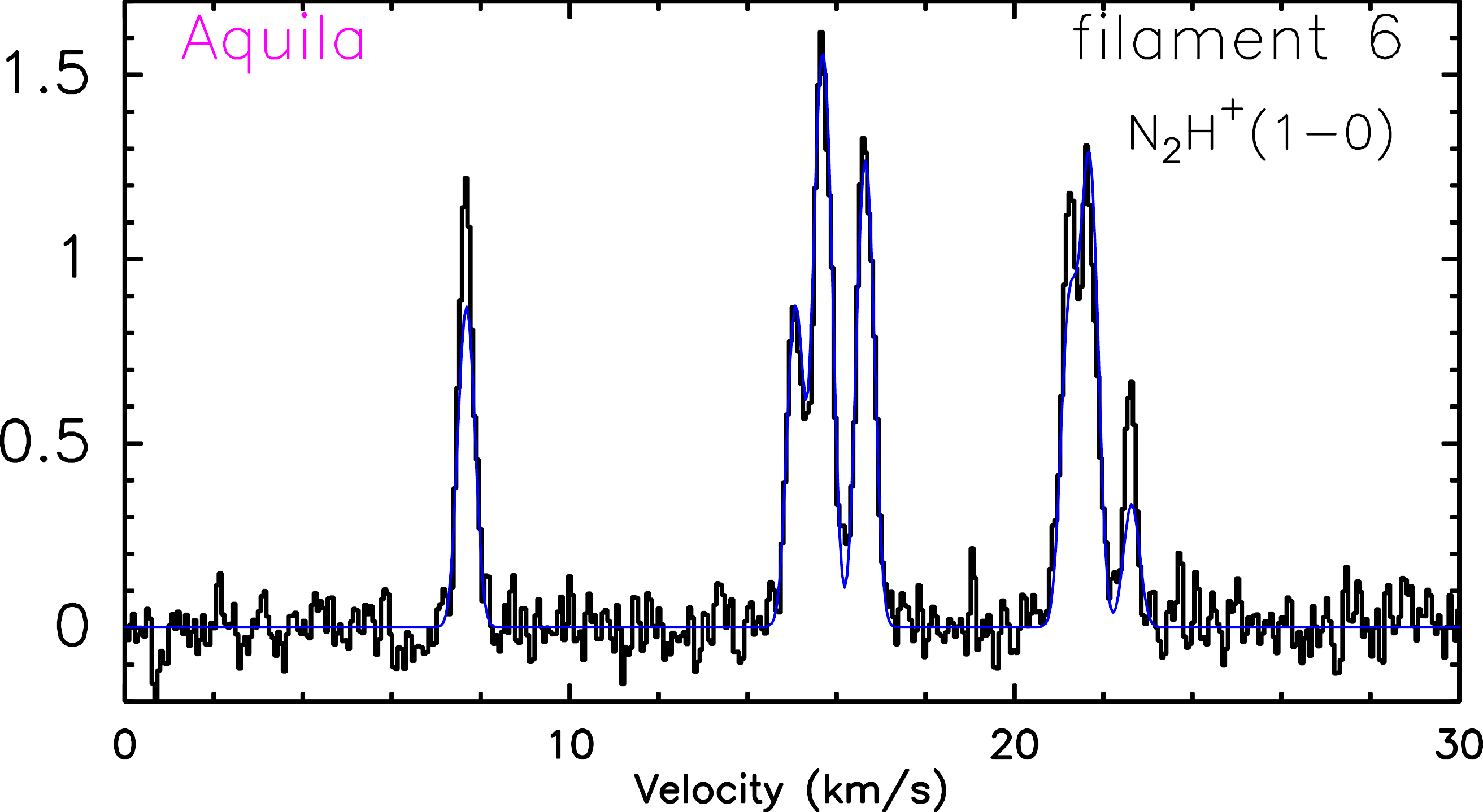}}  
                   \resizebox{6.cm}{!}{
     \includegraphics[angle=0]{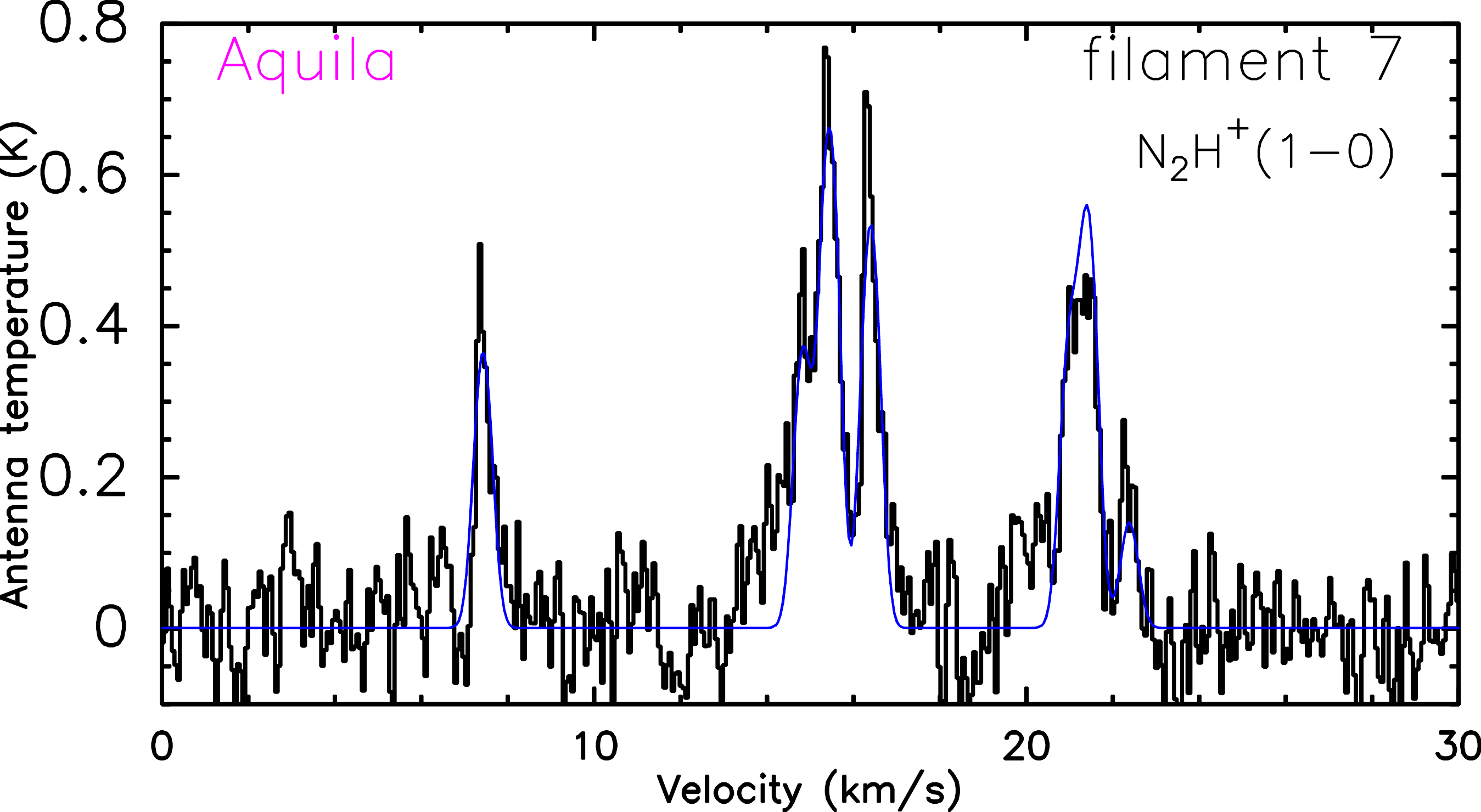}}  
                   \resizebox{6.cm}{!}{
     \includegraphics[angle=0]{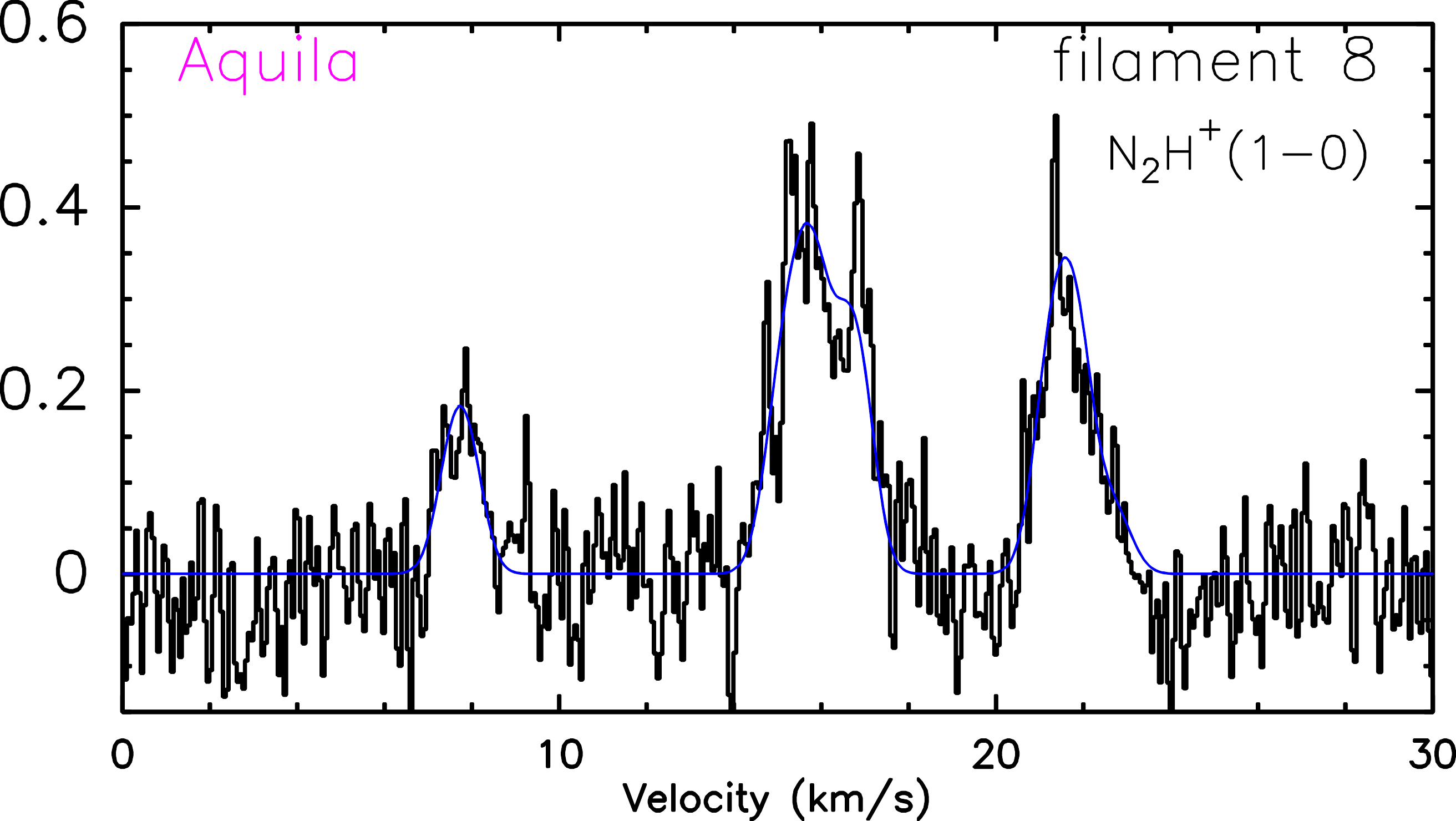}}  
                   \resizebox{6.cm}{!}{
     \includegraphics[angle=0]{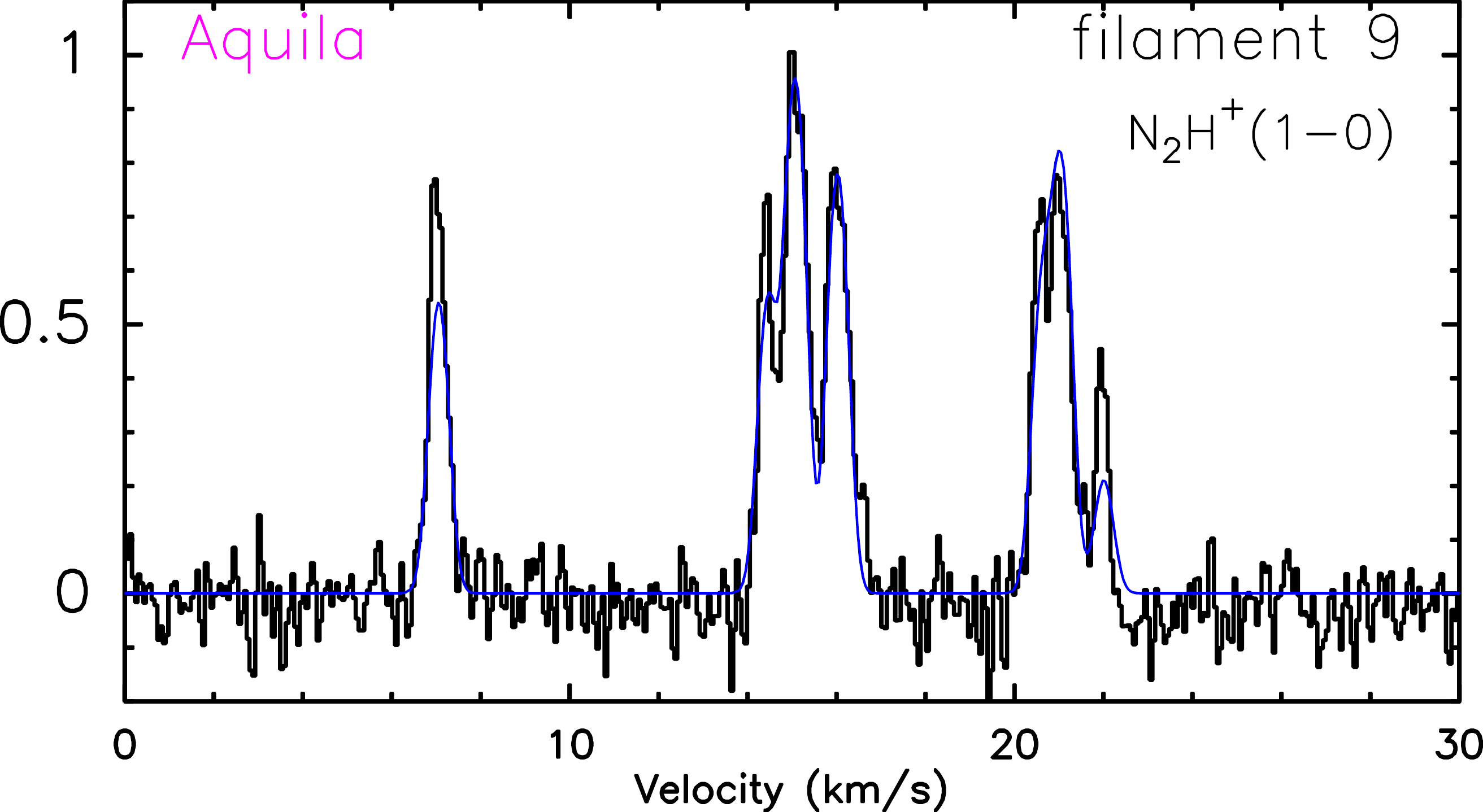}}   
  \caption{N$_{2}$H$^{+}$(1--0) spectra observed toward 9 filaments in Aquila. The  numbers indicated on the upper  right hand side of the plots,  correspond to the filaments marked on the column density map of Fig.~\ref{Aqu} and  listed in Table~1. The corresponding  single-component hyperfine structure Gaussian fits are highlighted in blue. Note that the y-axis scale is  not the same for all  spectra, but has been  adjusted to match the peak temperatures which    vary between   $\sim 0.2$ to $>$~3~K.}
              \label{AqN2H+spectra}
    \end{figure*}

\begin{figure*}
   \centering
     \resizebox{6.cm}{!}{
     \includegraphics[angle=0]{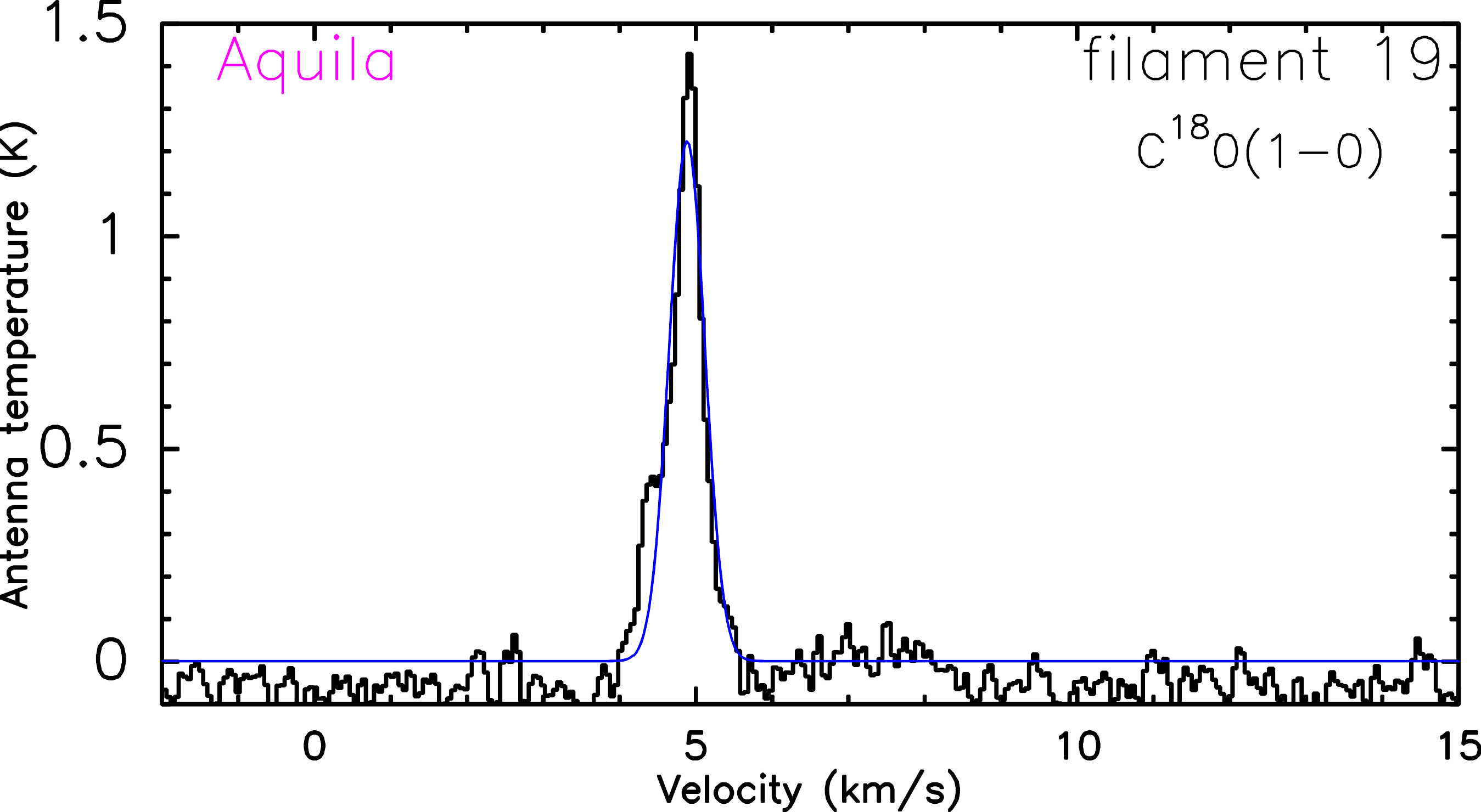}}  
         \resizebox{6.cm}{!}{
     \includegraphics[angle=0]{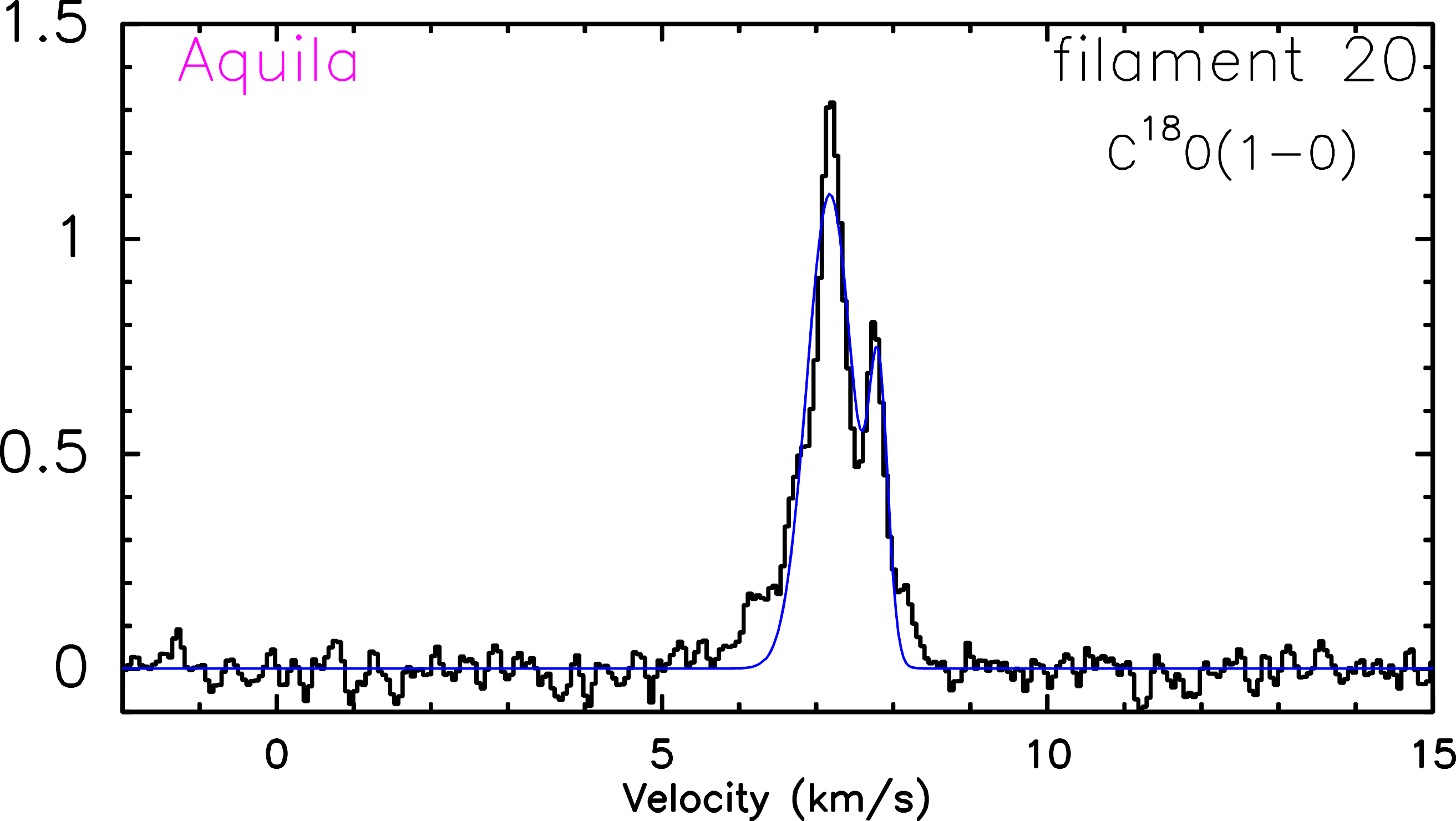}}   
              \resizebox{6.cm}{!}{
     \includegraphics[angle=0]{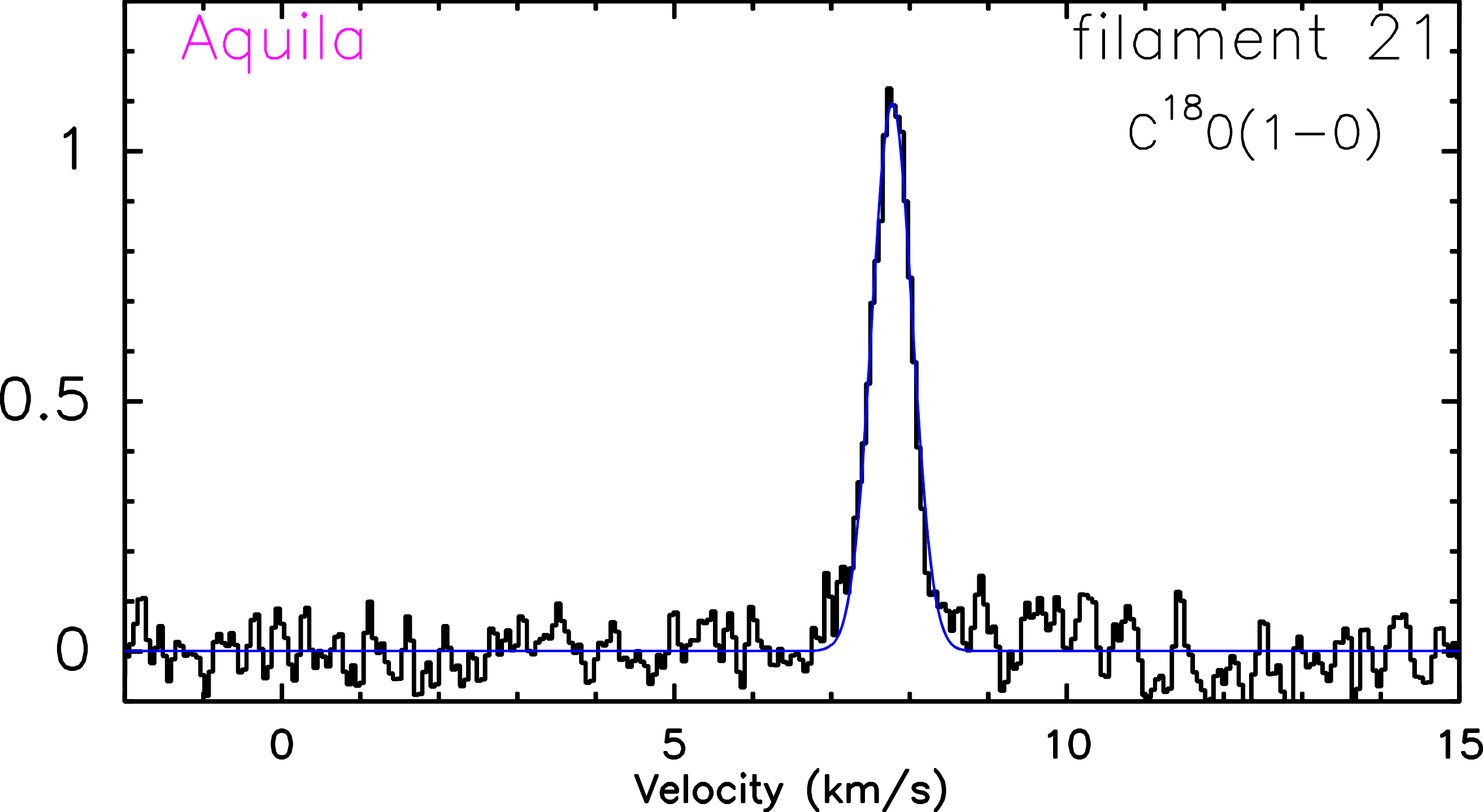}}  
        \resizebox{6.cm}{!}{
        \includegraphics[angle=0]{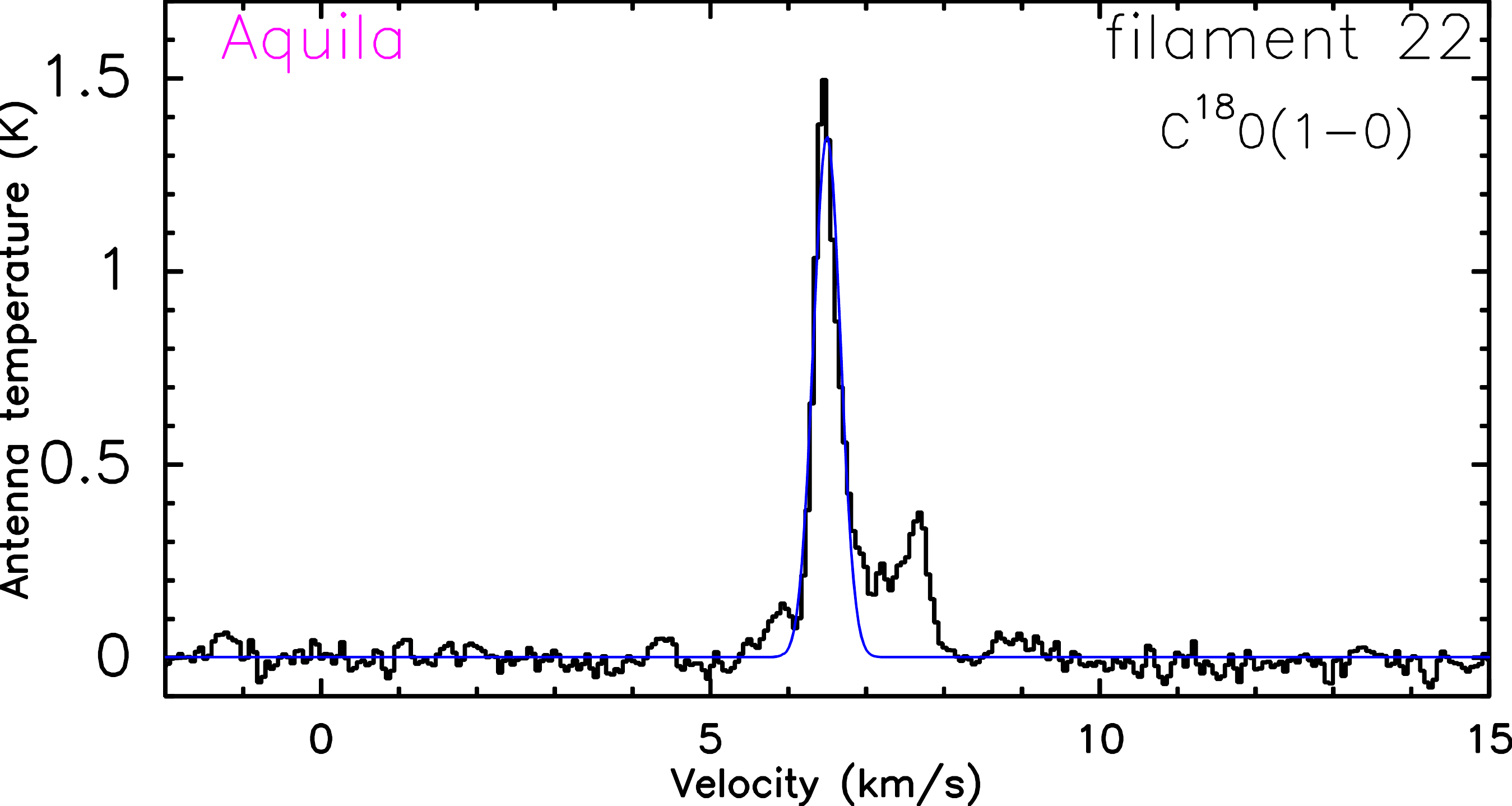}}  
                   \resizebox{6.cm}{!}{
     \includegraphics[angle=0]{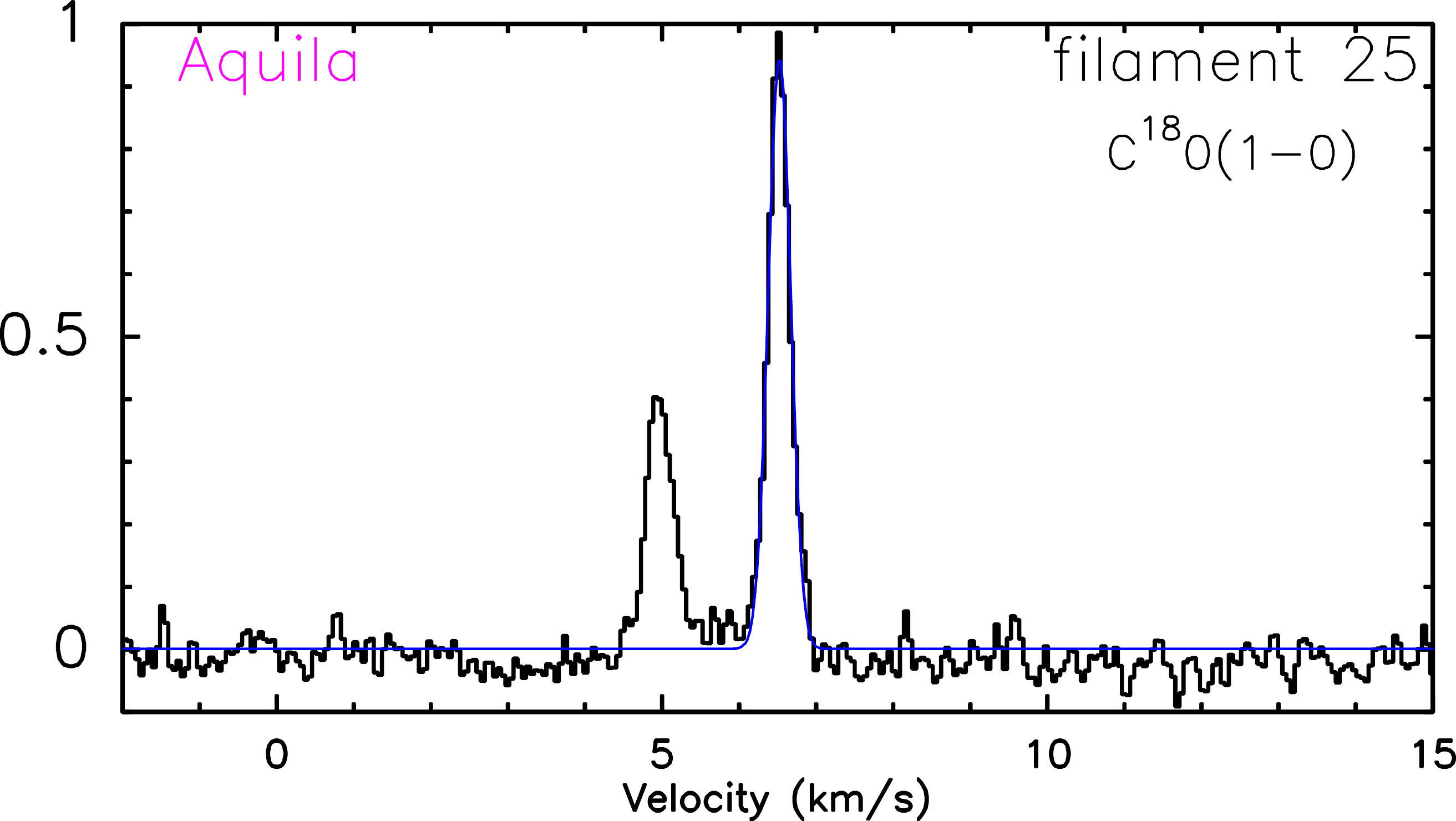}}  
                   \resizebox{6.cm}{!}{
     \includegraphics[angle=0]{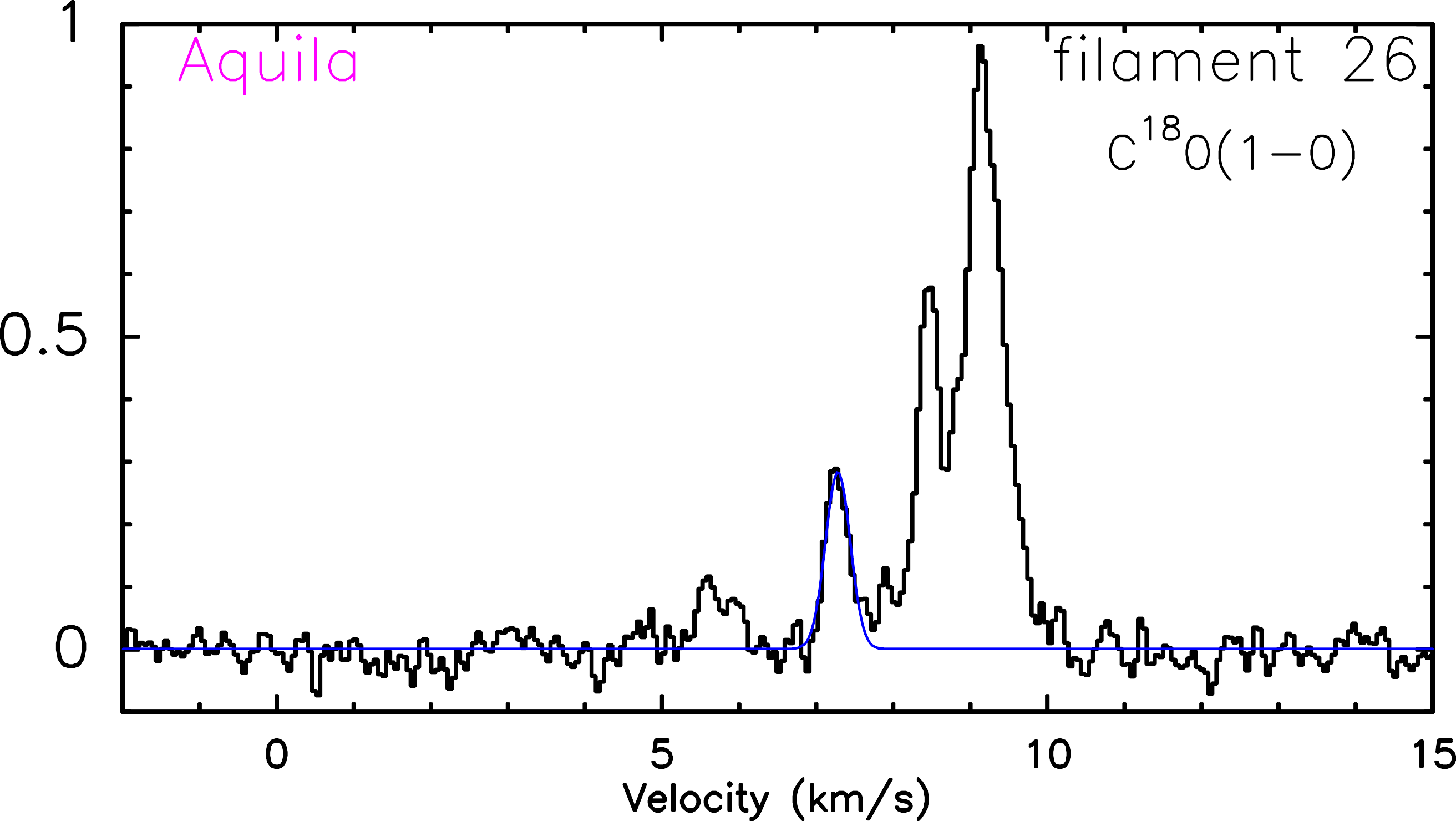}}  
  \caption{ C$^{18}$O(1--0) spectra observed toward 6 filaments in Aquila. The corresponding Gaussian fits are highlighted in blue. Note that the scale of the y-axis  has been  adjusted to match the peak temperature of each spectrum. {\bf Note 1:} Filament 20 shows two components separated by 0.63~km/s. The  component at V$_{\rm LSR}$~=~7.17~km/s have been chosen to be tracing  the filament, since this component has a velocity component which is closer to the average velocity ($\overline{\rm V}_{\rm LSR}~=~7.4$~km/s) of the filaments in Aquila which show one velocity  component. %
    {\bf Note 2:} The spectrum observed at the position of filament 22 shows two velocity components, one centered at 6.5~km/s and the other at 7.7~km/s. The strongest component has  been selected for similar reasons to that explained  for filament 20. {\bf Note 3:} The spectrum observed at the position of filament 25 has two velocity components (at  5.0~km/s and 6.52~km/s, respectively). The strongest component has  been selected for similar reasons to that explained  for filament 20. {\bf Note 4:} The spectrum observed at the position of filament 26 shows three  velocity components centered at 7.28~km/s, 8.45~km/s and 9.17~km/s, respectively. The component centered at 7.28~km/s has    been selected 
  because it is the component with  the closest V$_{\rm LSR}$ value  to the mean $\overline{\rm V}_{\rm LSR}$ of  neighboring filaments in Aquila. 
      }
              \label{Aq_C18Ospectra}
    \end{figure*}

\begin{figure*}
   \centering
    \resizebox{7.cm}{!}{
 \includegraphics[angle=0]{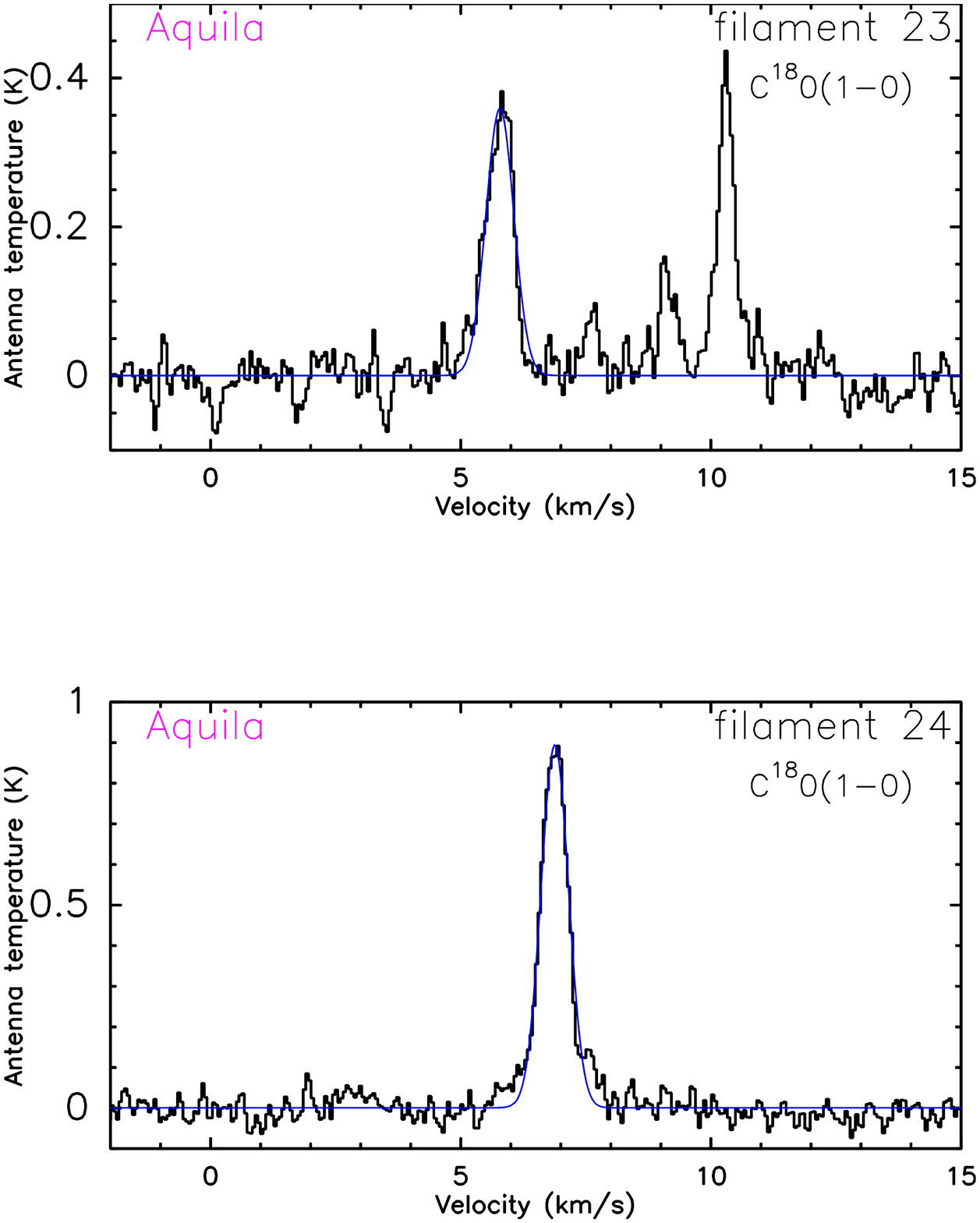}}  
                   \resizebox{10.cm}{!}{
     \includegraphics[angle=0]{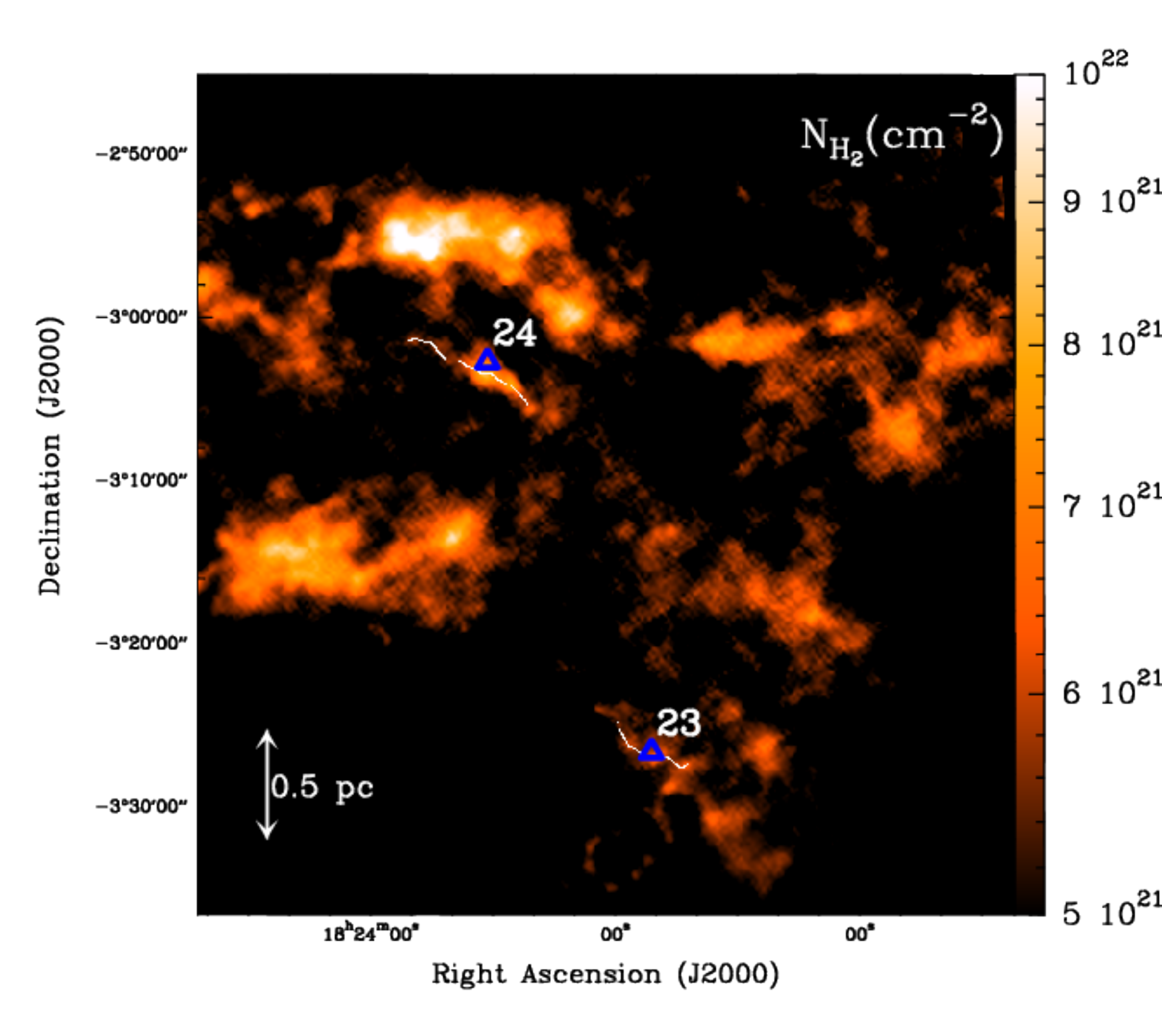}}  
  \caption{ C$^{18}$O(1--0) spectra  observed toward  filaments 23 and 24  in Aquila, located in the south west part of the region. The positions of the two filaments are shown on a  blow up of the column density map {\revised \citep[cf.,][]{Konyves2010} } on the right hand side. Filament 24 shows a single velocity C$^{18}$O(1--0) spectrum at V$_{\rm LSR}$~=~6.87~km/s, while  C$^{18}$O(1--0)  spectrum observed toward filament 23  shows three velocity components at V$_{\rm LSR}$~=~5.77~km/s, 9.13~km/s and 10.3~km/s.  The relevant velocity component associated to  filament 23 is most probably    the component  at V$_{\rm LSR}$~=~5.77~km/s (the closest to the LSR velocity of the neighboring filament 24), while the other components are likely  associated to  clouds/filaments observed on the same line-of-sight. %
  }
              \label{Aqu_SW}
    \end{figure*}

\begin{figure*}
   \centering
     \resizebox{7.cm}{!}{
     \includegraphics[angle=0]{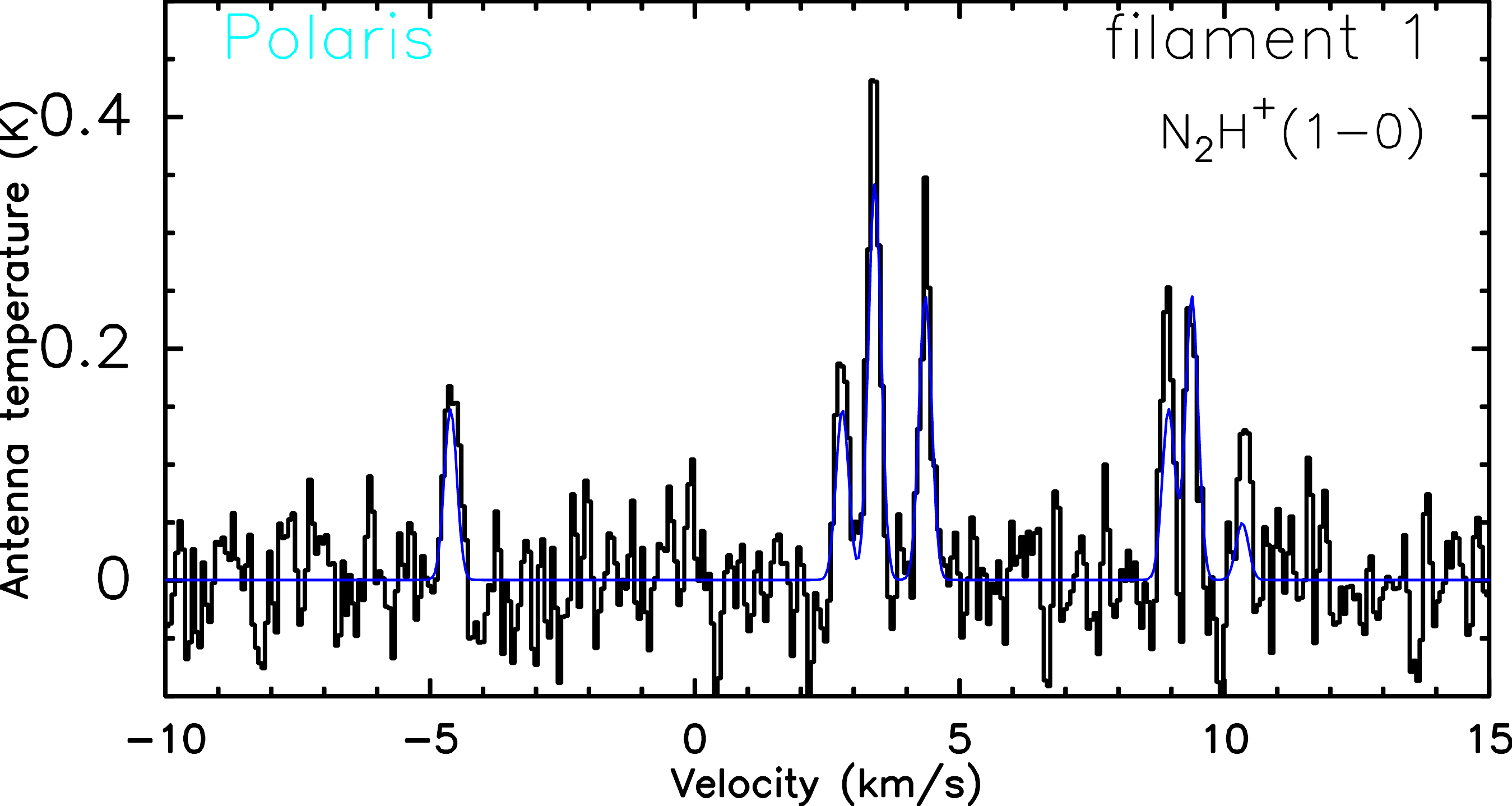}}  
  \caption{N$_{2}$H$^{+}$(1--0)  spectrum  observed toward filament 1 in Polaris. The corresponding hyperfine structure Gaussian   fit is highlighted in blue.}
              \label{PolN2H+spectra}
    \end{figure*}

\begin{figure*}
   \centering
      \resizebox{6.cm}{!}{
     \includegraphics[angle=0]{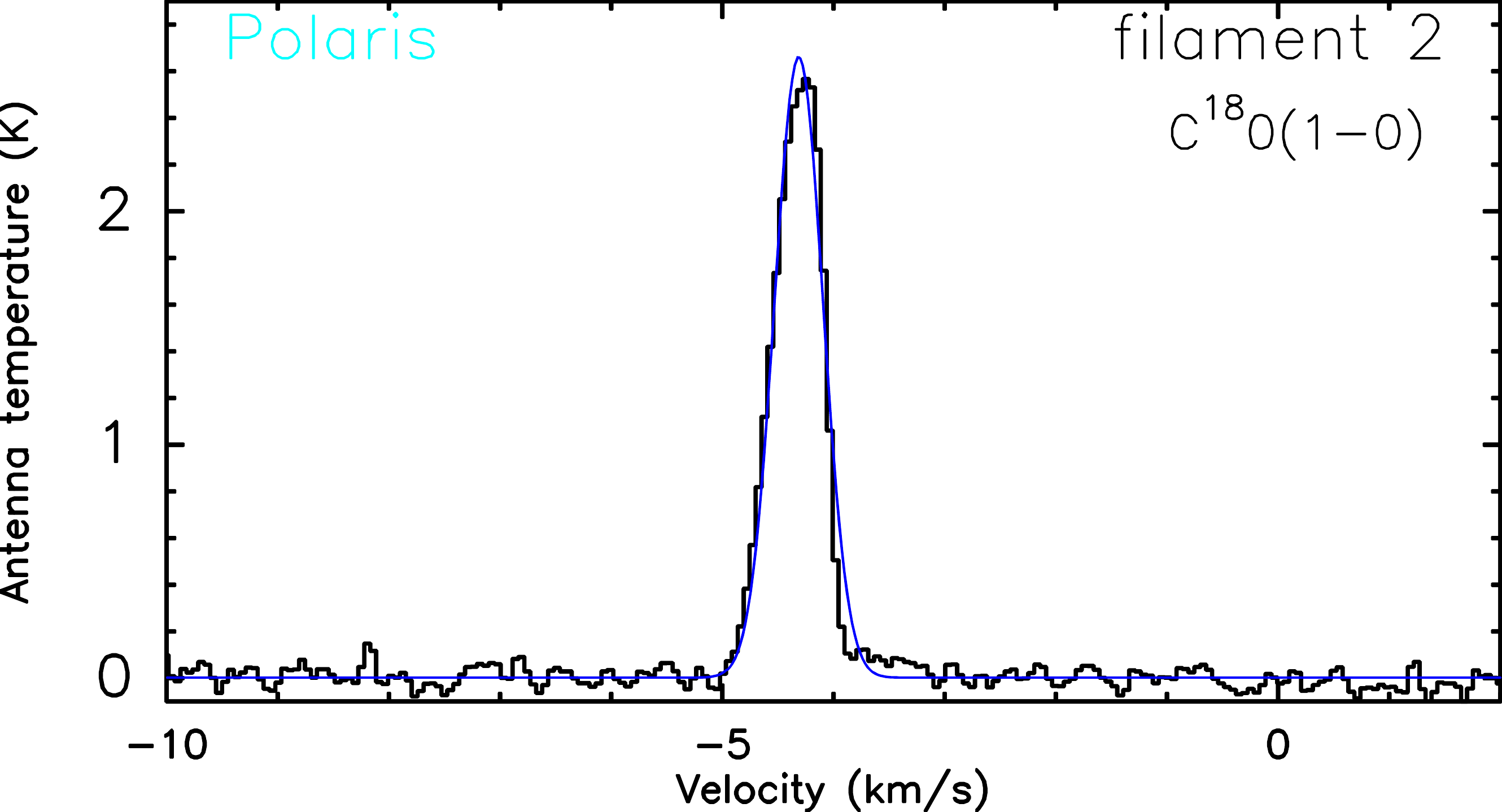}}   
              \resizebox{6.cm}{!}{
     \includegraphics[angle=0]{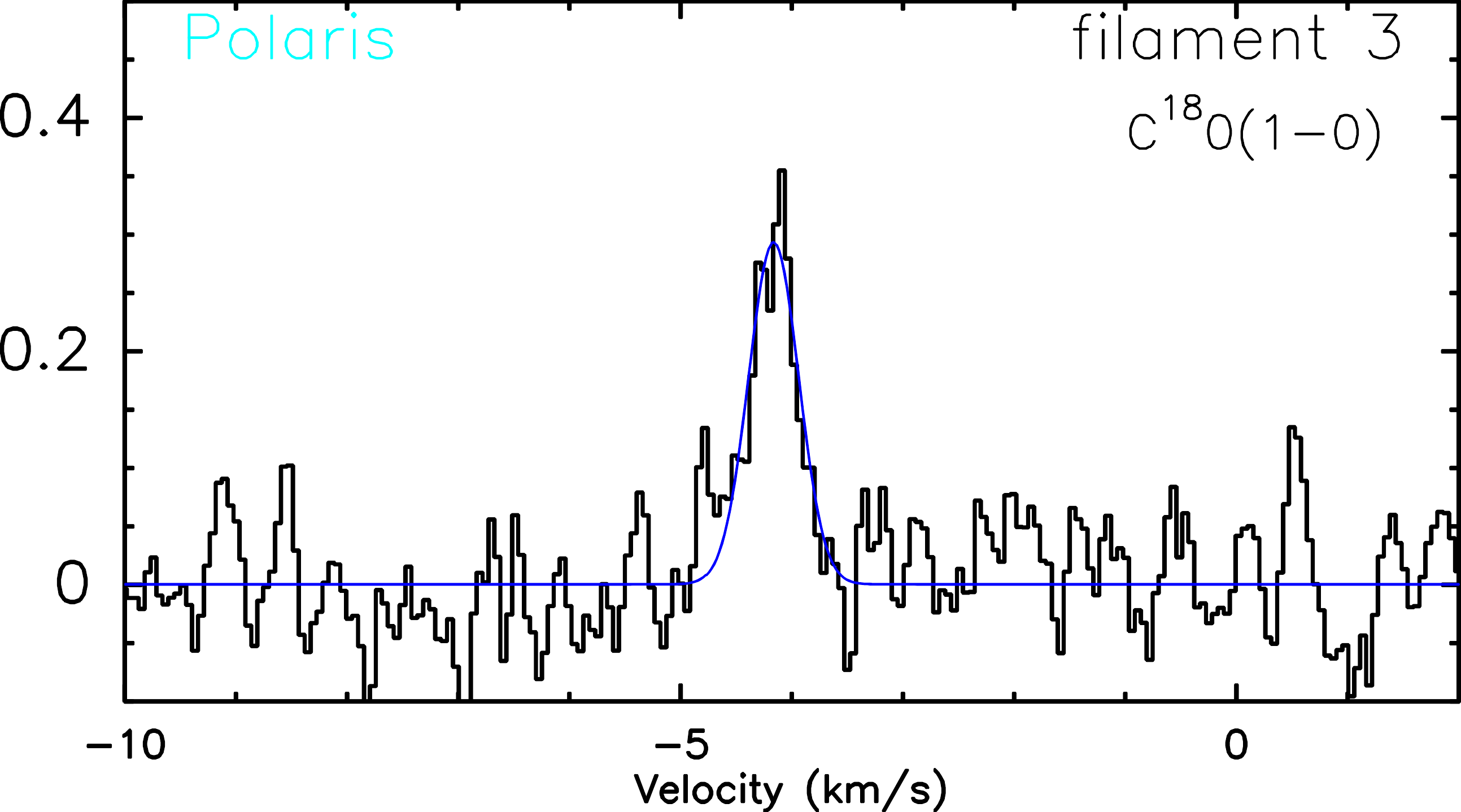}}  
                   \resizebox{6.cm}{!}{
     \includegraphics[angle=0]{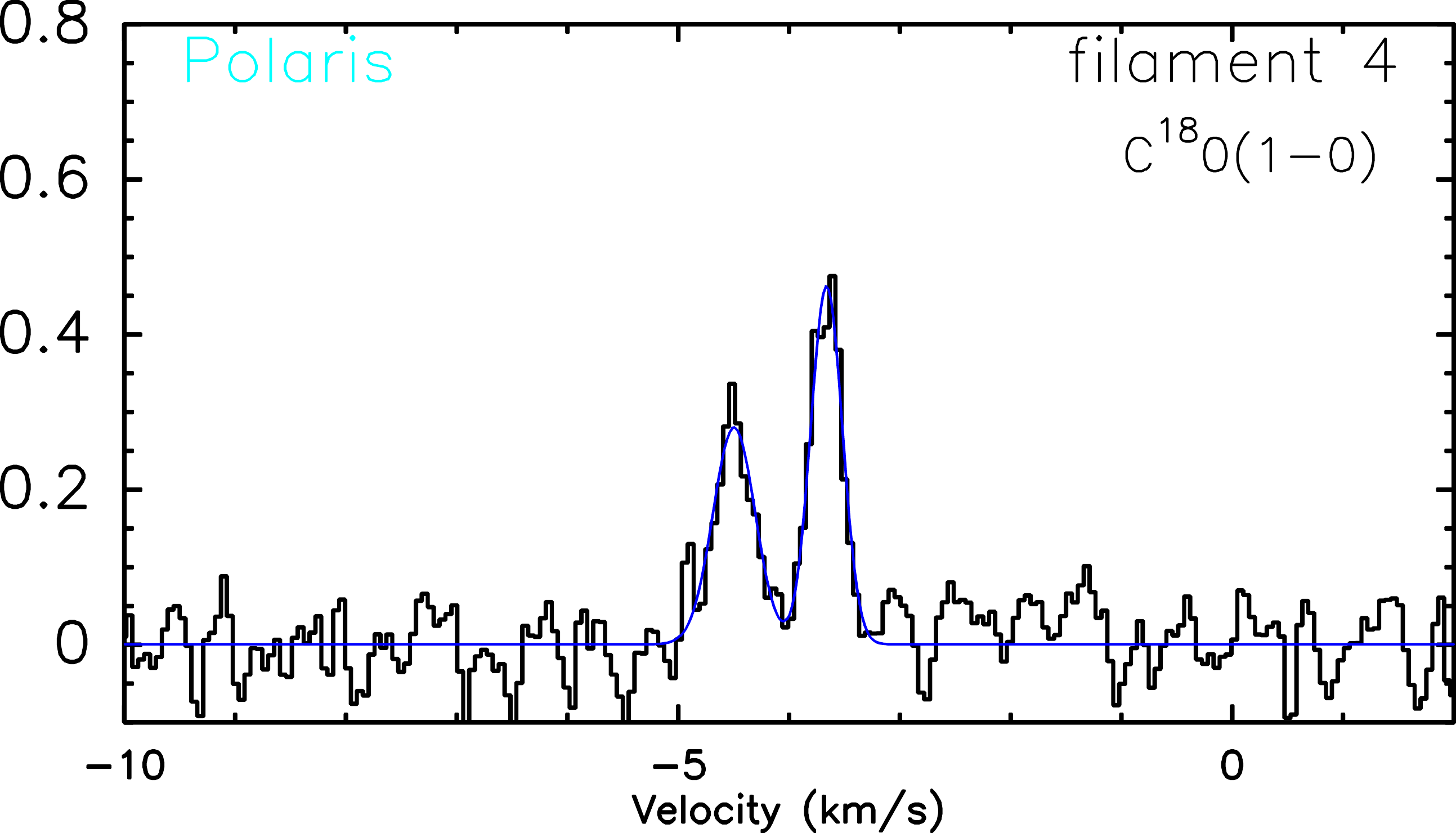}}  
       \resizebox{6.cm}{!}{
     \includegraphics[angle=0]{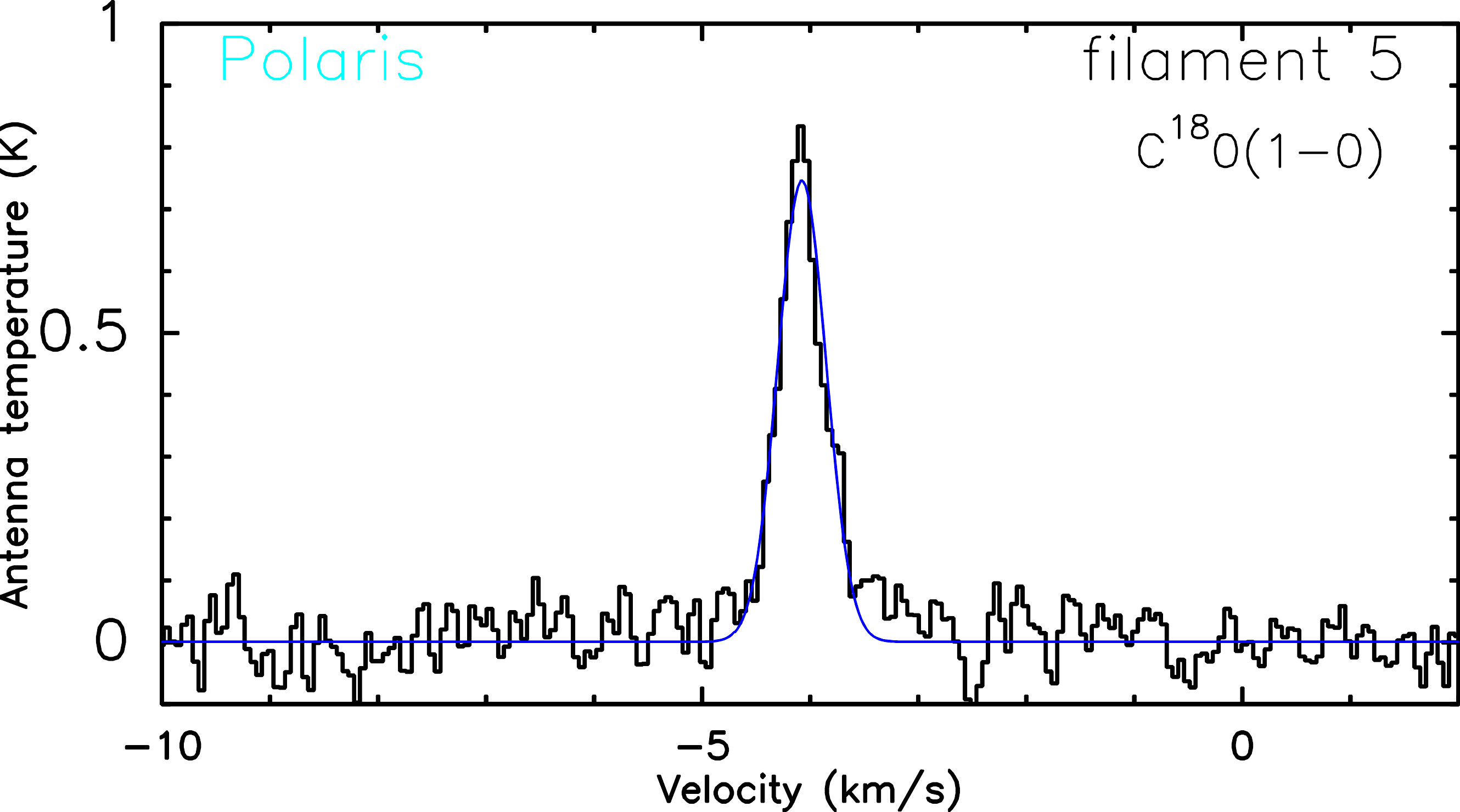}}   
              \resizebox{6.cm}{!}{
     \includegraphics[angle=0]{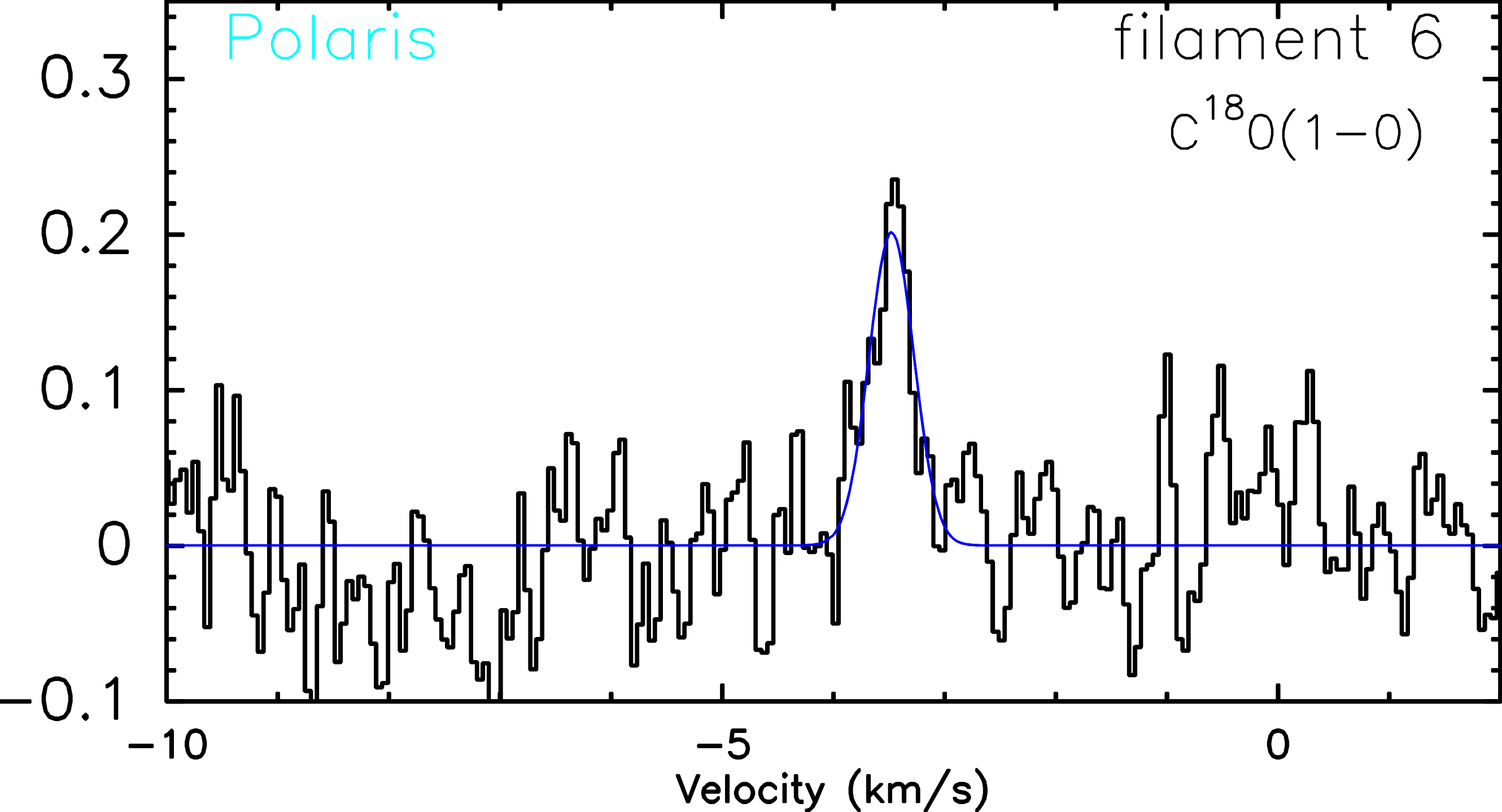}}  
                   \resizebox{6.cm}{!}{
     \includegraphics[angle=0]{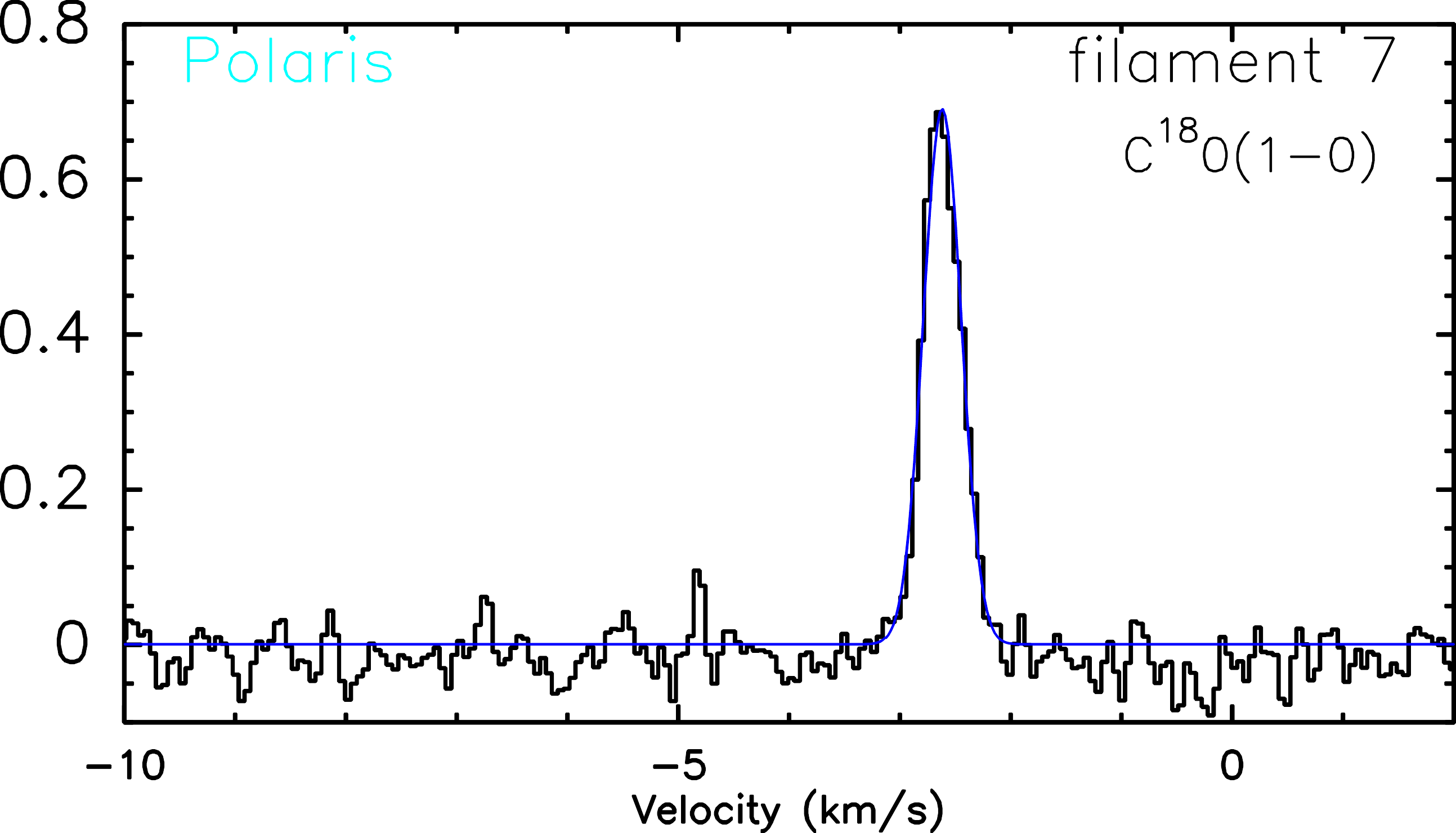}}  
  \caption{C$^{18}$O(1--0) spectra  observed toward 6 filaments in Polaris. The corresponding Gaussian fits are highlighted in blue. {\bf Note 1:} The spectrum observed toward filament 4 shows two velocity components. Both velocity components have similar linewidths (0.19~km/s and 0.14~km/s, respectively), they are separated by $\sim$0.8~km/s. They most probably belong to two filaments which are seen on the same line-of-sight (but a large map would be  needed to study the kinematics of the field). The  component observed at $-$4.5~km/s has been taken to be representative of filament 4.  }
              \label{PolC18Ospectra}
    \end{figure*}

    The  N$_{2}$H$^{+}$(1--0) and C$^{18}$O(1--0)/(2--1) spectra observed toward the sample of filaments studied in this work are shown in  Appendix~\ref{VelComp}. 
       The 23 filaments detected in  N$_{2}$H$^{+}$(1--0) in Aquila, IC5146, and Polaris all show a single velocity component 
     (cf. Fig.~\ref{ICN2H+spectra}, Fig.~\ref{AqN2H+spectra}/\ref{AqN2H+spectra2} and Fig.~\ref{PolN2H+spectra}). This is consistent with the view
     that the filaments are velocity coherent structures \citep[e.g.,][]{Hacar2011}. 
     While 
        most of the filaments detected  in  C$^{18}$O  show one velocity component,
    some of them  show multiple (two or three) C$^{18}$O velocity components (such as filaments 20, 22, 23, 25, 26 in Aquila and filament 4 in Polaris -- see Fig.~\ref{Aq_C18Ospectra}/\ref{Aqu_SW} and Fig.~\ref{PolC18Ospectra}).
    Some of these multiple C$^{18}$O velocity components  
     may be due to   line-of-sight mixing of emissions  not necessarily  tracing the selected filament, but  background and/or foreground structures (especially in the case of the Aquila Rift which is a region close to the Galactic plane).

     Assuming that interstellar  filaments are velocity-coherent structures,   the most likely velocity component associated with  each filament  was  considered  (selecting the velocity component with the velocity  closest  to that  of  neighboring filaments). Its properties were then used  in the  discussion    of the results presented in this paper (see detailed explanations in the captions of the figures presenting the observed spectra).

\begin{figure}[!h]
   \centering
     \resizebox{8.cm}{!}{
     \includegraphics[angle=0]{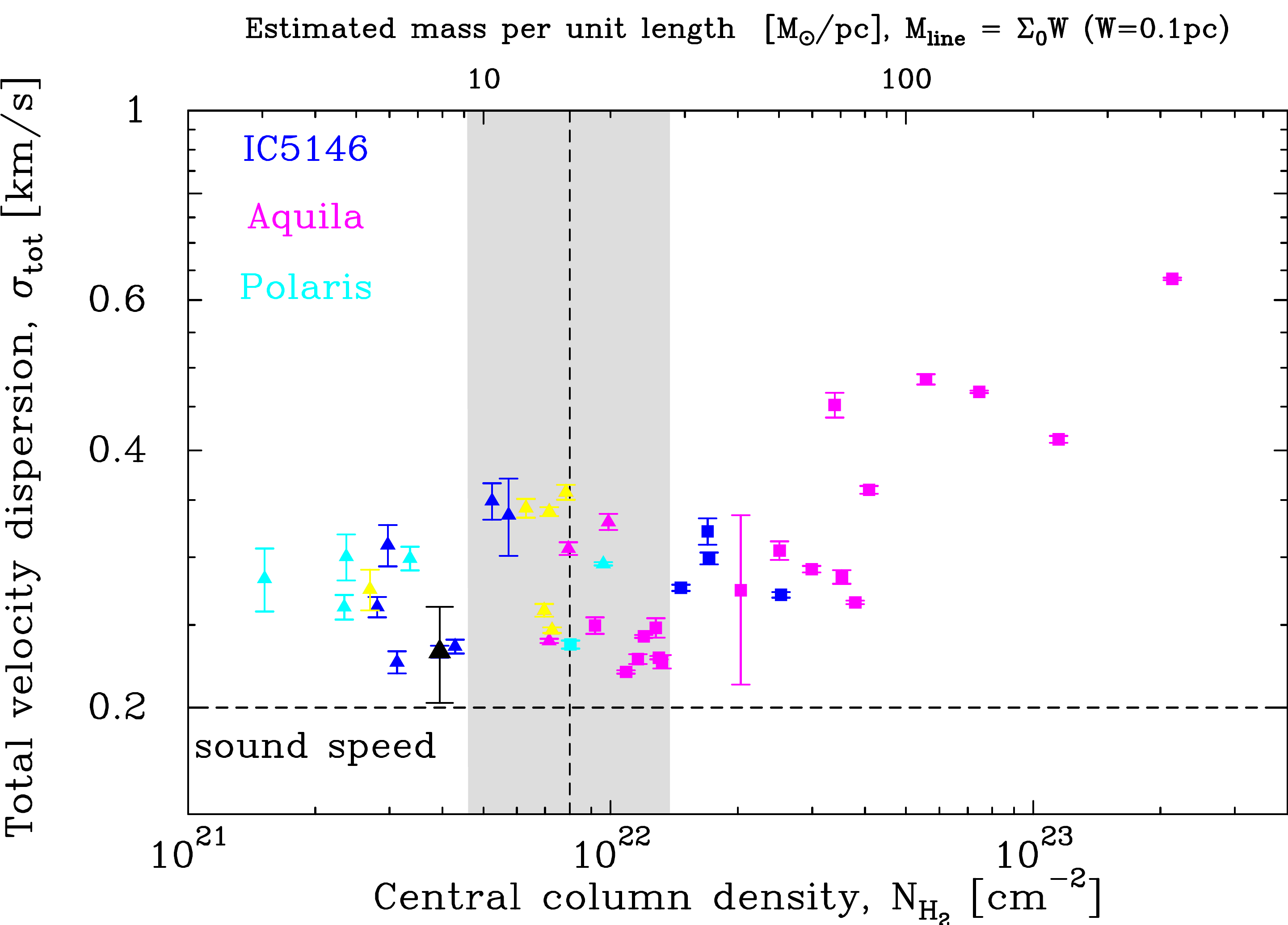}}
          \caption{Total velocity dispersion versus  observed central  column density. Similar  plot to Fig.~\ref{VelDisp_coldens}  for the  filaments observed in IC5146, Aquila and Polaris.  The  six C$^{18}$O spectra showing two or three velocity components are  flagged in yellow. } 
   \label{VelDisp_1comp}%
    \end{figure}

    In order to investigate how the uncertainty  in selecting the relevant velocity  component  affects our  results, the 6 filaments with spectra showing more than one C$^{18}$O velocity component were flagged in  the sample. %
    Fig.~\ref{VelDisp_1comp} shows the  total velocity dispersion  as a function of  central column density for the resulting  subsample of 38 filaments, where the velocity dispersions of the filaments inferred from the  C$^{18}$O spectra (which show multiple velocity components) corresponding to  filaments  20, 22, 23, 25, 26 in Aquila and 4 in Polaris are flagged in yellow. %
Our general     conclusions on filament properties, such as total velocity dispersion, central column density, and  mass per unit length 
remain unchanged whether or not we take the 6 flagged filaments into account.

     \vspace{2.cm}
    
    \begin{figure*}
   \centering
     \resizebox{6.cm}{!}{
     \includegraphics[angle=0]{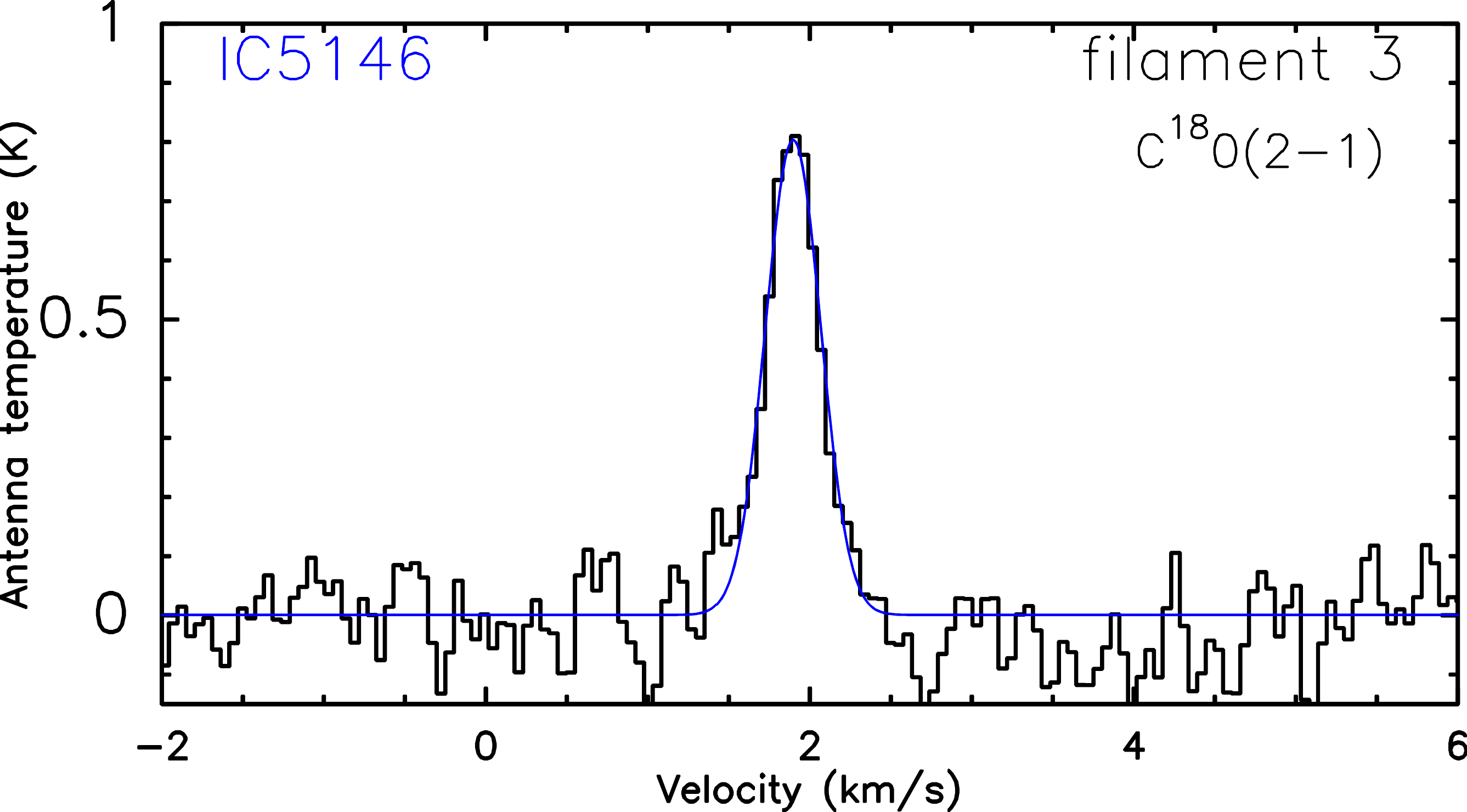}}  
         \resizebox{6.cm}{!}{
     \includegraphics[angle=0]{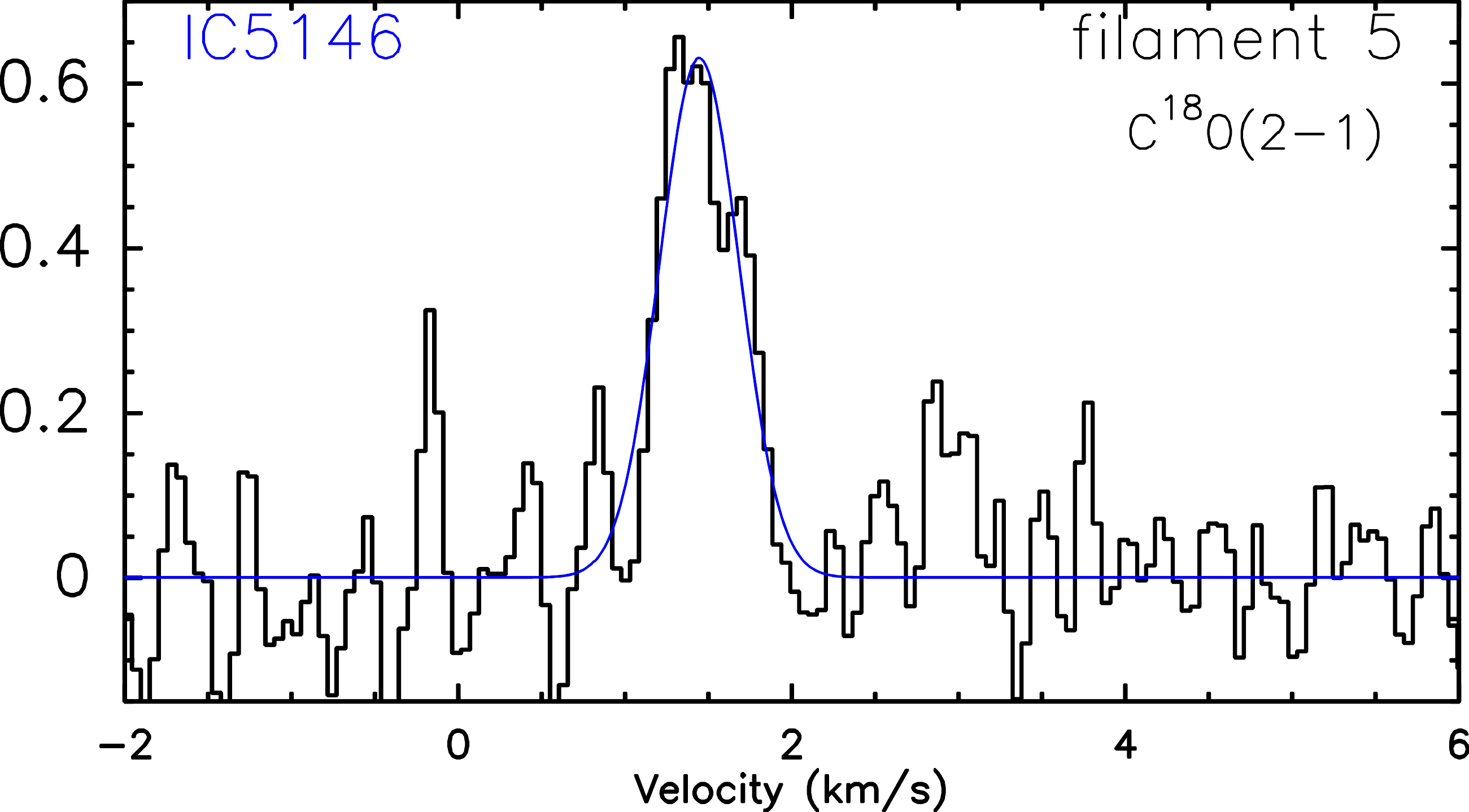}}   
              \resizebox{6.cm}{!}{
     \includegraphics[angle=0]{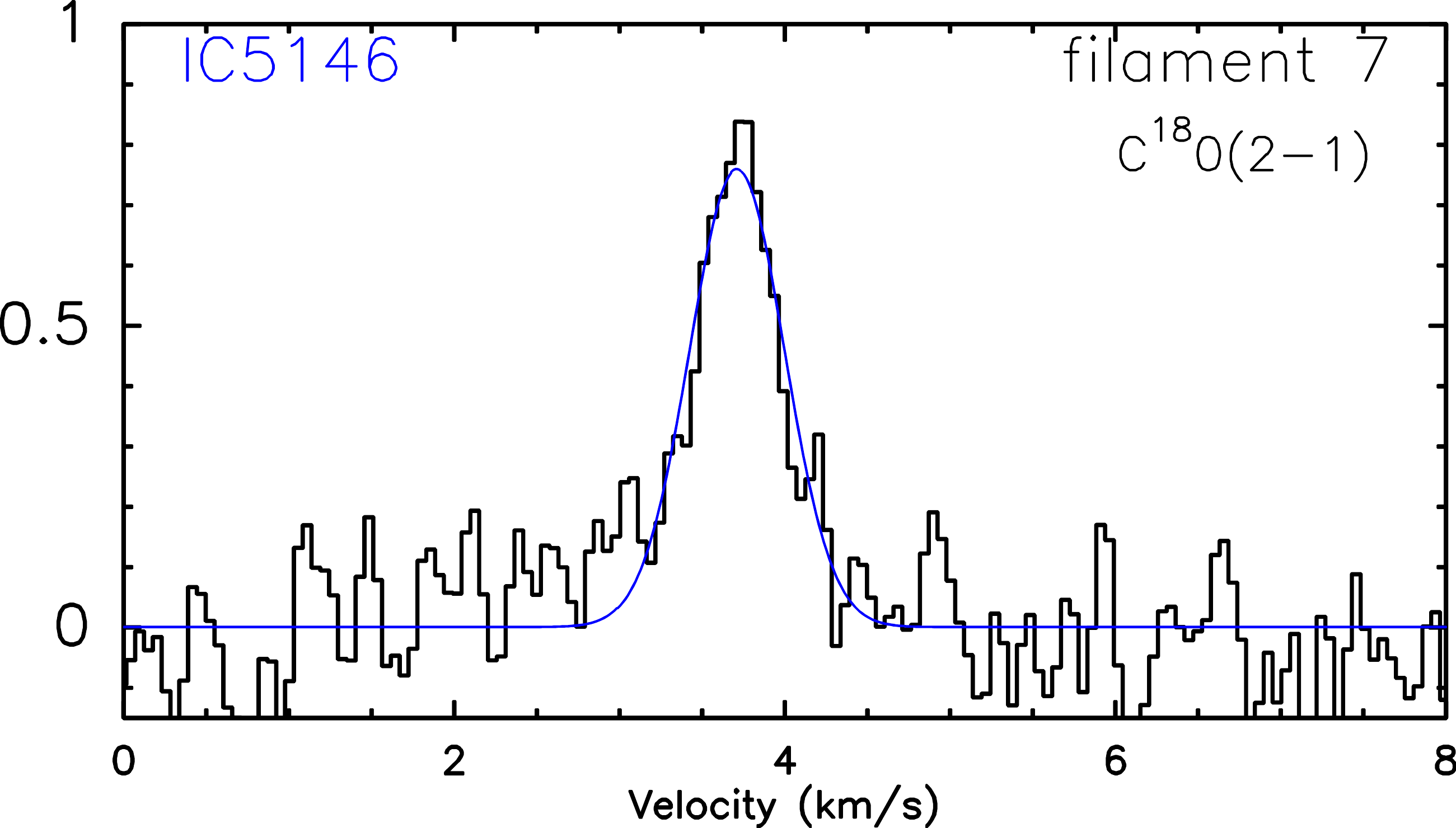}}  
        \resizebox{6.cm}{!}{
        \includegraphics[angle=0]{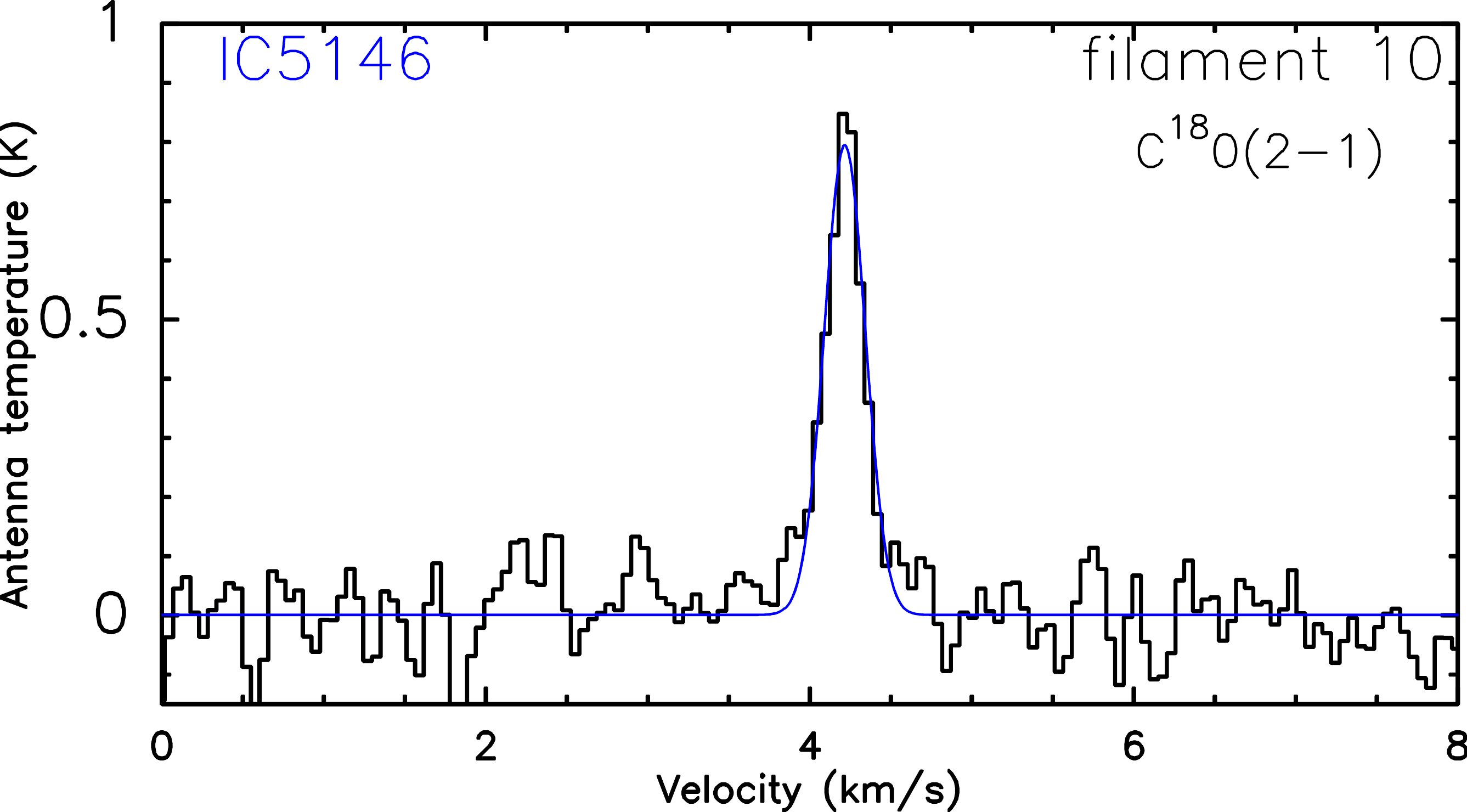}}  
         \resizebox{6.cm}{!}{
     \includegraphics[angle=0]{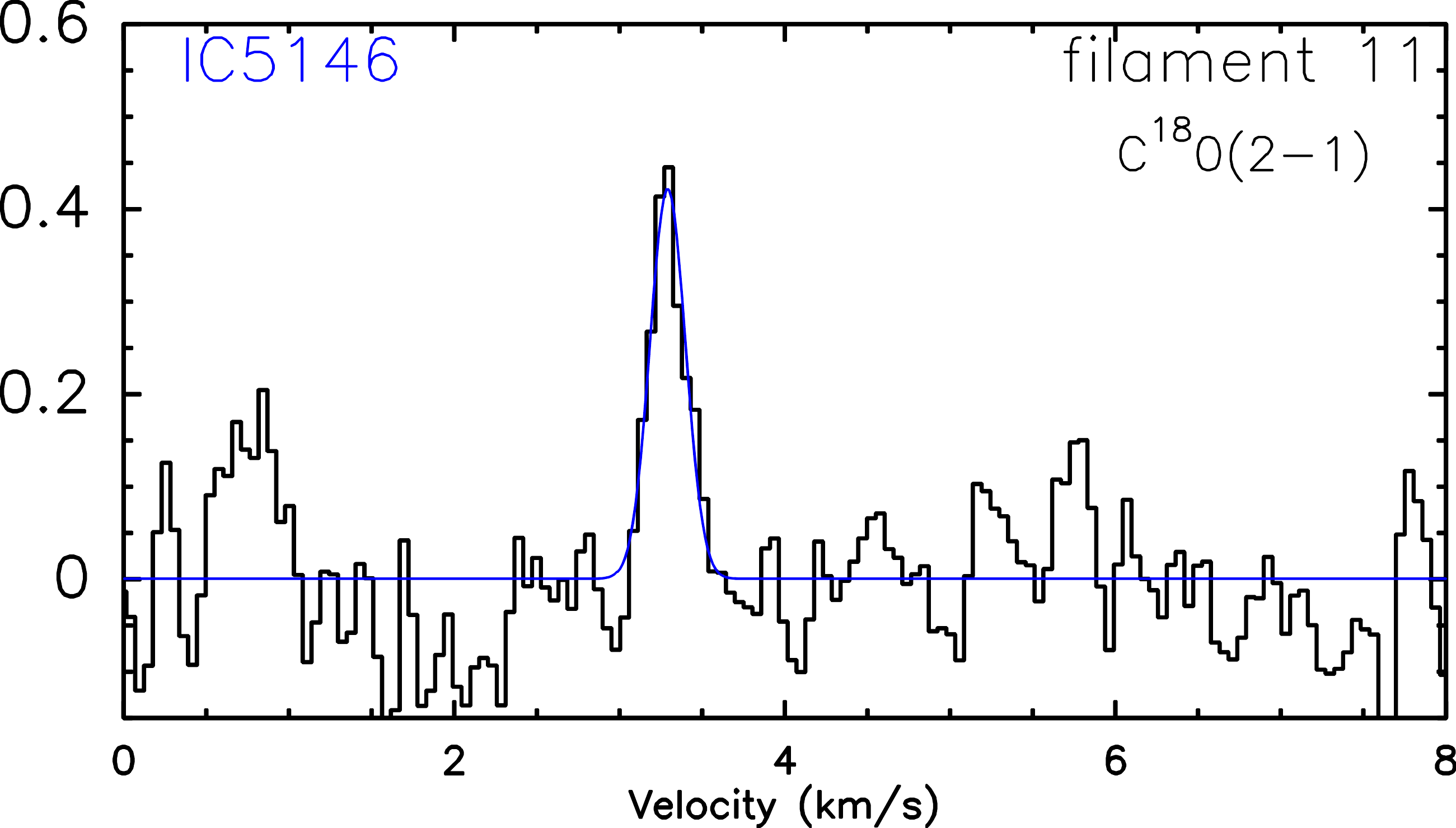}}   
              \resizebox{6.cm}{!}{
     \includegraphics[angle=0]{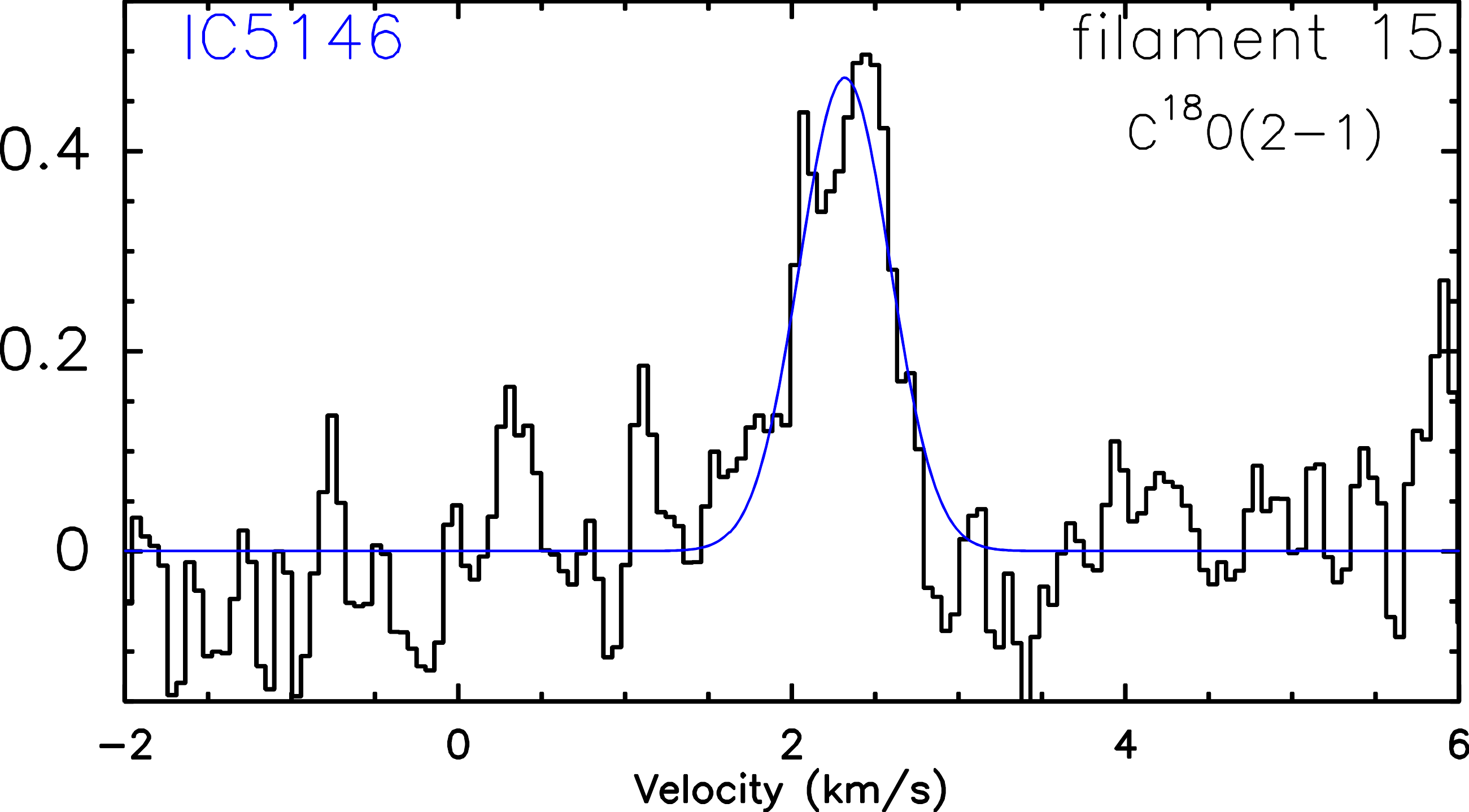}}  
  \caption{ C$^{18}$O(2--1) spectra  observed toward 6 filaments in IC5146. The filament numbers are indicated at the upper right  of each panel. The corresponding single-component Gaussian fits are highlighted in blue.   }
              \label{ICC18Ospectra}
    \end{figure*}


  \begin{figure*}
   \centering
        \resizebox{11.cm}{!}{
      \includegraphics[angle=0]{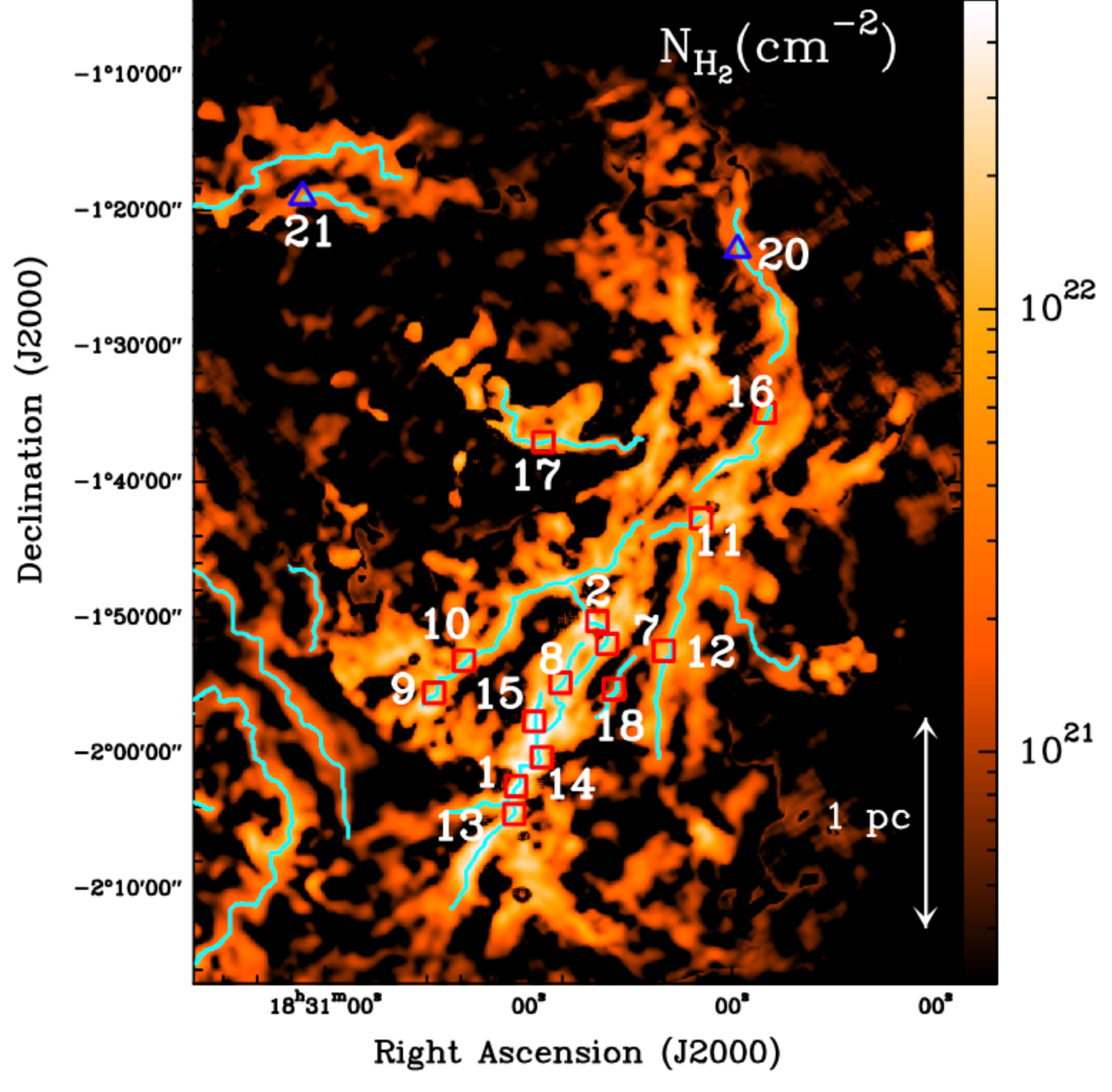}}
  \caption{ 
  Curvelet component of the column density map of the Aquila field around the Serpens South filaments taken from   \citet{Andre2010}. The curvelet component is obtained using a morphological component analysis algorithm \citep[MCA, from][]{Stark2003} which enhances the contrast of the filamentary structure against the more diffuse background of the cloud. The positions  of the observed spectra are plotted in red squares and blue triangles for N$_{2}$H$^{+}$ and C$^{18}$O,  respectively. The numbers correspond to the filaments listed in Table~1 and are given at the upper right of the spectra in Fig.~\ref{AqN2H+spectra} and some of the spectra of  Fig.~\ref{Aq_C18Ospectra}.  } 
         \label{Aqu}%
    \end{figure*}

    \begin{figure*}
   \centering
                   \resizebox{6.cm}{!}{
     \includegraphics[angle=0]{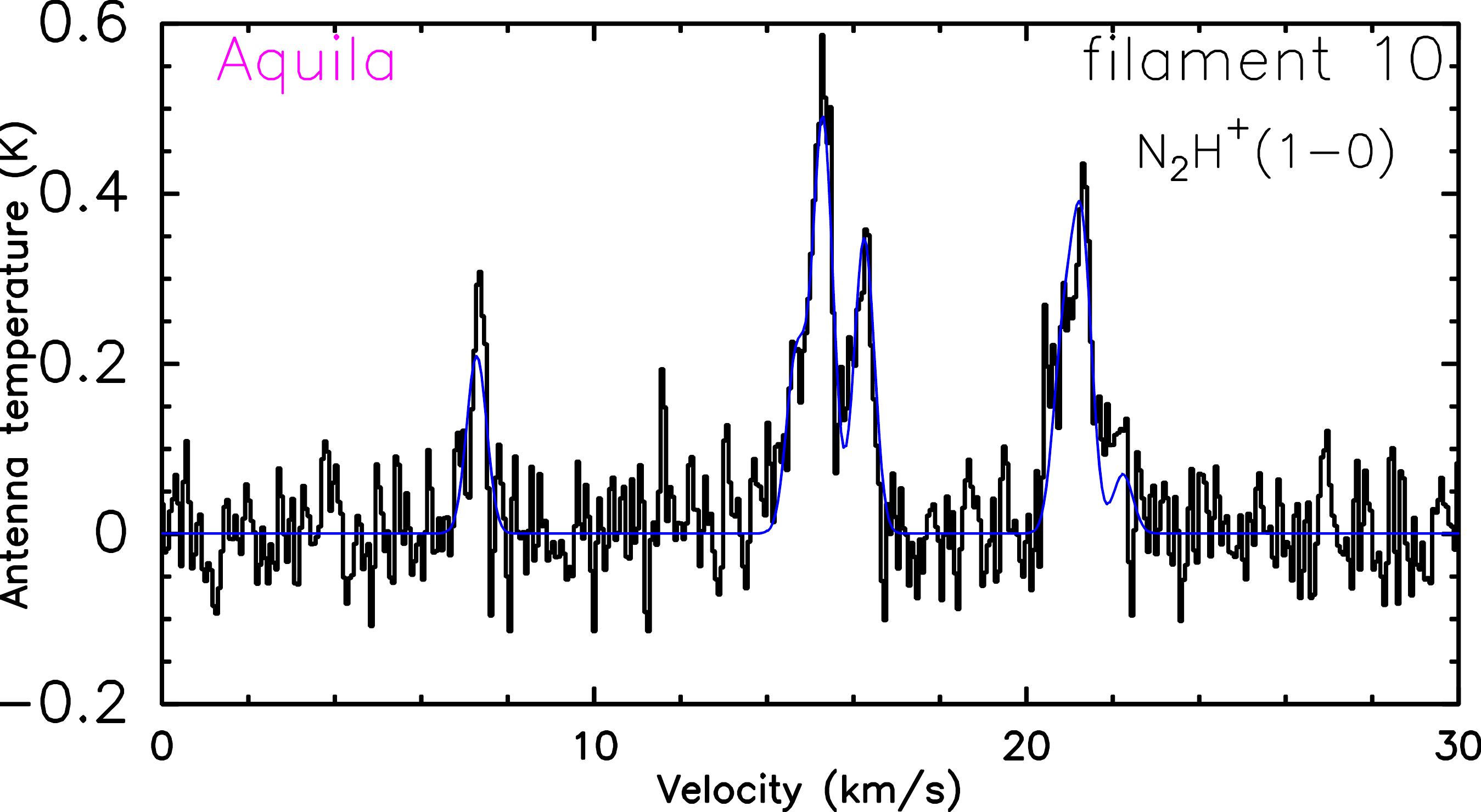}}
                   \resizebox{6.cm}{!}{
     \includegraphics[angle=0]{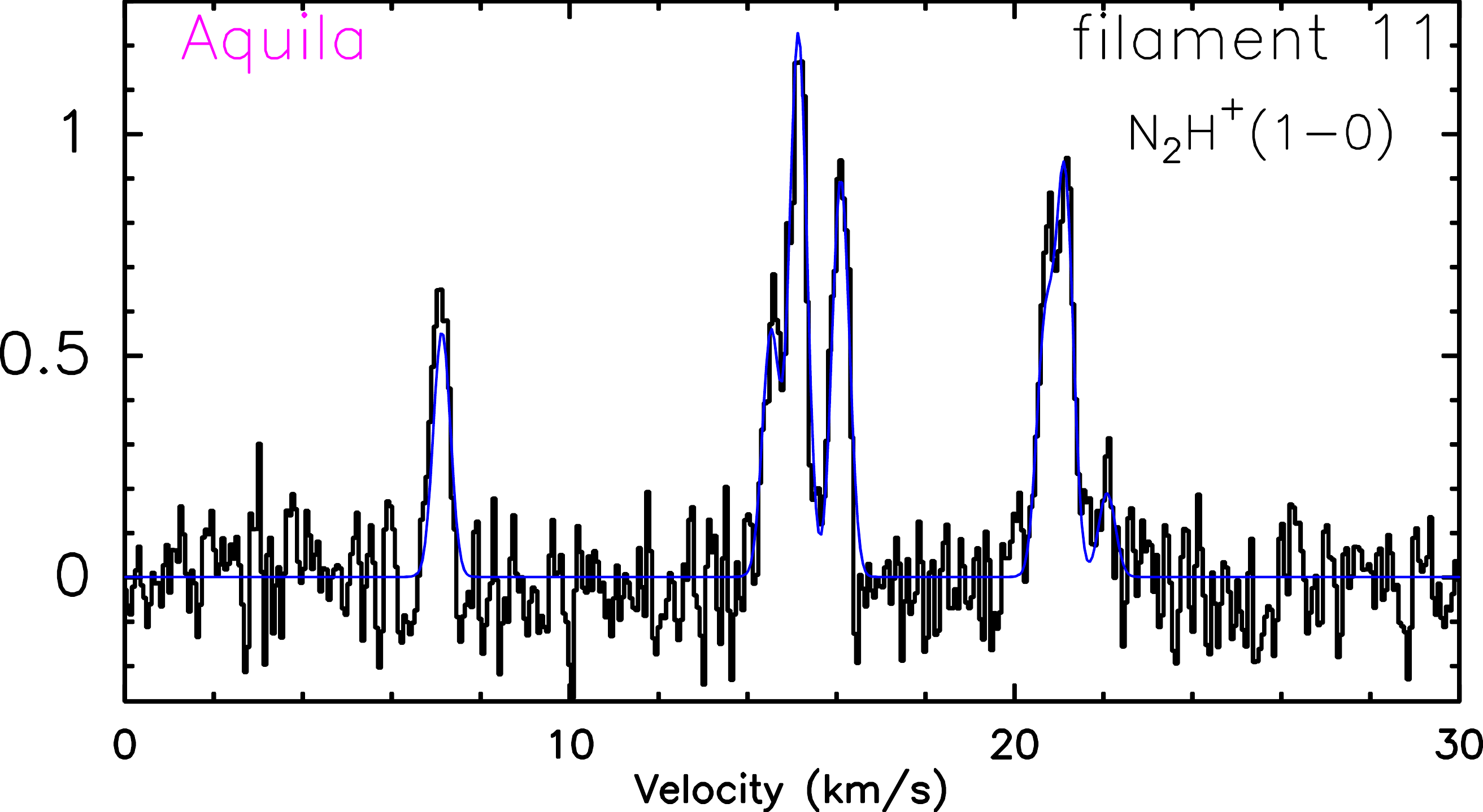}}  
                   \resizebox{6.cm}{!}{
     \includegraphics[angle=0]{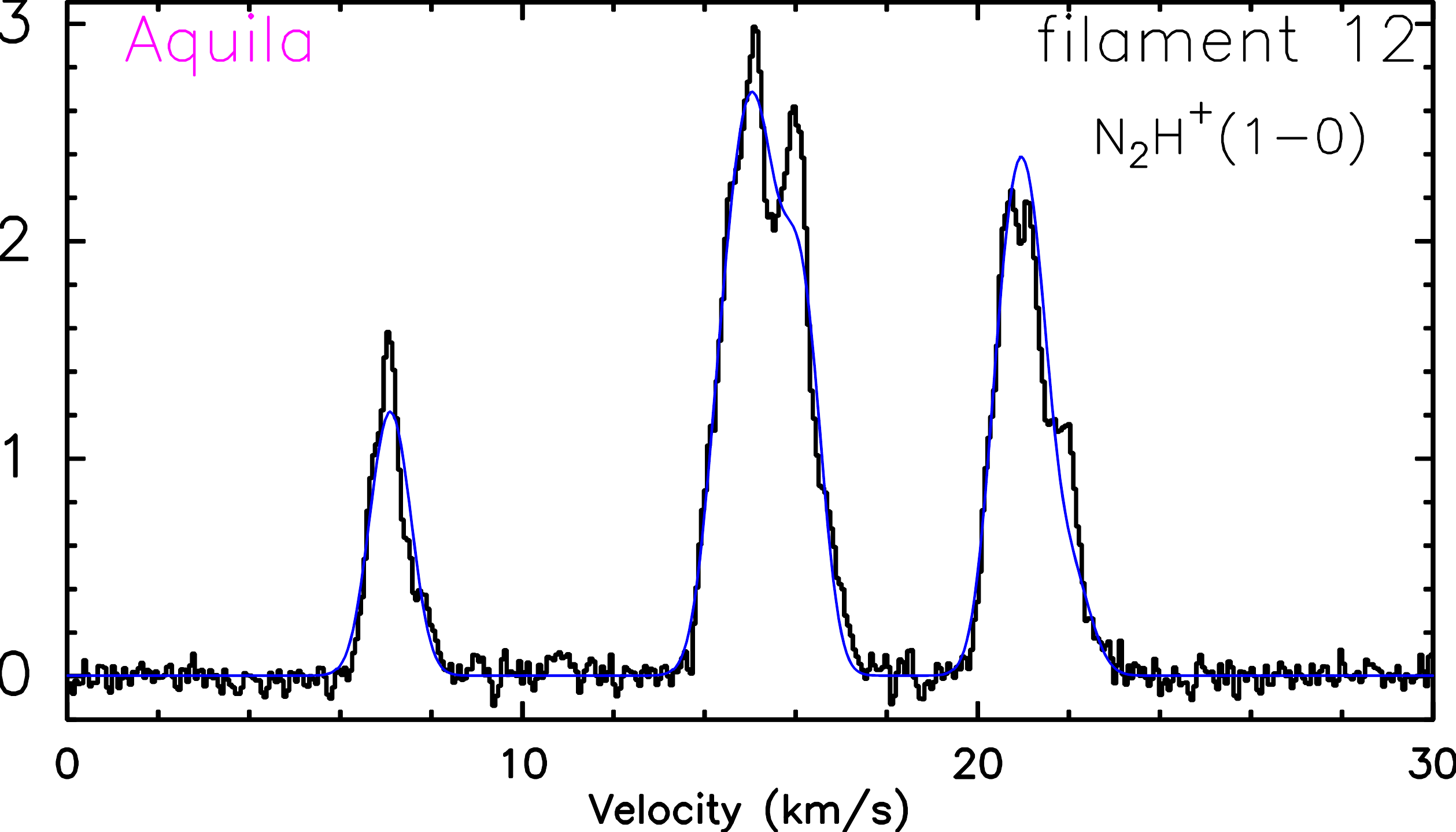}}  
                   \resizebox{6.cm}{!}{
     \includegraphics[angle=0]{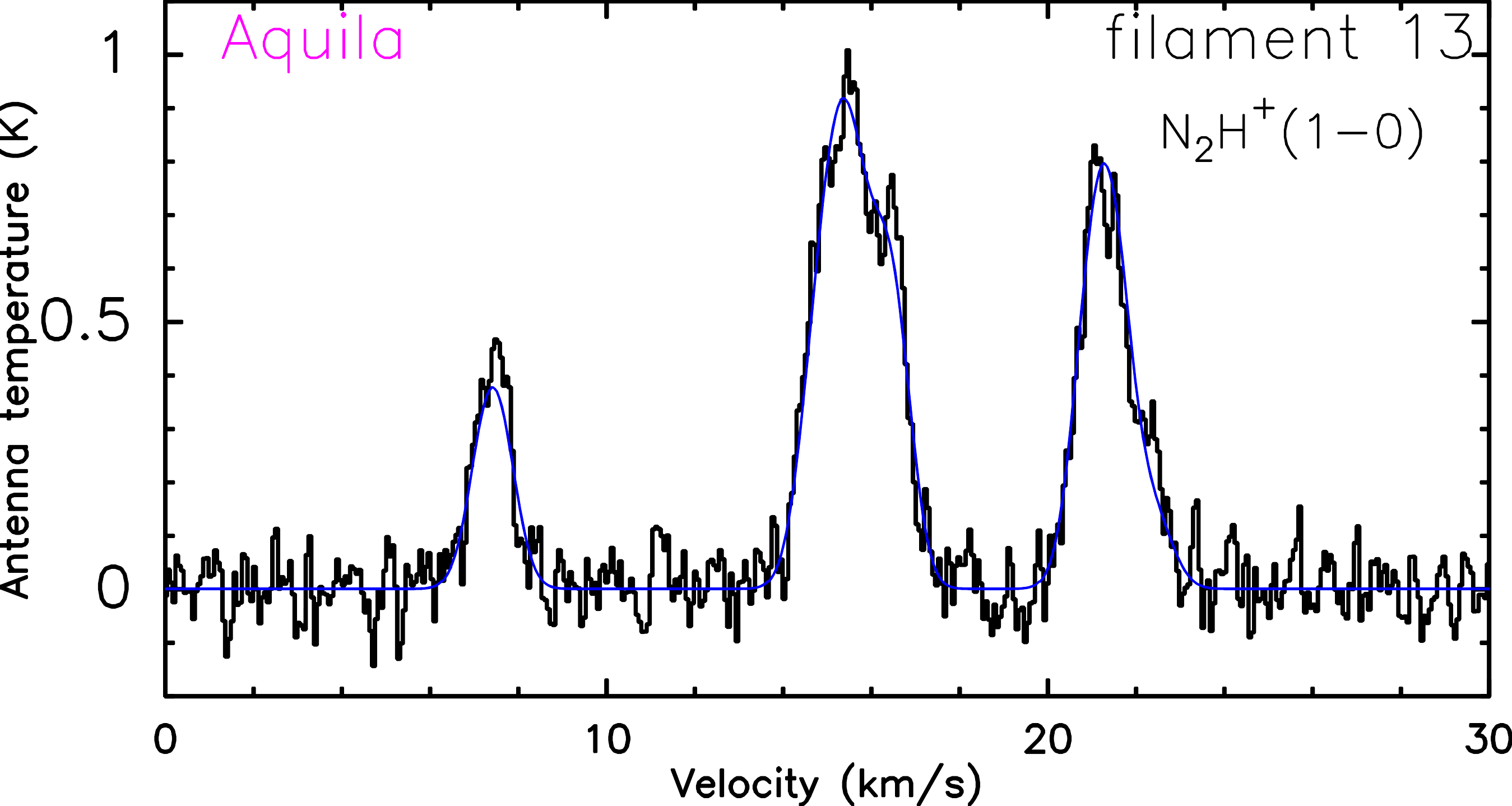}}  
                   \resizebox{6.cm}{!}{
     \includegraphics[angle=0]{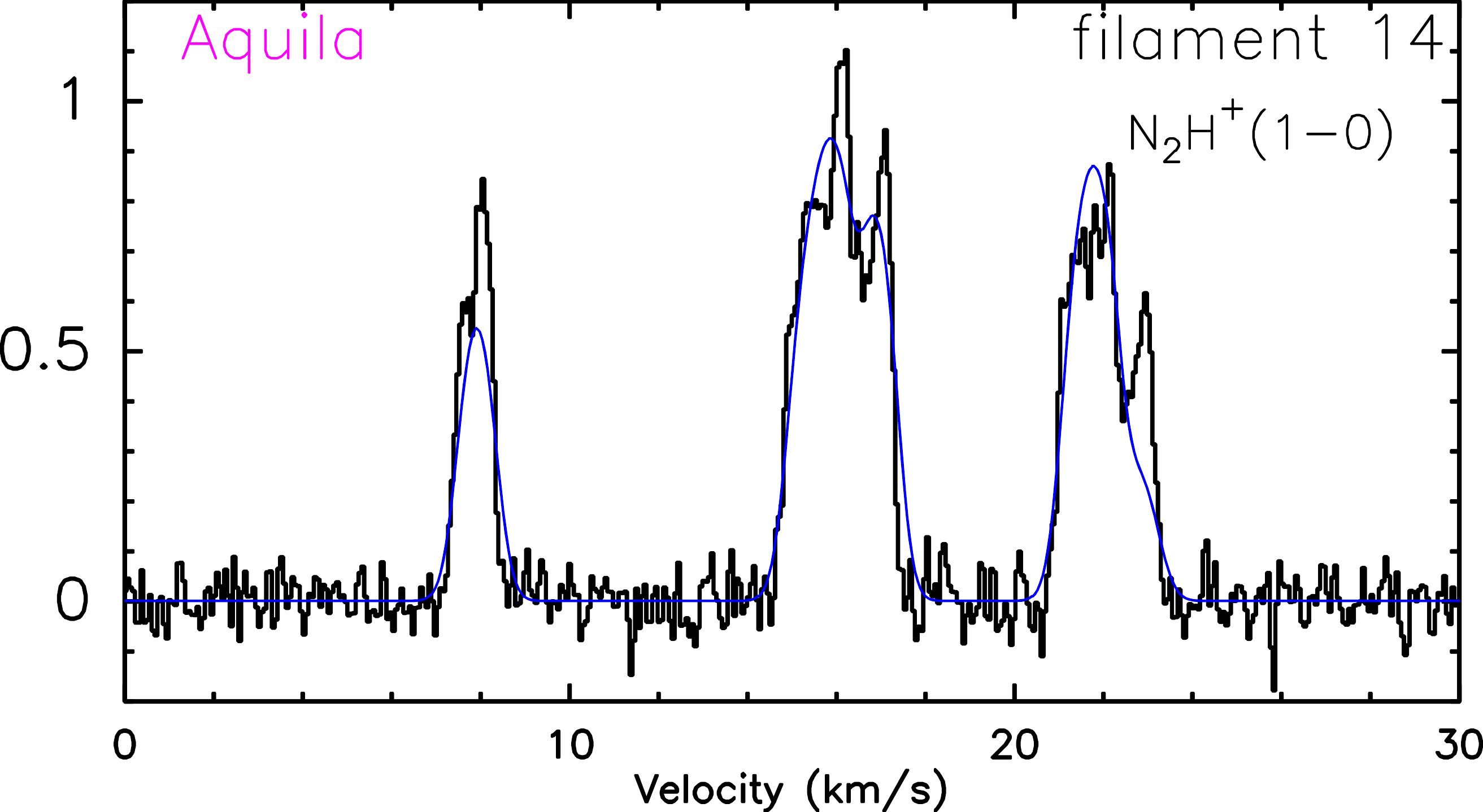}}  
                   \resizebox{6.cm}{!}{
     \includegraphics[angle=0]{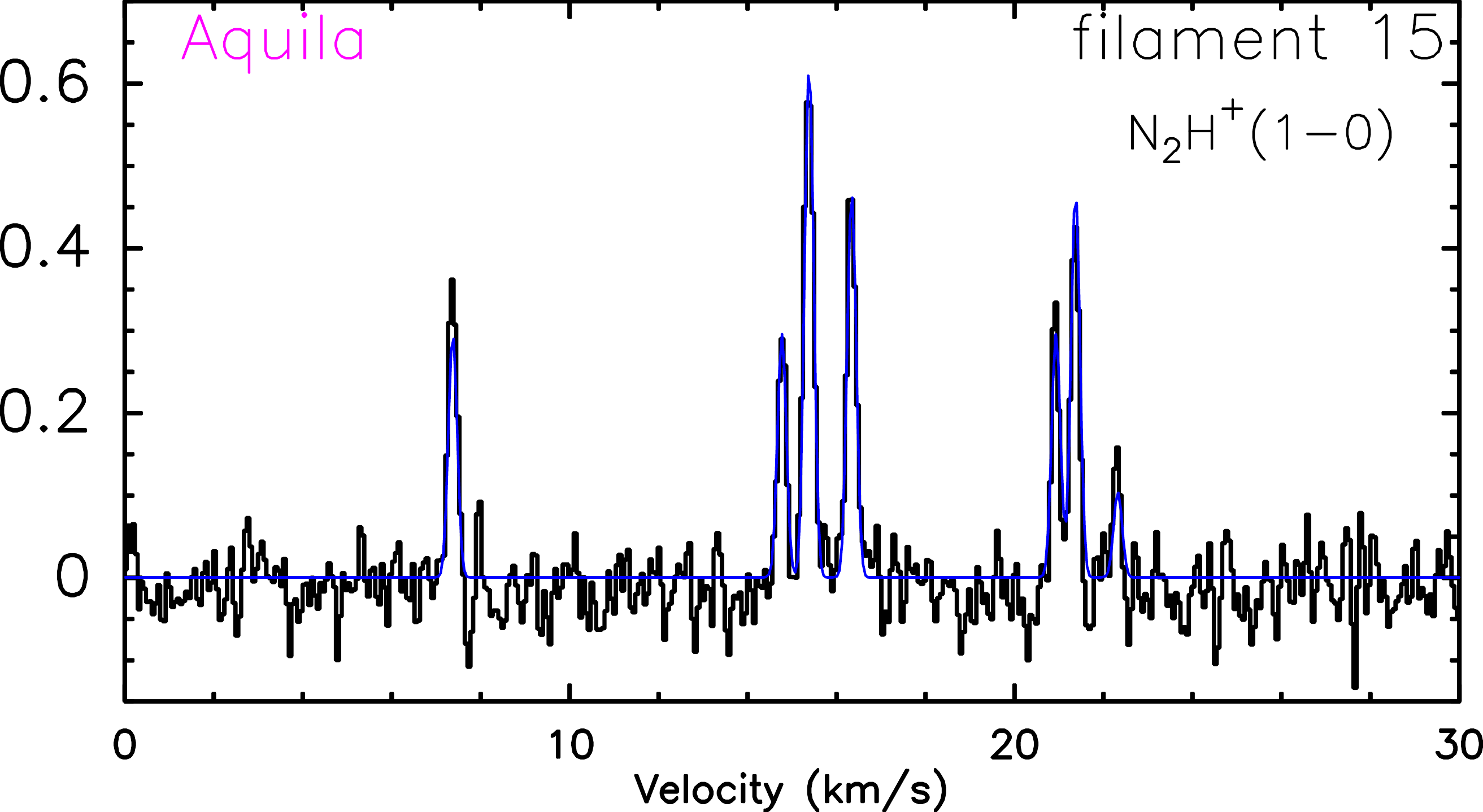}}  
                   \resizebox{6.cm}{!}{
     \includegraphics[angle=0]{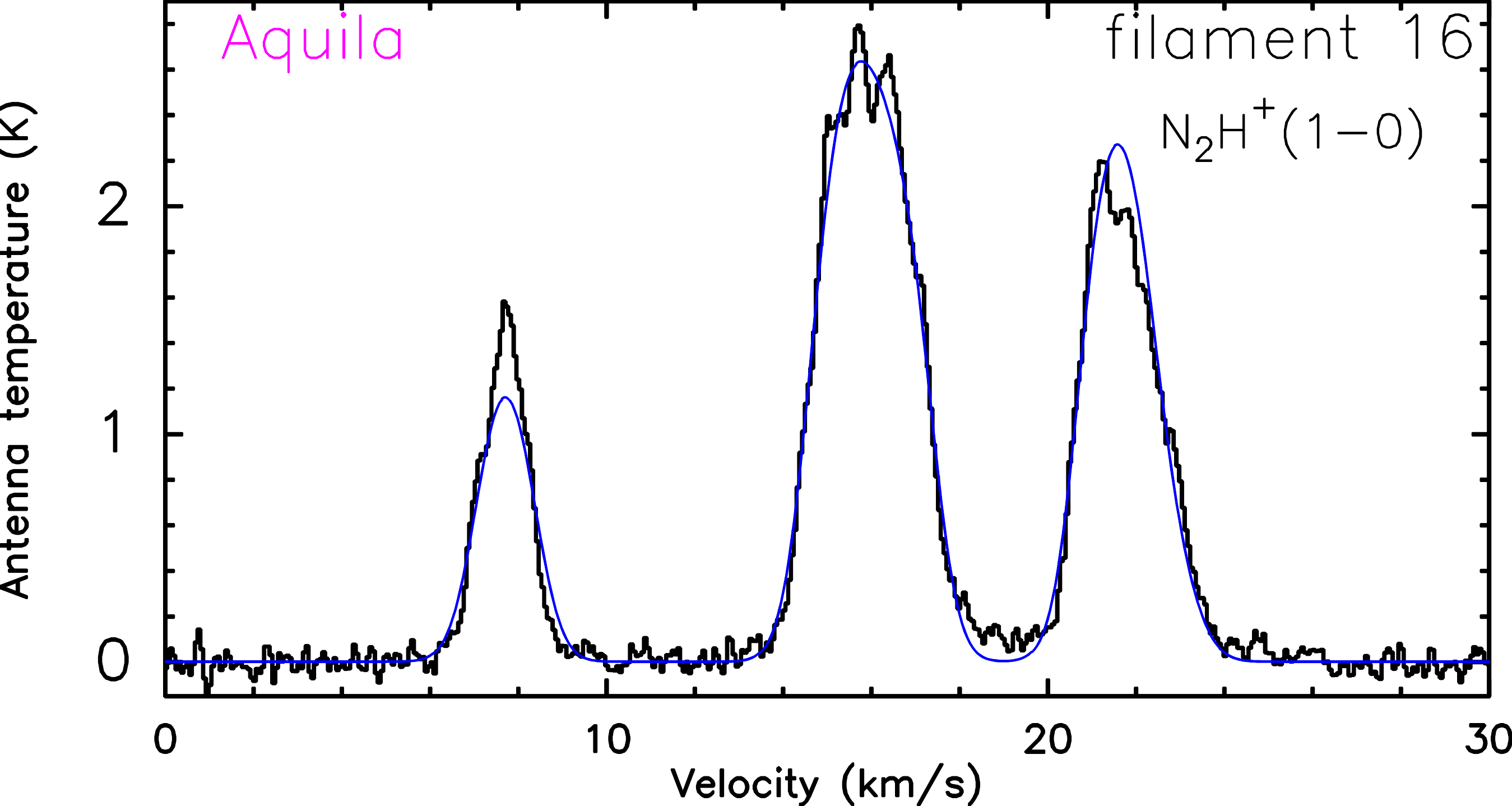}}  
                   \resizebox{6.cm}{!}{
     \includegraphics[angle=0]{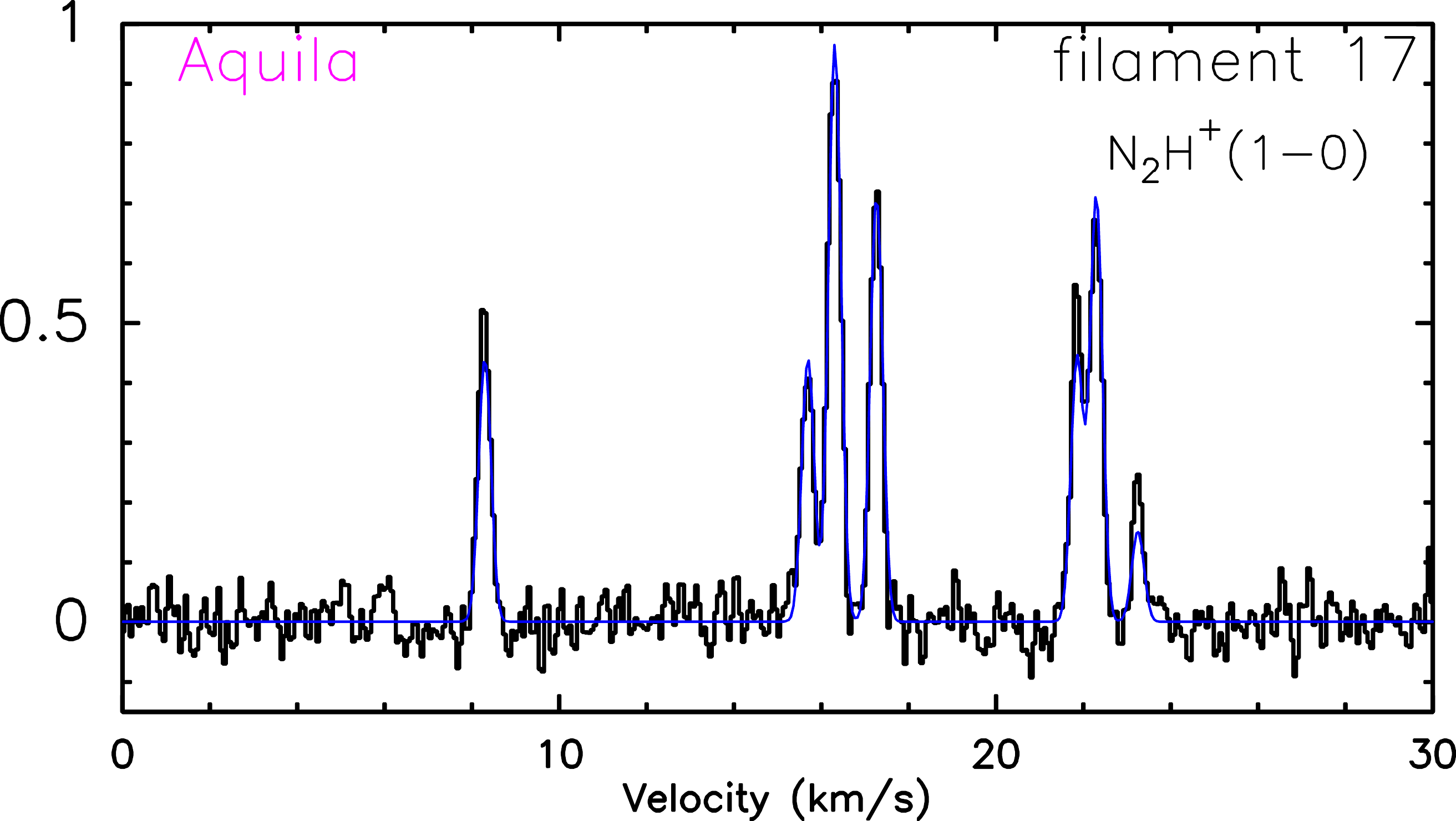}}  
                   \resizebox{6.cm}{!}{
     \includegraphics[angle=0]{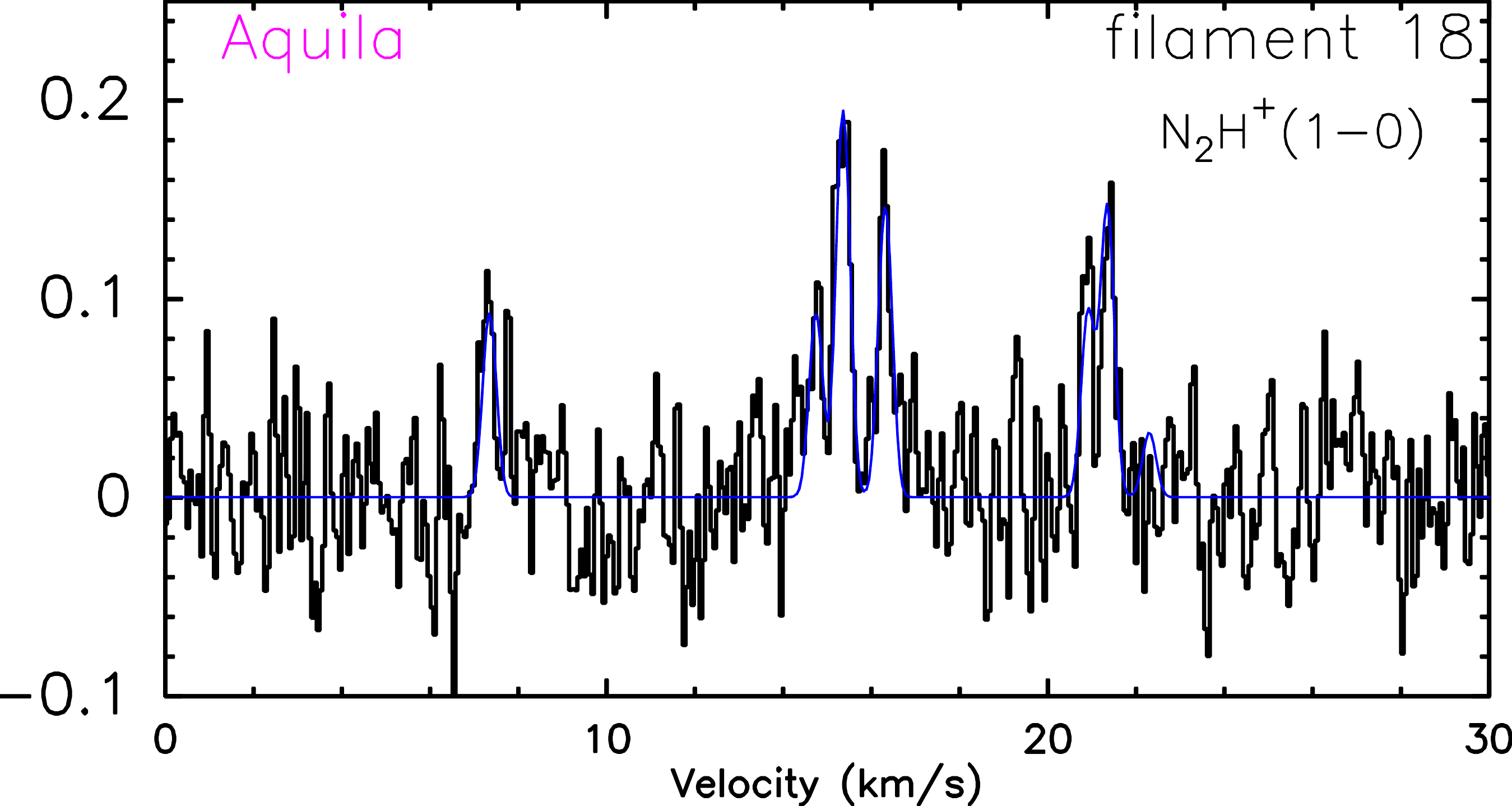}}    
  \caption{Continuation of Fig.~\ref{AqN2H+spectra} presenting the N$_{2}$H$^{+}$(1--0) spectra observed toward 9 filaments in Aquila. }
              \label{AqN2H+spectra2}
    \end{figure*}

  \begin{figure*}
   \centering
     \resizebox{8.cm}{!}{
   \vspace{-0.4cm}
    \includegraphics[angle=0]{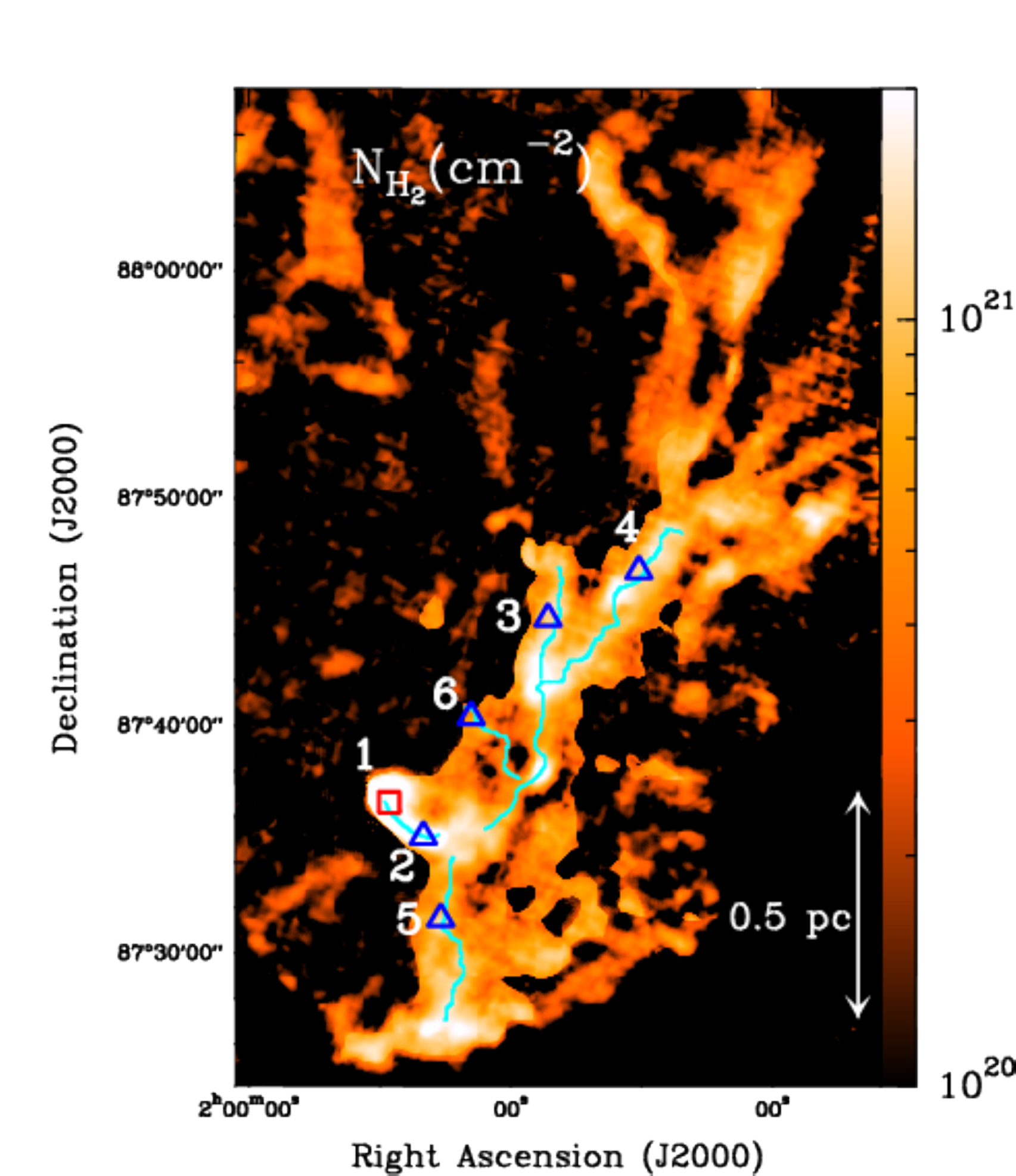}}
  \caption{ 
 Curvelet component of the column density map of a subregion in the Polaris Cloud  taken from   \citet{Andre2010}. The positions  of the observed spectra are plotted in red squares and blue triangles for N$_{2}$H$^{+}$ and C$^{18}$O,  respectively. The numbers correspond to the filaments listed in Table~1.   }
         \label{Pol}%
    \end{figure*}
    
       \end{appendix}

\end{document}